\documentclass[aps,prl,twocolumn,10pt,floatfix,superscriptaddress]{revtex4-2}

\usepackage{amsmath,amssymb}
\usepackage{graphicx}
\usepackage[hidelinks]{hyperref}

\usepackage{color}

\graphicspath{{./}}
\newcommand{\bk}{\mathbf{k}}
\newcommand{\B}{\mathbf{B}}

\newcommand{\EF}{E_F}

\makeatletter
\providecommand{\bstctlcite}[1]{%
  \@bsphack
  \@for\@citeb:=#1\do{\edef\@citeb{\expandafter\@firstofone\@citeb}%
    \if@filesw\immediate\write\@auxout{\string\citation{\@citeb}}\fi}%
  \@esphack}
\makeatother

\begin{document}
% References include titles (PRL strongly encourages consistency).

\title{Quantum Oscillation Signatures of \texorpdfstring{$\mathbb{Z}_2$}{Z2} Monopole Charge in Nodal-Ring Semimetals}

\author{Aravind Karthigeyan}
\affiliation{Department of Physics, The University of Texas at Austin, Austin, Texas 78712, USA}
\author{Yoonseok Hwang}
\affiliation{Blackett Laboratory, Imperial College London, London SW7 2AZ, United Kingdom}
\author{Bohm-Jung Yang}
\affiliation{Department of Physics and Astronomy, Seoul National University, Seoul 08826, Korea}
\affiliation{Center for Theoretical Physics (CTP), Seoul National University, Seoul 08826, Korea}
\affiliation{Institute of Applied Physics, Seoul National University, Seoul 08826, Korea}
\author{Junyeong Ahn}
\email[Contact author: ]{junyeong.ahn@austin.utexas.edu}
\affiliation{Department of Physics, The University of Texas at Austin, Austin, Texas 78712, USA}

\date{\today}

\begin{abstract}
Topological semimetals host band nodes characterized by quantized invariants that can appear in bulk responses, yet some invariants remain hidden from standard probes. In particular, band nodes can carry secondary topological charges whose transport signatures are still largely unexplored. Here we study three-dimensional nodal-line semimetals in which nodal rings carry both the Berry phase $w_1\pi$ and a $\mathbb{Z}_2$ monopole charge $w_2$. We show that magnetic quantum oscillations, usually treated as a probe of $w_1$, can directly diagnose $w_2$, with the relevant signal selected by the magnetic-field direction. For a field along the ring axis, the inner and outer extremal orbits of the toroidal Fermi surface both encircle the $w_2$-enforced thread and exhibit a topological phase shift $\nu w_2\pi$ in the $\nu$th harmonic, which is accessible through standard phase-resolved quantum-oscillation analysis. By contrast, for a field applied perpendicular to the ring axis, the relevant extremal orbit exhibits the usual $\pi$ phase shift associated with the Berry phase $w_1\pi$, independent of $w_2$. For weak doping, three-dimensional ABC-stacked graphdiyne is predicted to exhibit the proposed oscillations in a field range accessible with present-day high-field facilities.
\end{abstract}
\maketitle

{\it Introduction---}Magnetic quantum oscillations map extremal Fermi-surface areas and encode a phase offset determined by the Berry phase accumulated along the semiclassical cyclotron orbit~\cite{mikitik1999,shoenberg}.
Extracting this phase has become a standard tool for identifying topological semimetals~\cite{armitage2018weyl}.
In nodal-line semimetals, which are three-dimensional metals whose conduction and valence bands touch along closed loops in momentum space, a cyclotron orbit linking the nodal loop acquires a quantized Berry phase $\pi$, producing a robust half-period shift in magnetic quantum oscillations~\cite{fang2016topological,schoop2016,li2018,yang2018quantum,hu2016,vanDelft2018}.
Experimentally, this shift has been extracted from Landau-fan intercepts in the ZrSiS/HfSiS family and related compounds~\cite{schoop2016,ali2016,singha2017,hu2016,vanDelft2018}.

However, this $\pi$ Berry phase does not fully characterize nodal-line topology in systems with spacetime-inversion ($PT$) symmetry and negligible spin-orbit coupling. In such systems, a nodal line can also carry a $\mathbb{Z}_2$ monopole charge $w_2$, defined as the second Stiefel--Whitney class evaluated on a two-dimensional manifold enclosing it~\cite{ahn2019stiefel,ahn2018,fang2015,zhao2016,zhao2017}. The Berry phase on a loop linking the nodal line reflects only the first Stiefel--Whitney class $w_1$, with $\Phi=\pi w_1$ modulo $2\pi$. Nodal lines with $w_2=1$ can be created or annihilated only in pairs, unlike trivial nodal lines with $w_2=0$, which can disappear by shrinking to a point. A monopole nodal line therefore has additional topological robustness~\cite{fang2015,bzdusek2017}, analogous to the role of an integer monopole charge in stabilizing Weyl points. Despite extensive theoretical interest in the second Stiefel--Whitney class~\cite{ahn2019failure,bouhon2020geometric,unal2020topological,salerno2020floquet,lee2020two,liu2021nematic,chen2021graphyne,pan2022two,liu2022second,guan2022landau,qian2023stable,yu2023euler,kwon2024quantum,yue2024stability,shiozaki2024discrete,shiozaki2025z2,lee2025euler,sato2025three,kobayashi2026euler}, monopole nodal lines~\cite{wu2019non,tiwari2020,liu2025correspondence,li2025general,liu2025floquet}, their connection to higher-order topological phases~\cite{wang2019,bouhon2019,ahn2019symmetry,ahn2020higher,li2024phononic,wu2025breakdown}, and progress in experimental realizations~\cite{pan2023real,xue2023stiefel,ma2024observation,xiang2024}, no bulk transport signature has yet been identified that distinguishes a $w_2{=}1$ line from an ordinary $w_2{=}0$ line.

\begin{figure}[!htb]
\includegraphics[width=\linewidth]{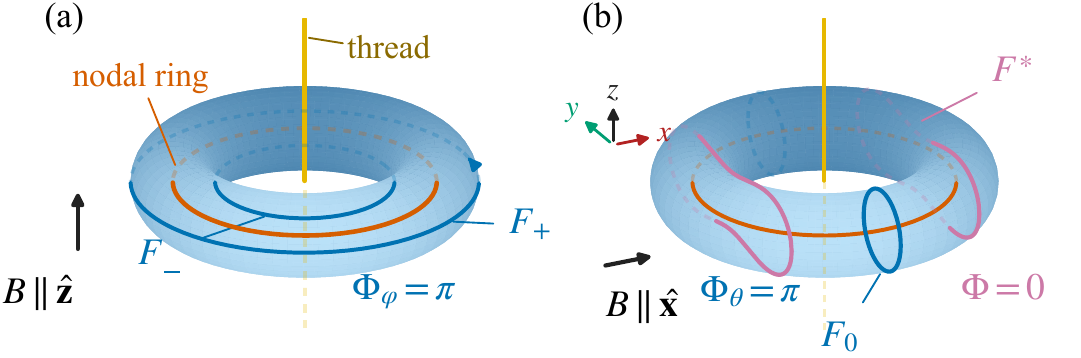}
\caption{Cyclotron-orbit geometry for a toroidal Fermi surface around a $\mathbb{Z}_2$ monopole nodal ring. (a) For $\B\parallel\hat{z}$, the inner and outer extremal orbits both encircle the thread and carry the toroidal Berry phase \(\Phi_{\phi}=\pi\), giving the two Onsager frequencies $F_{\pm}$. (b) For $\B\parallel\hat{x}$, linking orbits at $k_x=0$ carry $\Phi_{\theta}=\pi$, while nonlinking maximal orbits at $k_x=\pm k_x^{\ast}$ have $\Phi=0$ and give $F^{\ast}$.}
\label{fig:schematic}
\end{figure}
\vspace{-0.5em}
%letter
In this paper, we show that magnetic quantum oscillations provide a direct bulk diagnostic of this monopole charge. A monopole nodal ring has a richer topological structure than an ordinary nodal ring: $w_2=1$ enforces, within the occupied or unoccupied subspace, a nodal line that threads the torus hole, forming a linked ``ring-and-thread'' configuration~\cite{ahn2018,ahn2019failure} shown in Fig.~\ref{fig:schematic}. This thread produces an additional Berry phase along the toroidal direction. Hence, for a magnetic field along the ring axis, both inner and outer extremal orbits encircle the thread and acquire a $\pi$ phase shift, whereas the corresponding orbits of a trivial ring do not. An in-plane field gives a complementary check by probing the familiar Berry phase of orbits linking the nodal ring. We confirm both ring-axis and in-plane diagnostics analytically in a four-band $\bk\cdot\mathbf{p}$ model and numerically through lattice Kubo calculations. We also discuss three-dimensional ABC-stacked graphdiyne as a candidate material platform.

{\it Model---}We begin with a four-band $\bk\cdot\mathbf{p}$ model~\cite{fang2015,ahn2018}
\begin{equation}
H_0(\bk)=\hbar v[\big(k_x-m_0\,t_x\big)s_x+k_y\,t_y s_y+k_z\,s_z],
\label{eq:H0}
\end{equation}
where $s_i$ and $t_i$ are Pauli matrices acting in two independent internal subspaces, which may be viewed, for example, as orbital and sublattice degrees of freedom.
This Hamiltonian has $PT$ symmetry $PTH_0(\bk)(PT)^{-1}=H_0(\bk)$, where $PT=\mathcal{K}$ denotes complex conjugation.
Since the Hamiltonian matrix is real-valued, its eigenvectors can be chosen real, and the occupied sector is classified by two $\mathbb{Z}_2$ invariants of real vector bundles, the first and second Stiefel--Whitney classes $w_{i=1,2}$~\cite{fang2015,ahn2018}.

\begin{figure}
\includegraphics[width=\linewidth]{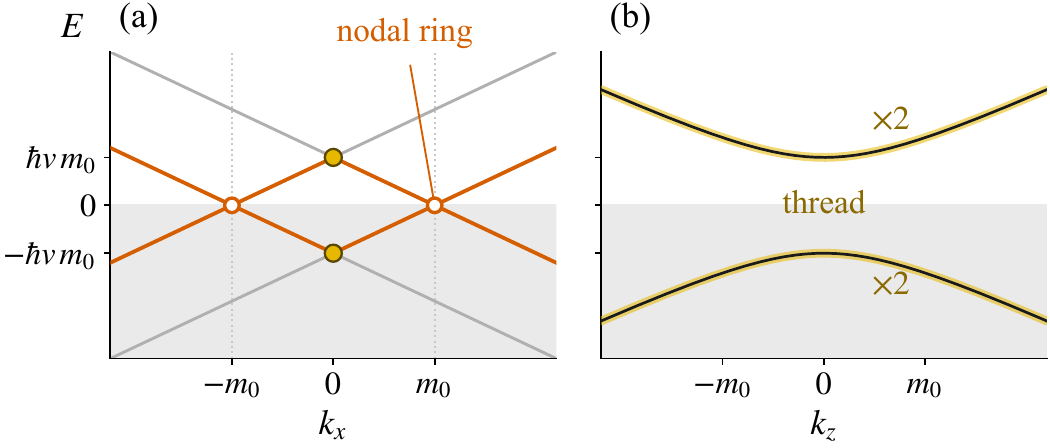}
\caption{Energy-momentum dispersion of the $\mathbb{Z}_2$ monopole nodal ring in Eq.~\eqref{eq:H0}.
(a) At $k_z=0$ and $k_y=0$, the conduction and valence branches meet at $|k_x|=m_0$, corresponding to the nodal ring at $k_\rho=m_0$.
(b) Along $k_x=k_y=0$, the occupied (and unoccupied) bands are twofold degenerate for all $k_z$, forming the linked ``thread'' in the occupied/unoccupied subspace.}
\label{fig:kp}
\end{figure}

In this model, the nodal ring of interest lies on $k_z=0$ and $k_\rho\equiv\sqrt{k_x^2+k_y^2}=m_0$ at energy $E=0$.
The topological charges of the nodal ring are therefore defined by the Stiefel--Whitney classes of the two bands below this energy.
Any loop linking this linearly dispersing nodal ring carries a nontrivial $w_1=1$.
It also carries a nontrivial monopole charge $w_2=1$~\cite{ahn2018}.
This can be understood from the band structure: the two occupied bands (and similarly the two unoccupied bands) are also degenerate along the line $k_x=k_y=0$ for all $k_z$, which threads the ring and realizes the linked ``ring-and-thread'' nodal structure.
Figure~\ref{fig:kp} shows representative dispersion cuts through the nodal ring and along the thread.

At finite doping away from the nodal ring with Fermi energy $0<\EF<\hbar v m_0$, the relevant low-energy bands have dispersions
$E(\bk)=\pm\hbar v\sqrt{k_z^2+(k_\rho-m_0)^2}$,
so the Fermi surface is a torus,
\begin{equation}
\left(\sqrt{k_x^2+k_y^2}-m_0\right)^2+k_z^2=k_F^2,\quad
k_F\equiv E_F/\hbar v,
\label{eq:torus}
\end{equation}
sketched in Fig.~\ref{fig:schematic}.
As a result of the linked nodal structure, the energy eigenstates at the Fermi level acquire nontrivial Berry phases along both the poloidal ($\theta$) and toroidal ($\phi$) directions, as illustrated in Fig.~\ref{fig:schematic}.
While $\Phi_{\theta}=\pi$ is common for stable nodal rings, $\Phi_{\phi}=\pi$ is the fingerprint of monopole nodal rings and can be probed in quantum oscillations.

For the lattice Kubo calculations below, we use the sine-regularized tight-binding model
\begin{align}
H_{\rm TB}(\bk)=&
\epsilon\big[\sin (k_xa)-m_0a\,t_x\big]s_x
+\epsilon\sin (k_ya)\,t_y s_y\notag\\
&+\epsilon\sin (k_za)\,s_z,
\label{eq:Hlat}
\end{align}
where $a$ is the lattice constant and $\epsilon=\hbar v/a$.
For $0<m_0a<1$, this model has eight isolated $\mathbb{Z}_2$ monopole nodal rings located near the eight time-reversal-invariant momenta.
Near each lattice copy, expanding about the corresponding time-reversal-invariant momentum gives Eq.~\eqref{eq:H0} up to sign changes of the local momenta and basis rotations.
These symmetry-related copies have the same $\mathbb{Z}_2$ charge and Berry phases modulo $2\pi$.
In the isolated-pocket regime relevant for the phase extraction below, they modify the overall oscillation amplitude and degeneracy factors but not the Onsager frequencies or phase offsets.

{\it Phase shift in magnetic quantum oscillations---}For the spinless $PT$-symmetric bands considered here, a closed cyclotron orbit in momentum space obeys the leading Onsager relation~\cite{shoenberg,mikitik1999}
\begin{equation}
A(\EF)=\frac{2\pi eB}{\hbar}\left(n+\gamma\right),\qquad
\gamma=\frac12-\frac{\Phi}{2\pi},
\label{eq:onsager}
\end{equation}
where $A(\EF)$ is the orbit area in the plane perpendicular to the applied magnetic field $\B$, and $\Phi$ is the quantized Berry phase of the Bloch state at the Fermi level in the spinless real-band setting.
This relation implies that the cyclotron orbit coincides with an extremal orbit only for discrete values of the magnetic field.
Accordingly, extremal cyclotron orbits recur periodically in $1/B$ with Onsager frequency $F\equiv (\hbar/2\pi e)A$.
Since $\Phi$ is quantized to $0$ or $\pi$ (mod $2\pi$) by $PT$ symmetry, $\gamma$ is pinned to $1/2$ or $0$.
In three dimensions, the oscillatory response of an extremal orbit additionally acquires the standard three-dimensional curvature phase $\varphi_{\rm 3D}=+\pi/4$ for a minimal extremal-area slice and $\varphi_{\rm 3D}=-\pi/4$ for a maximal extremal-area slice~\cite{li2018}.

\begin{figure*}[t]
\centering
\includegraphics[width=\textwidth]{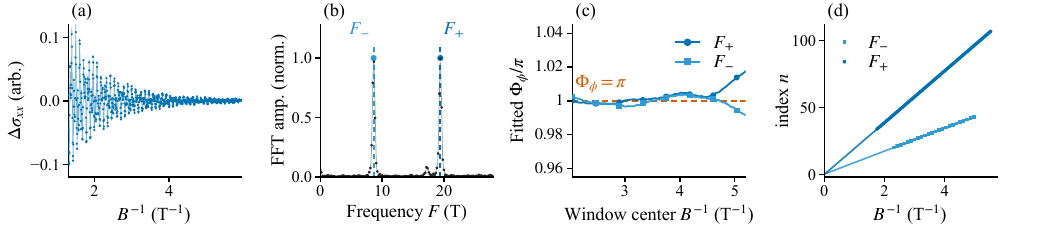}
\caption{Quantum oscillations of the electrical conductivity for $\B\parallel\hat{z}$ in the model of Eq.~\eqref{eq:Hlat} with $v=7.0\times10^5$~m/s, $m_0=0.0202~\text{\AA}^{-1}$, and $\EF=18.6$ meV. Here we use an enlarged lattice constant 
$a=4.95$ nm, ten times the physical value, which keeps the required magnetic supercells small enough for the lattice calculation to remain tractable. We reproduce the same phase offsets at the physical $a=4.95~\text{\AA}$ in the Supplemental Material~\cite{supp}.
(a) Background-subtracted $\Delta\sigma_{xx}$. (b) Hann-windowed FFT showing $F_{-}\simeq8.6~\mathrm T$ and $F_{+}\simeq19.4~\mathrm T$. (c) Sliding-window local-quadrature phases. (d) Raw conductivity-peak Landau fan before 3D curvature correction. The $y$-intercepts $0.1299\pm 0.0049$ and $-0.1300\pm 0.0039$ for $F_-$ and $F_+$ both support $\Phi_{\phi}=\pi$ within 1\% through the Onsager relation in Eq.~\eqref{eq:onsager} corrected by the 3D curvature phases $\pm 1/8$.}
\label{fig:Bz}
\end{figure*}

\begin{figure*}[t]
\centering
\includegraphics[width=\textwidth]{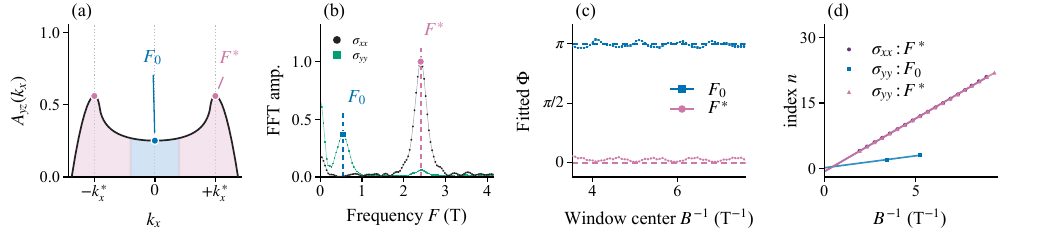}
\caption{Quantum oscillations of the electrical conductivity for $\B\parallel\hat{x}$ using the same model and parameters as in Fig.~\ref{fig:Bz}. (a) Total $yz$-section area $A_{yz}(k_x)$ (units of $m_0^2$); the $k_x=0$ section consists of two degenerate linking loops of area $A_{yz}(0)/2$ each, giving $F_0$, while $k_x=\pm k_x^\ast$ gives nonlinking maximal loops and $F^\ast$. (b) Hann-windowed FFTs of $\sigma_{xx}$ and $\sigma_{yy}$; $\sigma_{xx}$ is dominated by the nonlinking $F^\ast$ series, while $\sigma_{yy}$ shows the linking $F_0$ peak with a suppressed $F^\ast$ component. (c) Local phase fits to $\sigma_{yy}$ identify $\Phi_\theta=\pi$ for $F_0$ and $\Phi=0$ for $F^\ast$. (d) Raw conductivity-peak Landau fan; quantitative intercept fits are given in the Supplemental Material.}
\label{fig:Bx}
\end{figure*}

We analyze the oscillatory conductivity as a function of inverse magnetic field.
From the raw $\sigma_{ii}(B)$ we subtract an empirical smooth nonoscillatory background $\sigma_{ii}^{\rm smooth}(1/B)$, fitted over the inverse-field window used for the analysis, to obtain $\Delta\sigma_{ii}(1/B)$, and compute its Hann-windowed fast Fourier transform (FFT) in $1/B$ to identify the dominant Onsager frequencies.
To extract phase offsets, we fit $\Delta\sigma_{ii}(1/B)$ to a multifrequency Lifshitz--Kosevich (LK) form
\begin{equation}
\Delta\sigma_{ii}\propto\sum_{\alpha}C_{\alpha}\cos\!\Big[2\pi\big(F_{\alpha}/B-\gamma_{\alpha}\big)+\varphi^{\alpha}_{\rm 3D}\Big],
\label{eq:LKfit}
\end{equation}
with the curvature phase fixed to $\varphi^{\alpha}_{\rm 3D}=+\pi/4$ for minimal extremal orbits and $-\pi/4$ for maximal extremal orbits.
The fitted $\gamma_{\alpha}$ is then converted to $\Phi$ via Eq.~\eqref{eq:onsager}.
For lattice data, the smooth, nonuniversal envelope is absorbed into window-dependent quadrature amplitudes over sliding $1/B$ windows, as defined in the Supplemental Material~\cite{supp}\nocite{watanabe2019proof,lewenkopf2013recursive,drouvelis2006parallel,esteveparedes2023velocity,huertas2006,min2006,jennings1976,kiefer2010,wiggers2019,zhu2011,kormanyos2015}. This is the fitting procedure used for Figs.~\ref{fig:Bz} and \ref{fig:Bx}.

{\it Phase shifts for the out-of-plane magnetic field---}We first consider $\B$ along the ring axis $\hat{z}$, where cyclotron orbits trace $k_x$--$k_y$ loops at fixed $k_z$.
This field orientation directly probes the consequences of the $w_2$-enforced thread.

For an ordinary nodal ring without monopole charge ($w_2=0$), the toroidal
extremal orbits at $k_z=0$ do not enclose any band degeneracy and therefore carry
trivial toroidal Berry phase $\Phi_\phi=0$, yielding the standard phase shift
$\gamma=1/2$~\cite{li2018}. The inner and outer cross sections of the torus
behave, in this respect, like ordinary parabolic-band Fermi surfaces.
For the $\mathbb{Z}_2$ monopole ring, on the other hand, the $w_2$-enforced
thread at $k_x=k_y=0$ pierces the torus hole. As a result, both the inner and
outer toroidal extremal orbits encircle the thread and are pinned to the
toroidal Berry phase $\Phi_\phi=\pi$, hence $\gamma_\pm=0$.

The exact continuum Landau spectrum provides a direct check of this Berry-phase
pinning in the Landau fan. For $\B=B\hat z$, minimally coupling Eq.~\eqref{eq:H0} to a uniform magnetic field (see the Supplemental Material~\cite{supp} for the derivation and sign convention) gives
\begin{align}
    \varepsilon_{n,\lambda}^2(k_z)
=
k_z^2+\left(m_0+\lambda\sqrt{2eBn/\hbar}\right)^2,
\end{align}
for $n\ge1$ and $\lambda=\pm$,
together with the unpaired branch $\varepsilon_0^2(k_z)=k_z^2+m_0^2$,
where $\varepsilon\equiv E/(\hbar v)$.
For $0<k_F<m_0$, only the $\lambda=-$ branch crosses the Fermi
level. The condition $\varepsilon_{n,-}(k_z)=k_F$ gives
$
\sqrt{2eBn/\hbar}
=
m_0\pm\sqrt{k_F^2-k_z^2}.
$
Equivalently, using the torus cross section in Eq.~\eqref{eq:torus},
\begin{align}
    \frac{\hbar A_\pm(k_z)}{2\pi eB}=n,\quad
A_\pm(k_z)=
\pi\left(m_0\pm\sqrt{k_F^2-k_z^2}\right)^2.
\end{align}
At the extremal slice $k_z=0$, $A_\pm\equiv A_\pm(0)$ and
\begin{align}
\label{eq:continuum-fan}
    n=\frac{F_\pm}{B},
\qquad
F_\pm=\frac{\hbar A_\pm}{2\pi e}=\frac{\hbar}{e}\frac{(m_0\pm k_F)^2}{2}.
\end{align}
Comparing this exact fan with Eq.~\eqref{eq:onsager}, or equivalently with
$n=F_\pm/B-\gamma_\pm$, gives $\gamma_\pm=0$.
Thus the exact Landau fan is the Landau-level manifestation of the topological
statement above: both toroidal extremal orbits have $\Phi_\phi=\pi$. No weak-field or thin-torus expansion is involved in this fan relation. The condition $0<k_F<m_0$ only ensures that the Fermi surface is a closed torus on the low-energy ($\lambda=-$) branch.

The oscillatory conductivity waveform requires one further asymptotic step:
applying the usual many-Landau-level Lifshitz--Kosevich expansion to this exact
fan. For the transverse channel
$\sigma_\perp\equiv\sigma_{xx}=\sigma_{yy}$, which is the channel used in the
lattice calculation below, the leading asymptotic form is
\begin{align}
&\sigma_{\perp}^{\mathrm{osc}}(B)
\simeq
\frac{e^2}{\hbar}\frac{g_dv\,\tau_{\mathrm{tr}}\sqrt{k_F eB/\hbar}}{4\pi^2}\times\notag\\
&\sum_{\nu=1}^{\infty}\frac{1}{\sqrt{\nu}}
\Bigg[
\sqrt{Y_+}\,\mathcal R_{\nu,+}(B)
\cos\!\left(
2\pi \nu\left(\frac{F_+}{B}-\gamma_+\right)-\frac{\pi}{4}
\right)
\nonumber\\
&
+
\sqrt{Y_-}\,\mathcal R_{\nu,-}(B)
\cos\!\left(
2\pi \nu\left(\frac{F_-}{B}-\gamma_-\right)+\frac{\pi}{4}
\right)
\Bigg].
\label{eq:osc_Bz}
\end{align}
Here $Y_r\equiv m_0+r k_F$ with $r=\pm$.
The factor $g_d$ denotes any extra spin, valley, or symmetry-related pocket multiplicity not included in a single spinless continuum copy; for one such copy $g_d=1$. The parameter $\tau_{\rm tr}$ is a phenomenological transport lifetime. Both enter only the nonuniversal amplitude, not the frequency or the Onsager offset.
The damping factor $\mathcal R_{\nu,r}(B)=R^{(r)}_{T,\nu}(B)R^{(r)}_{D,\nu}(B)R^{(r)}_{S,\nu}$ is the standard LK reduction factor for the $\nu$th harmonic of orbit $r$, evaluated with cyclotron mass $m_{c,r}=\hbar e|\partial F_r/\partial E_F|=\hbar Y_r/v$. For our low-energy model, this gives $m_{c,\pm}=\hbar(m_0\pm k_F)/v$. The Zeeman reduction factor $R^{(r)}_{S,\nu}=\cos[\pi\nu g_r m_{c,r}/(2m_e)]$, with $g_r$ the effective $g$ factor of orbit $r$, is typically close to unity~\cite{supp}, and we set $R^{(r)}_{S,\nu}=1$ in our spinless calculation.
Equation~\eqref{eq:osc_Bz} should therefore be read as a leading LK/constant-relaxation-time expression for the
amplitude. The frequencies $F_r$ and the absence of the Onsager offsets,
$\gamma_r=0$, are exact consequences of the continuum Landau fan in Eq.~\eqref{eq:continuum-fan}, whereas the
envelope, damping factors, and overall normalization require the usual
many-Landau-level condition $F_r/B\gg1$.

To corroborate the single-monopole continuum prediction in a lattice setting, we compute the in-plane magnetoconductivity of the sine-regularized tight-binding model in Eq.~\eqref{eq:Hlat} using a Kubo calculation at small rational flux~\cite{akkermans2007mesoscopic,rammer1986quantum,mahan2000manyparticle}.
The magnetic-supercell, broadening, and phase-extraction protocols are described in the Supplemental Material~\cite{supp}.
%The lattice signal is used to extract frequencies and phase offsets, not an absolute calibrated conductivity.
Figure~\ref{fig:Bz} shows two dominant oscillation frequencies associated with the inner and outer $k_z=0$ extremal contours, in agreement with the Onsager relation.
With the curvature phases fixed, sliding-window local-quadrature fits give $\Phi_\phi=\pi$ for both toroidal series at about the $1\%$ level across the stable fitting-window range.
The raw conductivity-peak Landau fan in Fig.~\ref{fig:Bz}(d), after the 3D curvature correction is applied to its intercepts, gives a consistent visual check.

The longitudinal component $\sigma_{zz}(B)$ has exactly the same frequencies and Onsager offsets but a different LK envelope because the group velocity $v_z(\bk)=\hbar^{-1}\partial E(\bk)/\partial k_z\propto k_z$ vanishes pointwise on the $k_z=0$ extremal orbits due to $z$-flipping mirror symmetry~\cite{supp}.

{\it Phase shifts for the in-plane magnetic field---}For $\B\parallel\hat{x}$,
the $k_y$--$k_z$ cyclotron sections contain linking $F_0$ and nonlinking
$F^\ast$ extremal families [Fig.~\ref{fig:Bx}(a)], allowing the ordinary
poloidal Berry phase to be separated using conductivity components.
The $k_x=0$ loops carry $\Phi_\theta=\pi$ and give
$F_0=\hbar k_F^2/(2e)$, while the nonlinking maximal loops carry
$\Phi=0$ and give $F^\ast=\hbar A_{\max}/(2\pi e)$.
The $x$-flipping mirror symmetry makes $v_x(\bk)=0$ on the linking orbits, so
$\sigma_{xx}$ is dominated by $F^\ast$, whereas $\sigma_{yy}$ reveals
both $F_0$ and $F^\ast$.
Figure~\ref{fig:Bx}(c) gives $\Phi_\theta=\pi$ for $F_0$ and
$\Phi=0$ for $F^\ast$; Fig.~\ref{fig:Bx}(d) is a visual fan check,
with quantitative extraction from FFT-identified local-quadrature fits.

{\it Discussion---}We close by examining the conditions under which the proposed diagnostics apply. First, our analysis assumes closed nodal rings with toroidal Fermi surfaces. It does not directly apply to open Fermi surfaces or saddle-point regimes, since closed semiclassical orbits are ill-defined there and magnetic tunneling becomes essential~\cite{shoenberg,kaganov1979,ahn2017,cohen1961,blount1962,alexandradinata2017,alexandradinata2018}. Open nodal lines may carry $\mathbb{Z}_2$ monopole charge, but at small doping they usually form Brillouin-zone-wrapping cylindrical Fermi surfaces, so a closed orbit along the nodal-line direction is unavailable. Once doping produces closed pockets in that direction, however, the same Berry-phase diagnostic we use should apply.

Second, the proposed diagnostic is most directly applicable when spin-orbit coupling (SOC) is negligible or weak. In the spinless limit considered here, $PT$ satisfies $(PT)^2=+1$, the Hamiltonian can be chosen real, and the Berry phase on a closed cyclotron orbit is quantized to $0$ or $\pi$. Finite SOC changes this symmetry setting because ordinary time reversal flips spins for spinful electrons, giving $(PT)^2=-1$, and the nodal ring and thread are generally gapped unless protected by additional crystalline symmetries. If the resulting SOC-induced gap $\Delta_{\rm SOC}$ is small compared with the Fermi energy scale, the geometric Berry phase of a local gapped orbit remains close to the spinless value. The measured LK offset, however, can also contain orbital-magnetic-moment, Zeeman, and spinful-band corrections; in the pure massive-Dirac reduction the orbital moment cancels the linear Berry-only shift. The signatures proposed here are therefore best controlled in light-element systems where $\Delta_{\rm SOC}\ll \EF$ and material-specific spinful corrections are negligible.

These conditions point to light-element candidate materials with isolated monopole nodal rings and simple toroidal Fermi pockets. Three-dimensional ABC-stacked graphdiyne is a particularly attractive candidate. This material is composed only of carbon, so SOC is expected to be very weak, and it was predicted to host the required linked ring-and-thread $\mathbb{Z}_2$ monopole nodal rings~\cite{nomura2018,ahn2018}. Using the local parameters obtained from density-functional-theory fits in Ref.~\cite{nomura2018}, we find simple toroidal Fermi surfaces at low doping, before tilt-induced reconstruction sets in. For example, at $\EF\simeq 8\,\mathrm{meV}$ one obtains $F_+\simeq 15.8\,\mathrm{T}$ and $F_-\simeq 11.2\,\mathrm{T}$. Other proposed candidates, including Si/Ge structures and transition-metal dichalcogenides~\cite{li2025general}, provide useful secondary targets, although many require additional material-specific validation or lie farther from the weak-SOC limit~\cite{li2025general,wang2019,liu2011}. We therefore regard ABC-stacked graphdiyne as the strongest current candidate for observing the proposed quantum-oscillation signatures. Observing this pinned $\pi$ phase in both toroidal series would amount to reading off the second Stiefel--Whitney class directly from a transport experiment.

\begin{acknowledgments}
This work was primarily supported by J.A.'s faculty start-up grant from The University of Texas at Austin.
Y.H. was supported by a UKRI Future Leaders Fellowship MR/Y017331/1.
B.-J.Y. was supported by the Samsung Science and Technology Foundation under Projects No. SSTF-BA2002-06 and No. SSTF-BA2601-02; by the National Research Foundation of Korea (NRF), funded by the Korean government (MSIT), under Grants No. RS-2021-NR060087 and No. RS-2025-00562579; by the Global Research Development Center (GRDC) Cooperative Hub Program through the NRF, funded by the MSIT, under Grant No. RS-2023-00258359; and by the Global-LAMP program of the NRF, funded by the Ministry of Education, under Grant No. RS-2023-00301976.
\end{acknowledgments}

% Enable titles in references via apsrev4-2 CONTROL entry (PRL recommendation).
\bstctlcite{apsrev4-2Control}
\bibliographystyle{apsrev4-2}
\bibliography{refs}

@CONTROL{apsrev4-2Control,
  title = {1},
}

@misc{supp,
  note = {See Supplemental Material at [URL will be inserted by publisher] for detailed derivations of the Landau spectrum and Lifshitz--Kosevich formulas, the numerical protocol, and material estimates, including Refs.~\cite{watanabe2019proof,lewenkopf2013recursive,drouvelis2006parallel,esteveparedes2023velocity,huertas2006,min2006,jennings1976,kiefer2010,wiggers2019,zhu2011,kormanyos2015}.}
}

@article{huertas2006,
  author  = {Huertas-Hernando, Daniel and Guinea, Francisco and Brataas, Arne},
  title   = {Spin-Orbit Coupling in Curved Graphene, Fullerenes, Nanotubes, and Nanotube Caps},
  journal = {Phys. Rev. B},
  volume  = {74},
  number  = {15},
  pages   = {155426},
  year    = {2006},
  doi     = {10.1103/PhysRevB.74.155426}
}

@article{min2006,
  author  = {Min, Hongki and Hill, Jesse E. and Sinitsyn, N. A. and Sahu, B. R. and Kleinman, Leonard and MacDonald, A. H.},
  title   = {Intrinsic and {Rashba} Spin-Orbit Interactions in Graphene Sheets},
  journal = {Phys. Rev. B},
  volume  = {74},
  number  = {16},
  pages   = {165310},
  year    = {2006},
  doi     = {10.1103/PhysRevB.74.165310}
}

@article{zhu2011,
  author  = {Zhu, Z. Y. and Cheng, Y. C. and Schwingenschl{\"o}gl, Udo},
  title   = {Giant Spin-Orbit-Induced Spin Splitting in Two-Dimensional Transition-Metal Dichalcogenide Semiconductors},
  journal = {Phys. Rev. B},
  volume  = {84},
  number  = {15},
  pages   = {153402},
  year    = {2011},
  doi     = {10.1103/PhysRevB.84.153402}
}

@article{kormanyos2015,
  author  = {Korm{\'a}nyos, Andor and Burkard, Guido and Gmitra, Martin and Fabian, Jaroslav and Z{\'o}lyomi, Viktor and Drummond, Neil D. and Fal'ko, Vladimir},
  title   = {{$k\cdot p$} theory for two-dimensional transition metal dichalcogenide semiconductors},
  journal = {2D Mater.},
  volume  = {2},
  number  = {2},
  pages   = {022001},
  year    = {2015},
  doi     = {10.1088/2053-1583/2/2/022001}
}

@article{watanabe2019proof,
  author        = {Watanabe, Haruki},
  title         = {A Proof of the {Bloch} Theorem for Lattice Models},
  journal       = {J. Stat. Phys.},
  volume        = {177},
  pages         = {717--726},
  year          = {2019},
  doi           = {10.1007/s10955-019-02386-1},
  eprint        = {1904.02700},
  archivePrefix = {arXiv},
  primaryClass  = {cond-mat.stat-mech}
}

@article{lewenkopf2013recursive,
  author        = {Lewenkopf, Caio H. and Mucciolo, Eduardo R.},
  title         = {The Recursive {Green}'s Function Method for Graphene},
  journal       = {J. Comput. Electron.},
  volume        = {12},
  pages         = {203--231},
  year          = {2013},
  doi           = {10.1007/s10825-013-0458-7},
  eprint        = {1304.3934},
  archivePrefix = {arXiv},
  primaryClass  = {cond-mat.mes-hall}
}

@article{drouvelis2006parallel,
  author        = {Drouvelis, P. S. and Schmelcher, P. and Bastian, P.},
  title         = {Parallel Implementation of the Recursive {Green}'s Function Method},
  journal       = {J. Comput. Phys.},
  volume        = {215},
  pages         = {741--756},
  year          = {2006},
  doi           = {10.1016/j.jcp.2005.11.010},
  eprint        = {cond-mat/0507415},
  archivePrefix = {arXiv}
}

@article{esteveparedes2023velocity,
  author        = {Esteve-Paredes, J. J. and Palacios, J. J.},
  title         = {A Comprehensive Study of the Velocity, Momentum and Position Matrix Elements for {Bloch} States Using a Local Orbital Basis},
  journal       = {SciPost Phys. Core},
  volume        = {6},
  pages         = {002},
  year          = {2023},
  doi           = {10.21468/SciPostPhysCore.6.1.002},
  eprint        = {2201.12290},
  archivePrefix = {arXiv},
  primaryClass  = {cond-mat.mes-hall}
}

@article{wiggers2019,
  author  = {Wiggers, F. B. and Fleurence, A. and Aoyagi, K. and Yonezawa, T. and Yamada-Takamura, Y. and Feng, H. and Zhuang, J. and Du, Y. and Kovalgin, A. Y. and de Jong, M. P.},
  title   = {{Van der Waals} integration of silicene and hexagonal boron nitride},
  journal = {2D Mater.},
  volume  = {6},
  number  = {3},
  pages   = {035001},
  year    = {2019},
  doi     = {10.1088/2053-1583/ab0a29}
}

@article{jennings1976,
  author  = {Jennings, H. M. and Richman, M. H.},
  title   = {A hexagonal ({Wurtzite}) form of silicon},
  journal = {Science},
  volume  = {193},
  number  = {4259},
  pages   = {1242--1243},
  year    = {1976},
  doi     = {10.1126/science.193.4259.1242}
}

@article{kiefer2010,
  author  = {Kiefer, Florian and Hlukhyy, Viktor and Karttunen, Antti J. and F{\"a}ssler, Thomas F. and Gold, Christian and Scheidt, Ernst-Wilhelm and Scherer, Wolfgang and Nyl{\'e}n, Johanna and H{\"a}ussermann, Ulrich},
  title   = {Synthesis, structure, and electronic properties of {4H}-germanium},
  journal = {J. Mater. Chem.},
  volume  = {20},
  pages   = {1780--1786},
  year    = {2010},
  doi     = {10.1039/B921575A}
}

@book{shoenberg,
  author    = {Shoenberg, David},
  title     = {Magnetic Oscillations in Metals},
  publisher = {Cambridge University Press},
  address   = {Cambridge},
  year      = {1984},
}

@book{akkermans2007mesoscopic,
  author    = {Akkermans, Eric and Montambaux, Gilles},
  title     = {Mesoscopic Physics of Electrons and Photons},
  publisher = {Cambridge University Press},
  address   = {Cambridge},
  year      = {2007},
  doi       = {10.1017/CBO9780511618833}
}

@article{rammer1986quantum,
  title = {Quantum field-theoretical methods in transport theory of metals},
  author = {Rammer, J. and Smith, H.},
  journal = {Rev. Mod. Phys.},
  volume = {58},
  pages = {323--359},
  year = {1986},
  publisher = {American Physical Society}
}

@book{mahan2000manyparticle,
  title = {Many-Particle Physics},
  author = {Mahan, Gerald D.},
  edition = {3},
  publisher = {Kluwer Academic/Plenum Publishers},
  address = {New York},
  year = {2000}
}

@article{mikitik1999,
  author  = {Mikitik, G. P. and Sharlai, Yu. V.},
  title   = {Manifestation of {Berry}'s Phase in Metal Physics},
  journal = {Phys. Rev. Lett.},
  volume  = {82},
  number  = {10},
  pages   = {2147--2150},
  year    = {1999},
  doi     = {10.1103/PhysRevLett.82.2147},
}

@article{li2018,
  author  = {Li, Cequn and Wang, C. M. and Wan, Bo and Wan, Xiangang and Lu, Hai-Zhou and Xie, X. C.},
  title   = {Rules for Phase Shifts of Quantum Oscillations in Topological Nodal-Line Semimetals},
  journal = {Phys. Rev. Lett.},
  volume  = {120},
  number  = {14},
  pages   = {146602},
  year    = {2018},
  doi     = {10.1103/PhysRevLett.120.146602},
}

@article{yang2018quantum,
  title = {Quantum oscillations in nodal line systems},
  author = {Yang, Hui and Moessner, Roderich and Lim, Lih-King},
  journal = {Phys. Rev. B},
  volume = {97},
  number = {16},
  pages = {165118},
  year = {2018},
  publisher = {American Physical Society}
}

@article{hu2016,
  author  = {Hu, Jin and Tang, Zhijie and Liu, Jinyu and Liu, Xue and Zhu, Yanglin and Graf, David and Myhro, Kevin and Tran, Son and Lau, Chun Ning and Wei, Jiang and Mao, Zhiqiang},
  title   = {Evidence of Topological Nodal-Line Fermions in {ZrSiSe} and {ZrSiTe}},
  journal = {Phys. Rev. Lett.},
  volume  = {117},
  number  = {1},
  pages   = {016602},
  year    = {2016},
  doi     = {10.1103/PhysRevLett.117.016602},
}

@article{vanDelft2018,
  author  = {van Delft, M. R. and Pezzini, S. and Khouri, T. and M{\"u}ller, C. S. A. and Breitkreiz, M. and Schoop, L. M. and Carrington, A. and Hussey, N. E. and Wiedmann, S.},
  title   = {Electron-Hole Tunneling Revealed by Quantum Oscillations in the Nodal-Line Semimetal {HfSiS}},
  journal = {Phys. Rev. Lett.},
  volume  = {121},
  number  = {25},
  pages   = {256602},
  year    = {2018},
  doi     = {10.1103/PhysRevLett.121.256602},
}

@article{fang2015,
  author  = {Fang, Chen and Chen, Yige and Kee, Hae-Young and Fu, Liang},
  title   = {Topological Nodal Line Semimetals with and without Spin-Orbital Coupling},
  journal = {Phys. Rev. B},
  volume  = {92},
  number  = {8},
  pages   = {081201(R)},
  year    = {2015},
  doi     = {10.1103/PhysRevB.92.081201},
}

@article{bzdusek2017,
  title={Robust doubly charged nodal lines and nodal surfaces in centrosymmetric systems},
  author={Bzdu{\v{s}}ek, Tom{\'a}{\v{s}} and Sigrist, Manfred},
  journal={Phys. Rev. B},
  volume={96},
  number={15},
  pages={155105},
  year={2017},
  publisher={APS}
}

@article{ahn2018,
  author  = {Ahn, Junyeong and Kim, Dongwook and Kim, Youngkuk and Yang, Bohm-Jung},
  title   = {Band Topology and Linking Structure of Nodal Line Semimetals with {$Z_2$} Monopole Charges},
  journal = {Phys. Rev. Lett.},
  volume  = {121},
  number  = {10},
  pages   = {106403},
  year    = {2018},
  doi     = {10.1103/PhysRevLett.121.106403},
}

@article{ahn2019failure,
  title={Failure of {Nielsen--Ninomiya} theorem and fragile topology in two-dimensional systems with space-time inversion symmetry: application to twisted bilayer graphene at magic angle},
  author={Ahn, Junyeong and Park, Sungjoon and Yang, Bohm-Jung},
  journal={Phys. Rev. X},
  volume={9},
  number={2},
  pages={021013},
  year={2019},
  publisher={APS}
}

@article{ahn2019stiefel,
  title={{Stiefel--Whitney} classes and topological phases in band theory},
  author={Ahn, Junyeong and Park, Sungjoon and Kim, Dongwook and Kim, Youngkuk and Yang, Bohm-Jung},
  journal={Chin. Phys. B},
  volume={28},
  number={11},
  pages={117101},
  year={2019},
  publisher={Chinese Physical Society and IOP Publishing Ltd}
}

@article{ahn2019symmetry,
  title={Symmetry representation approach to topological invariants in {$C_{2z}T$}-symmetric systems},
  author={Ahn, Junyeong and Yang, Bohm-Jung},
  journal={Phys. Rev. B},
  volume={99},
  number={23},
  pages={235125},
  year={2019},
  publisher={APS}
}

@article{ahn2020higher,
  title={Higher-order topological superconductivity of spin-polarized fermions},
  author={Ahn, Junyeong and Yang, Bohm-Jung},
  journal={Phys. Rev. Res.},
  volume={2},
  number={1},
  pages={012060},
  year={2020},
  publisher={APS}
}

@article{lee2020two,
  title={Two-dimensional higher-order topology in monolayer graphdiyne},
  author={Lee, Eunwoo and Kim, Rokyeon and Ahn, Junyeong and Yang, Bohm-Jung},
  journal={npj Quantum Mater.},
  volume={5},
  number={1},
  pages={1},
  year={2020},
  publisher={Nature Publishing Group UK London}
}

@article{shiozaki2024discrete,
  author        = {Shiozaki, Ken and Chen, Jing-Yuan},
  title         = {A Discrete Formulation of Second {Stiefel-Whitney} Class for Band Theory},
  journal       = {arXiv e-prints},
  year          = {2024},
  eprint        = {2412.18796},
  archivePrefix = {arXiv},
  primaryClass  = {cond-mat.mes-hall}
}

@article{shiozaki2025z2,
  author        = {Shiozaki, Ken},
  title         = {{$\mathbb{Z}_2$} topological invariant in three-dimensional {P}{T}- and {P}{C}-symmetric class {CI} band structures},
  journal       = {SciPost Phys. Core},
  volume        = {9},
  pages         = {029},
  year          = {2026},
  doi           = {10.21468/SciPostPhysCore.9.2.029},
  eprint        = {2509.19825},
  archivePrefix = {arXiv},
  primaryClass  = {cond-mat.mes-hall}
}

@article{wu2019non,
  title={{Non-Abelian} band topology in noninteracting metals},
  author={Wu, QuanSheng and Soluyanov, Alexey A and Bzdu{\v{s}}ek, Tom{\'a}{\v{s}}},
  journal={Science},
  volume={365},
  number={6459},
  pages={1273--1277},
  year={2019},
  publisher={American Association for the Advancement of Science}
}

@article{tiwari2020,
  title={{Non-Abelian} topology of nodal-line rings in {PT}-symmetric systems},
  author={Tiwari, Apoorv and Bzdu{\v{s}}ek, Tom{\'a}{\v{s}}},
  journal={Phys. Rev. B},
  volume={101},
  number={19},
  pages={195130},
  year={2020},
  publisher={APS}
}

@article{unal2020topological,
  title={Topological {Euler} class as a dynamical observable in optical lattices},
  author={{\"U}nal, F Nur and Bouhon, Adrien and Slager, Robert-Jan},
  journal={Phys. Rev. Lett.},
  volume={125},
  number={5},
  pages={053601},
  year={2020},
  publisher={APS}
}

@article{bouhon2020geometric,
  title={Geometric approach to fragile topology beyond symmetry indicators},
  author={Bouhon, Adrien and Bzdu{\v{s}}ek, Tom{\'a}{\v{s}} and Slager, Robert-Jan},
  journal={Phys. Rev. B},
  volume={102},
  number={11},
  pages={115135},
  year={2020},
  publisher={APS}
}

@article{chen2021graphyne,
  title={Graphyne as a second-order and real {Chern} topological insulator in two dimensions},
  author={Chen, Cong and Wu, Weikang and Yu, Zhi-Ming and Chen, Ziyu and Zhao, YX and Sheng, Xian-Lei and Yang, Shengyuan A},
  journal={Phys. Rev. B},
  volume={104},
  number={8},
  pages={085205},
  year={2021},
  publisher={APS}
}

@article{pan2022two,
  title={Two-dimensional {Stiefel-Whitney} insulators in liganded {Xenes}},
  author={Pan, Mingxiang and Li, Dexin and Fan, Jiahao and Huang, Huaqing},
  journal={npj Comput. Mater.},
  volume={8},
  number={1},
  pages={1},
  year={2022},
  publisher={Nature Publishing Group UK London}
}

@article{kwon2024quantum,
  title={Quantum geometric bound and ideal condition for {Euler} band topology},
  author={Kwon, Soonhyun and Yang, Bohm-Jung},
  journal={Phys. Rev. B},
  volume={109},
  number={16},
  pages={L161111},
  year={2024},
  publisher={APS}
}

@article{lee2025euler,
  title={Euler band topology in spin-orbit coupled magnetic systems},
  author={Lee, Seung Hun and Qian, Yuting and Yang, Bohm-Jung},
  journal={Phys. Rev. B},
  volume={111},
  number={24},
  pages={245127},
  year={2025},
  publisher={APS}
}

@article{liu2021nematic,
  title={Nematic topological semimetal and insulator in magic-angle bilayer graphene at charge neutrality},
  author={Liu, Shang and Khalaf, Eslam and Lee, Jong Yeon and Vishwanath, Ashvin},
  journal={Phys. Rev. Res.},
  volume={3},
  number={1},
  pages={013033},
  year={2021},
  publisher={APS}
}

@article{liu2022second,
  title={Second-order and real {Chern} topological insulator in twisted bilayer $\alpha$-graphyne},
  author={Liu, Bin-Bin and Zeng, Xu-Tao and Chen, Cong and Chen, Ziyu and Sheng, Xian-Lei},
  journal={Phys. Rev. B},
  volume={106},
  number={3},
  pages={035153},
  year={2022},
  publisher={APS}
}

@article{guan2022landau,
  title={Landau levels of the {Euler} class topology},
  author={Guan, Yifei and Bouhon, Adrien and Yazyev, Oleg V},
  journal={Phys. Rev. Res.},
  volume={4},
  number={2},
  pages={023188},
  year={2022},
  publisher={APS}
}

@article{yu2023euler,
  title={Euler-obstructed nematic nodal superconductivity in twisted bilayer graphene},
  author={Yu, Jiabin and Xie, Ming and Wu, Fengcheng and Das Sarma, Sankar},
  journal={Phys. Rev. B},
  volume={107},
  number={20},
  pages={L201106},
  year={2023},
  publisher={APS}
}

@article{liu2025correspondence,
  title={Correspondence between {Euler} charges and nodal-line topology in {Euler} semimetals},
  author={Liu, Wenwen and Wang, Hanyu and Yang, Biao and Zhang, Shuang},
  journal={Sci. Adv.},
  volume={11},
  number={2},
  pages={eads5081},
  year={2025},
  publisher={American Association for the Advancement of Science}
}

@article{li2025general,
  title={General construction of three-dimensional {$Z_2$} monopole charge nodal line semimetals and prediction of abundant candidate materials},
  author={Li, Yongpan and Qian, Shifeng and Liu, Cheng-Cheng},
  journal={Phys. Rev. B},
  volume={111},
  number={12},
  pages={125101},
  year={2025},
  publisher={APS}
}

@article{yue2024stability,
  title = {Stability and noncentered {PT} symmetry of real topological phases},
  author = {Yue, S. J. and Liu, Q. and Yang, S. A. and Zhao, Y. X.},
  journal = {Phys. Rev. B},
  volume = {109},
  number = {19},
  pages = {195116},
  year = {2024},
  publisher = {American Physical Society}
}

@article{salerno2020floquet,
  title={Floquet-engineering of nodal rings and nodal spheres and their characterization using the quantum metric},
  author={Grazia Salerno and Nathan Goldman and Giandomenico Palumbo},
  journal={Phys. Rev. Res.},
  volume={2},
  number={1},
  pages={013224},
  year={2020},
  publisher={APS}
}

@article{sato2025three,
  title={Three-dimensional spinless {Euler} insulators with rotational symmetry},
  author={Sato, Manabu and Kobayashi, Shingo and Hirayama, Motoaki and Furusaki, Akira},
  journal={Phys. Rev. B},
  volume={112},
  number={19},
  pages={195108},
  year={2025},
  publisher={APS}
}

@article{kobayashi2026euler,
  title={Euler band topology in superfluids and superconductors},
  author={Kobayashi, Shingo and Sato, Manabu and Furusaki, Akira},
  journal={Phys. Rev. B},
  volume={113},
  number={6},
  pages={L060501},
  year={2026},
  publisher={APS}
}

@article{qian2023stable,
  title={Stable higher-order topological {Dirac} semimetals with {$Z_2$} monopole charge in alternating-twist multilayer graphene and beyond},
  author={Qian, Shifeng and Li, Yongpan and Liu, Cheng-Cheng},
  journal={Phys. Rev. B},
  volume={108},
  number={24},
  pages={L241406},
  year={2023},
  publisher={APS}
}

@article{li2024phononic,
  title={Phononic {Stiefel--Whitney} Topology with Hinge Vibrational Modes in {3D} Carbon Allotrope {$4^3$T57-CA}},
  author={Li, Yang},
  journal={{ACS Omega}},
  volume={9},
  number={46},
  pages={46610--46614},
  year={2024},
  publisher={ACS Publications},
  doi={10.1021/acsomega.4c08904}
}

@article{wu2025breakdown,
  title={Breakdown of boundary criticality and exotic topological semimetals in {PT}-invariant systems},
  author={Wu, Hong and An, Jun-Hong},
  journal={Phys. Rev. B},
  volume={112},
  number={4},
  pages={L041101},
  year={2025},
  publisher={APS}
}

@article{liu2025floquet,
  title={Floquet control of topological phases and {Hall} effects in {$Z_2$} nodal line semimetals},
  author={Liu, Pu and Cui, Chaoxi and Li, Lei and Li, Runze and Xu, Dong-Hui and Yu, Zhi-Ming},
  journal={Phys. Rev. B},
  volume={111},
  number={23},
  pages={235105},
  year={2025},
  publisher={APS}
}

@article{xue2023stiefel,
  title={{Stiefel-Whitney} topological charges in a three-dimensional acoustic nodal-line crystal},
  author={Xue, Haoran and Chen, ZY and Cheng, Zheyu and Dai, JX and Long, Yang and Zhao, YX and Zhang, Baile},
  journal={Nat. Commun.},
  volume={14},
  number={1},
  pages={4563},
  year={2023},
  publisher={Nature Publishing Group UK London}
}

@article{pan2023real,
  title={Real higher-order {Weyl} photonic crystal},
  author={Pan, Yuang and Cui, Chaoxi and Chen, Qiaolu and Chen, Fujia and Zhang, Li and Ren, Yudong and Han, Ning and Li, Wenhao and Li, Xinrui and Yu, Zhi-Ming and others},
  journal={Nat. Commun.},
  volume={14},
  number={1},
  pages={6636},
  year={2023},
  publisher={Nature Publishing Group UK London}
}

@article{ma2024observation,
  title={Observation of higher-order nodal-line semimetal in phononic crystals},
  author={Ma, Qiyun and Pu, Zhenhang and Ye, Liping and Lu, Jiuyang and Huang, Xueqin and Ke, Manzhu and He, Hailong and Deng, Weiyin and Liu, Zhengyou},
  journal={Phys. Rev. Lett.},
  volume={132},
  number={6},
  pages={066601},
  year={2024},
  publisher={APS}
}

@article{xiang2024,
  title={Demonstration of acoustic higher-order topological {Stiefel-Whitney} semimetal},
  author={Xiang, Xiao and Peng, Yu-Gui and Gao, Feng and Wu, Xiaoxiao and Wu, Peng and Chen, Zhaoxian and Ni, Xiang and Zhu, Xue-Feng},
  journal={Phys. Rev. Lett.},
  volume={132},
  number={19},
  pages={197202},
  year={2024},
  publisher={APS}
}

@article{nomura2018,
  author  = {Nomura, Takafumi and Habe, Tetsuro and Sakamoto, Ryota and Koshino, Mikito},
  title   = {Three-dimensional graphdiyne as a topological nodal-line semimetal},
  journal = {Phys. Rev. Mater.},
  volume  = {2},
  number  = {5},
  pages   = {054204},
  year    = {2018},
  doi     = {10.1103/PhysRevMaterials.2.054204},
}

@article{wang2019,
  author  = {Wang, Zhijun and Wieder, Benjamin J. and Li, Jian and Yan, Binghai and Bernevig, B. Andrei},
  title   = {Higher-Order Topology, Monopole Nodal Lines, and the Origin of Large {Fermi} Arcs in Transition Metal Dichalcogenides {$X$Te$_2$} ({$X$} = {Mo}, {W})},
  journal = {Phys. Rev. Lett.},
  volume  = {123},
  number  = {18},
  pages   = {186401},
  year    = {2019},
  doi     = {10.1103/PhysRevLett.123.186401},
}

@article{armitage2018weyl,
  author       = {Armitage, N. P. and Mele, E. J. and Vishwanath, A.},
  title        = {Weyl and {Dirac} semimetals in three-dimensional solids},
  journal      = {Rev. Mod. Phys.},
  volume       = {90},
  number       = {1},
  pages        = {015001},
  year         = {2018},
  doi          = {10.1103/RevModPhys.90.015001},
  url          = {https://doi.org/10.1103/RevModPhys.90.015001}
}

@article{bouhon2019,
  author  = {Bouhon, Adrien and Black-Schaffer, Annica M. and Slager, Robert-Jan},
  title   = {Wilson loop approach to fragile topology of split elementary band representations and topological crystalline insulators with time-reversal symmetry},
  journal = {Phys. Rev. B},
  volume  = {100},
  pages   = {195135},
  year    = {2019},
  doi     = {10.1103/PhysRevB.100.195135}
}

@article{fang2016topological,
  author       = {Fang, Chen and Weng, Hongming and Dai, Xi and Fang, Zhong},
  title        = {Topological nodal line semimetals},
  journal      = {Chin. Phys. B},
  volume       = {25},
  number       = {11},
  pages        = {117106},
  year         = {2016},
  doi          = {10.1088/1674-1056/25/11/117106},
  url          = {https://doi.org/10.1088/1674-1056/25/11/117106}
}

@article{schoop2016,
  author  = {Schoop, Leslie M. and Ali, Mazhar N. and Strasser, Carola and Topp, Andreas and Varykhalov, Andrei and Marchenko, Dmytro and Duppel, Volker and Parkin, Stuart S. P. and Lotsch, Bettina V. and Ast, Christian R.},
  title   = {Dirac cone protected by non-symmorphic symmetry and three-dimensional {Dirac} line node in {ZrSiS}},
  journal = {Nat. Commun.},
  volume  = {7},
  pages   = {11696},
  year    = {2016},
  doi     = {10.1038/ncomms11696}
}

@article{zhao2016,
  author  = {Zhao, Y. X. and Schnyder, Andreas P. and Wang, Z. D.},
  title   = {Unified Theory of {P}{T} and {C}{P} Invariant Topological Metals and Nodal Superconductors},
  journal = {Phys. Rev. Lett.},
  volume  = {116},
  pages   = {156402},
  year    = {2016},
  doi     = {10.1103/PhysRevLett.116.156402},
  
}

@article{zhao2017,
  author  = {Zhao, Y. X. and Lu, Y.},
  title   = {{P}{T}-Symmetric Real {Dirac} Fermions and Semimetals},
  journal = {Phys. Rev. Lett.},
  volume  = {118},
  pages   = {056401},
  year    = {2017},
  doi     = {10.1103/PhysRevLett.118.056401},
  
}

@article{ali2016,
  title = {Butterfly magnetoresistance, quasi-{2D} {Dirac} {Fermi} surface and topological phase transition in {ZrSiS}},
  author = {Ali, Mazhar N. and Schoop, Leslie M. and Garg, Chirag and Lippmann, Judith M. and Lara, Erik and Lotsch, Bettina and Parkin, Stuart S. P.},
  journal = {Sci. Adv.},
  volume = {2},
  number = {12},
  pages = {e1601742},
  year = {2016},
  doi = {10.1126/sciadv.1601742}
}

@article{singha2017,
  title = {Large nonsaturating magnetoresistance and signature of nondegenerate {Dirac} nodes in {ZrSiS}},
  author = {Singha, Ratnadwip and Pariari, Arnab Kumar and Satpati, Biswarup and Mandal, Prabhat},
  journal = {Proc. Natl. Acad. Sci. U.S.A.},
  volume = {114},
  number = {10},
  pages = {2468--2473},
  year = {2017},
  doi = {10.1073/pnas.1618004114}
}

@article{kaganov1979,
  title={Electron theory of metals and geometry},
  author={Kaganov, Moisei I and Lifshits, Il'ya Mikhailovich},
  journal={Sov. Phys. Usp.},
  volume={22},
  number={11},
  pages={904--927},
  year={1979}
}

@article{ahn2017,
  title={Electrodynamics on {Fermi} cyclides in nodal line semimetals},
  author={Ahn, Seongjin and Mele, EJ and Min, Hongki},
  journal={Phys. Rev. Lett.},
  volume={119},
  number={14},
  pages={147402},
  year={2017},
  publisher={APS}
}

@article{cohen1961,
  title={Magnetic breakdown in crystals},
  author={Cohen, Morrel H and Falicov, LM},
  journal={Phys. Rev. Lett.},
  volume={7},
  number={6},
  pages={231},
  year={1961},
  publisher={APS}
}

@article{blount1962,
  title={Bloch electrons in a magnetic field},
  author={Blount, EI},
  journal={Phys. Rev.},
  volume={126},
  number={5},
  pages={1636},
  year={1962},
  publisher={APS}
}

@article{alexandradinata2017,
  title={Geometric phase and orbital moment in quantization rules for magnetic breakdown},
  author={Alexandradinata, A and Glazman, Leonid},
  journal={Phys. Rev. Lett.},
  volume={119},
  number={25},
  pages={256601},
  year={2017},
  publisher={APS}
}

@article{alexandradinata2018,
  title={Semiclassical theory of {Landau} levels and magnetic breakdown in topological metals},
  author={Alexandradinata, A and Glazman, Leonid},
  journal={Phys. Rev. B},
  volume={97},
  number={14},
  pages={144422},
  year={2018},
  publisher={APS}
}

@article{liu2011,
  author  = {Liu, Cheng-Cheng and Jiang, Hua and Yao, Yugui},
  title   = {Low-Energy Effective {Hamiltonian} Involving Spin-Orbit Coupling in Silicene and Two-Dimensional Germanium and Tin},
  journal = {Phys. Rev. B},
  volume  = {84},
  number  = {19},
  pages   = {195430},
  year    = {2011},
  doi     = {10.1103/PhysRevB.84.195430},
}

@article{xiao2010berry,
  author        = {Xiao, Di and Chang, Ming-Che and Niu, Qian},
  title         = {{Berry} Phase Effects on Electronic Properties},
  journal       = {Reviews of Modern Physics},
  volume        = {82},
  pages         = {1959--2007},
  year          = {2010},
  doi           = {10.1103/RevModPhys.82.1959},
  eprint        = {0907.2021},
  archivePrefix = {arXiv},
  primaryClass  = {cond-mat.mes-hall}
}

@article{fuchs2010topological,
  author        = {Fuchs, J. N. and Pi{\'e}chon, F. and Goerbig, M. O. and Montambaux, G.},
  title         = {Topological {Berry} Phase and Semiclassical Quantization of Cyclotron Orbits for Two Dimensional Electrons in Coupled Band Models},
  journal       = {European Physical Journal B},
  volume        = {77},
  pages         = {351--362},
  year          = {2010},
  doi           = {10.1140/epjb/e2010-00259-2},
  eprint        = {1006.5632},
  archivePrefix = {arXiv},
  primaryClass  = {cond-mat.mes-hall}
}
\end{document}

\title{Supplemental Material for\texorpdfstring{\\}{: }
``Quantum Oscillation Signatures of \texorpdfstring{$\mathbb{Z}_2$}{Z2} Monopole Charge in Nodal-Ring Semimetals''}

\author{Aravind Karthigeyan}
\affiliation{Department of Physics, The University of Texas at Austin, Austin, Texas 78712, USA}
\author{Yoonseok Hwang}
\affiliation{Blackett Laboratory, Imperial College London, London SW7 2AZ, United Kingdom}
\author{Bohm-Jung Yang}
\affiliation{Department of Physics and Astronomy, Seoul National University, Seoul 08826, Korea}
\affiliation{Center for Theoretical Physics (CTP), Seoul National University, Seoul 08826, Korea}
\affiliation{Institute of Applied Physics, Seoul National University, Seoul 08826, Korea}
\author{Junyeong Ahn}
\affiliation{Department of Physics, The University of Texas at Austin, Austin, Texas 78712, USA}

\date{\today}
\maketitle

% Production version: omit the printed table of contents.
% The section map is already in the headings and in Sec.~\ref{sec:roadmap}.
% Keeping \tableofcontents causes the PDF text layer to flatten the title,
% author block, affiliations, and contents into one long boilerplate paragraph.
\clearpage

%%%%%%%%%%%%%%%%%%%%%%%%%%%%%%%%%%%%%%%%%%%%%%%%%%%%%%%%%%%%%%%%%%%%%%%%%%%%%%
\section{Roadmap, notation, and status of results}
\label{sec:roadmap}

This Supplemental Material collects the derivations and implementation details that underlie the main paper.
The central claim of the paper is that magnetic quantum oscillations diagnose the $\mathbb{Z}_2$ monopole charge $\wtwo$ of a nodal ring.
The most direct diagnostic uses $\B\parallel \hat z$, where the toroidal extremal orbits detect the $\wtwo$-enforced occupied-band thread piercing the torus hole. The complementary $\B\parallel \hat x$ geometry separates linking and nonlinking series by conductivity component.

Throughout, we distinguish the following topological invariants.
The invariants $\wone$ and $\wtwo$ are defined for the bands below the nodal-ring energy and are therefore independent of the Fermi level.
By contrast, $\Phitheta$ and $\Phiphi$ denote Berry phases of the \emph{Fermi-level cyclotron orbits}: $\Phitheta$ for the poloidal loop that links the nodal ring, and $\Phiphi$ for the toroidal loop that winds along the nodal-ring direction.
The Onsager offset is denoted by $\gamma$, and the universal three-dimensional stationary-phase contribution by $\varphi_{\rm 3D}=\pm \pi/4$.
For the nonlinking in-plane family, analytic formulas use the subscript notation $F_\star$ and $\gamma_\star$, while the paper and figure labels use $F^\ast$; these denote the same orbit family.

The main results have the following logical status.
\begin{enumerate}
  \item \emph{Exact continuum-model results:} the $\B\parallel\hat z$ Landau spectrum, the frequency formulas $F_\pm=(m_0\pm k_F)^2/2$ with $k_F=\EF$ in the $\hbar=e=v=1$ units used in the continuum derivation, the absence of a $1/2$ Onsager offset for the monopole ring, and the operator identities used in the transverse Kubo reduction.
  \item \emph{Asymptotic but controlled analysis:} the Lifshitz--Kosevich (LK) forms obtained from Poisson resummation and endpoint stationary phase for $\B\parallel\hat z$, and the semiclassical Lifshitz--Kosevich treatment for $\B\parallel\hat x$.
  \item \emph{Thin-torus asymptotics:} the universal constants controlling the maximal $k_x$ slices for $\B\parallel\hat x$, including $c\simeq0.65223$, $K\simeq6.19$, $c_0\equiv 3K/(4\pi)\simeq1.48$, the maximal-orbit velocity coefficient $\tilde c_y\simeq0.60875$, and the curvature coefficient $\kappa\simeq8.07888$.
  \item \emph{Numerical analysis:} the lattice regularization, broadened Kubo conductivity, fast-Fourier-transform (FFT) analysis, and multi-frequency phase extraction used to corroborate the $\B\parallel\hat z$ phase pinning and the $\B\parallel\hat x$ component-resolved phase separation.
\end{enumerate}

The remainder of the supplement is organized as follows.
Section~\ref{sec:model-topology} collects the continuum model, the toroidal Fermi surface, and the real-Bloch-state topology argument for the ``ring-and-thread'' structure.
Section~\ref{sec:phase-conventions} fixes the phase conventions and makes the harmonic-by-harmonic $\nu\pi$ phase shift between $\wtwo=1$ and $\wtwo=0$ explicit.
Section~\ref{sec:Bz-LL} derives the exact $\B\parallel\hat z$ Landau levels and the exact Onsager frequencies.
Section~\ref{sec:Bz-sigma} gives the transverse Kubo/Lifshitz--Kosevich result for $\sigma_{xx}=\sigma_{yy}$ at $\B\parallel\hat z$ and contrasts it with the longitudinal channel.
Section~\ref{sec:Bx} treats the $\B\parallel\hat x$ torus geometry, orbit averages, and component-resolved Lifshitz--Kosevich series, and also records why an exact $2\times2$ Landau-block reduction is obstructed in this geometry.
Section~\ref{sec:lattice} documents the lattice regularization and numerical analysis pipeline.
Section~\ref{sec:units} converts the continuum frequencies to SI units and tabulates representative graphdiyne estimates.
Section~\ref{sec:soc} derives the geometric Berry phase of a weakly gapped Dirac orbit, distinguishes Berry-only and measured LK offsets by keeping track of orbital-magnetic-moment and spinful corrections, and separates orbit-existence, frequency-shift, phase-robustness, and field-window criteria.
Section~\ref{sec:master} then compiles a candidate-material master reference, including an explicit derivation of the graphdiyne parameters from the density-functional-theory (DFT)-fitted model and a screening table for the broader Si/Ge/transition-metal-dichalcogenide (TMD) families.

%%%%%%%%%%%%%%%%%%%%%%%%%%%%%%%%%%%%%%%%%%%%%%%%%%%%%%%%%%%%%%%%%%%%%%%%%%%%%%
\section{Continuum model, toroidal Fermi surface, and real-Bloch-state topology}
\label{sec:model-topology}

\subsection{Four-band continuum model and exact dispersion}

The main text studies the four-band real Hamiltonian~\cite{fang2015,ahn2018}
\begin{equation}
H_0(\bk)=
\hbar v\bigl[
(k_x-m_0t_x)s_x+k_y t_y s_y+k_zs_z
\bigr],
\label{eq:H0}
\end{equation}
where $s_i$ and $t_i$ are Pauli matrices acting in two independent internal subspaces.
In most parts of this work, we take the units
\begin{equation}
    \hbar=e=v=1,
\end{equation}
where $-e$ is the electron charge.
In a basis with $PT=\mathcal K$ (complex conjugation), all matrix elements are real and the Bloch eigenvectors can be chosen real away from degeneracies~\cite{fang2015,zhao2016,zhao2017}.

Define
\begin{equation}
A_x\equiv k_x-m_0 t_x,\qquad A_y\equiv k_y t_y,\qquad A_z\equiv k_z,
\end{equation}
so that $H_0=A_x s_x+A_y s_y+A_z s_z$.
Using $s_i s_j=\delta_{ij}\I_2+i \epsilon_{ijk}s_k$ and $[s_i,t_j]=0$, one finds
\begin{equation}
H_0^2=(A_x^2+A_y^2+A_z^2)\I_4+i  s_z[A_x,A_y].
\end{equation}
Since $[A_x,A_y]=-2i  m_0 k_y t_z$, this becomes
\begin{equation}
H_0^2=(k_z^2+k_\rho^2+m_0^2)\I_4-2m_0 M_0,
\qquad
M_0\equiv k_x t_x-k_y t_z s_z,
\label{eq:H02}
\end{equation}
with $k_\rho\equiv\sqrt{k_x^2+k_y^2}$.
Because $M_0^2=k_\rho^2\I_4$, the eigenvalues of $M_0$ are $\pm k_\rho$, and hence the four-band spectrum is
\begin{equation}
E_\pm^2(\bk)=k_z^2+\bigl(m_0\pm k_\rho\bigr)^2.
\label{eq:dispersion}
\end{equation}
The semimetallic nodal ring lies on the $E_-$ branch at $k_z=0$ and $k_\rho=m_0$.
The two occupied bands (and, likewise, the two unoccupied bands) are degenerate along the line $k_x=k_y=0$ for all $k_z$.
This line is the occupied-band ``thread'' that links the nodal ring once.

\subsection{Toroidal Fermi surface at moderate doping}

For doping levels $0<\EF<m_0$ measured from the nodal energy, only the $E_-$ branch crosses the Fermi level.
Setting $E_-(\bk)=\pm \EF$ gives an exact torus,
\begin{equation}
\left(\sqrt{k_x^2+k_y^2}-m_0\right)^2+k_z^2=\EF^2.
\label{eq:torus}
\end{equation}
At fixed $k_z$, the field-perpendicular sections for $\B\parallel \hat z$ are circles of radii
\begin{equation}
k_\rho^{\pm}(k_z)=m_0\pm\sqrt{\EF^2-k_z^2},
\qquad |k_z|\le \EF,
\label{eq:kperppm-kz}
\end{equation}
with areas
\begin{equation}
A_\pm(k_z)=\pi\Bigl(m_0\pm\sqrt{\EF^2-k_z^2}\Bigr)^2.
\label{eq:Azpm}
\end{equation}
At $k_z=0$ these reduce to the two extremal toroidal orbits with areas $A_\pm\equiv A_\pm(0)=\pi(m_0\pm \EF)^2$ and Onsager frequencies $F_\pm=A_\pm/2\pi$ in the same dimensionless units.

\subsection{Real-Bloch-state topology, the ring-and-thread structure, and why the toroidal Berry phase is \texorpdfstring{$\pi$}{pi}}
\label{subsec:ring-thread}

The distinction between an ordinary Berry-phase nodal ring and a $\mathbb Z_2$-monopole nodal ring is encoded in the real occupied-band bundle.
For a loop $\ell$ linking a nodal line that separates occupied and unoccupied bands, the first Stiefel--Whitney class is $\wone(\ell)=\Phi_{\rm tot}(\ell)/\pi\;\mathrm{mod}\;2$, where $\Phi_{\rm tot}$ is the determinant Berry phase of the occupied subspace on the loop.
By contrast, the thread appearing below is a nodal line internal to the occupied-band subspace; it controls the $\wtwo$ linking formula and gives a $\pi$ Berry phase to the individual Fermi band encircling it, even though the determinant phase of the full occupied subspace on that small loop is trivial.
On a torus $T^2$ wrapping a nodal ring $\gamma$, the second Stiefel--Whitney class is related to the Berry phases on the toroidal and poloidal cycles and can be written as a sum of linking numbers with occupied-band nodal lines~\cite{ahn2018}:
\begin{equation}
\wtwo(T^2)=\sum_j \mathrm{Lk}\bigl(\gamma,\widetilde\gamma_j\bigr)\mod 2.
\label{eq:w2linking}
\end{equation}
Here $\widetilde\gamma_j$ denotes the $j$th nodal line formed within the occupied-band subspace, while
$\mathrm{Lk}(C_1,C_2)$ denotes the integer linking number of two closed curves $C_1$ and $C_2$ in momentum space.
Only the parity of the sum enters Eq.~\eqref{eq:w2linking}.
In a local continuum description, the straight thread should be understood as the local representative of the closed occupied-band nodal line in the full Brillouin zone.
In the present model there is one such occupied-band line, namely the thread $k_x=k_y=0$, and it links the semimetallic ring once.
Therefore $\wtwo=1$.

The consequence for the Fermi surface is immediate.
A toroidal cyclotron orbit on Eq.~\eqref{eq:torus} winds around the nodal-ring direction and is homotopic to a small loop encircling the occupied-band thread.
For hole doping, the individual topmost occupied band on this loop acquires a quantized $\pi$ Berry phase from the enclosed thread.
Thus the toroidal Berry phase of the Fermi surface at small hole doping is quantized to
\begin{equation}
\Phiphi=\pi .
\label{eq:Phiphi-pi}
\end{equation}
Equation~\eqref{eq:Phiphi-pi} is therefore the Fermi-surface manifestation of $\wtwo(T^2)=1$ for the monopole ring; it is not a second definition of $\wtwo$ itself.
The same argument for the unoccupied bands shows that electron-doped Fermi surfaces also carry the same Berry phase.
Likewise, the poloidal loop links the semimetallic nodal ring itself, so
\begin{equation}
\Phitheta=\pi
\label{eq:Phitheta-pi}
\end{equation}
for the same Fermi surfaces.
The $\B\parallel\hat z$ geometry probes Eq.~\eqref{eq:Phiphi-pi}, while the $\B\parallel\hat x$ geometry probes Eq.~\eqref{eq:Phitheta-pi}.

\subsection{Control model for a trivial \texorpdfstring{$\wtwo=0$}{w2=0} nodal ring}
\label{subsec:trivial-control}

The $\wtwo$ dependence is exposed by contrasting Eq.~\eqref{eq:H0} with a local two-band control model for an ordinary $PT$-symmetric nodal ring,
\begin{equation}
H_{\rm triv}(\bk)=\bigl(k_\rho-m_0\bigr)\sigma_x+k_z\sigma_z.
\label{eq:Htriv}
\end{equation}
This model has the same local ring dispersion near $k_\rho=m_0$, $k_z=0$, but there is no occupied-band thread piercing the torus hole.
The poloidal loop still links the nodal ring and therefore carries $\Phitheta=\pi$.
By contrast, the toroidal loop does not link any band degeneracy and can be contracted on the torus without crossing a node, so
\begin{equation}
\Phiphi=0 .
\label{eq:Phiphi-zero}
\end{equation}
The vanishing toroidal Berry phase is the corresponding Fermi-surface result for this $\wtwo=0$ control ring.
This simple control model is enough to establish the qualitative difference between the monopole and trivial cases in the $\B\parallel\hat z$ oscillations.

%%%%%%%%%%%%%%%%%%%%%%%%%%%%%%%%%%%%%%%%%%%%%%%%%%%%%%%%%%%%%%%%%%%%%%%%%%%%%%
\section{Phase conventions and the harmonic \texorpdfstring{$\nu\pi$}{nu pi} shift}
\label{sec:phase-conventions}

\subsection{Onsager quantization, Berry phase, and 3D curvature phase}

For a closed cyclotron orbit $\ell$ in the plane perpendicular to the magnetic field,
semiclassical quantization gives~\cite{shoenberg,mikitik1999}
\begin{equation}
A(\EF)=\frac{2\pi eB}{\hbar}\bigl(n+\gamma\bigr),
\qquad
\gamma=\frac12-\frac{\Phib}{2\pi}.
\label{eq:Onsager}
\end{equation}
The Onsager frequency is $F\equiv (\hbar/2\pi e)A$.
In a $PT$-symmetric system, $\Phib\in\{0,\pi\}$ mod $2\pi$, so the offset is pinned to $\gamma\in\{1/2,0\}$.

In three dimensions one must also integrate over the field-parallel momentum near an extremal slice.
This contributes the standard stationary-phase factor depending on the 3D curvature (i.e., on the Morse index $m$)~\cite{shoenberg,li2018}
\begin{equation}
\varphi_{\rm 3D}=\frac{\pi}{4}(1-2m)=
\begin{cases}
+\pi/4, & \text{minimal extremal slice},\\
-\pi/4, & \text{maximal extremal slice}.
\end{cases}
\label{eq:phi3D}
\end{equation}
Accordingly, a generic oscillatory conductivity component can be parametrized as
\begin{equation}
\Delta\sigma_{ii}(B)\propto
\sum_\alpha A_\alpha(B)
\cos\!\left[2\pi\left(\frac{F_\alpha}{B}-\gamma_\alpha\right)+\varphi_{\rm 3D,\alpha}\right],
\label{eq:genericLK}
\end{equation}
where $\alpha$ labels extremal orbit families.

Equation~\eqref{eq:genericLK} assumes that the transport amplitude multiplying
the stationary-phase integral is smooth and nonzero at the extremal slice. If the
conductivity component vanishes on the extremal orbit, the first nonzero term in
the amplitude expansion can change both the power-law envelope and the
stationary-phase contribution. This occurs for the longitudinal
$\B\parallel\hat z$ channel discussed in Sec.~\ref{subsec:Bz-sigma-scaling}.

\subsection{Explicit \texorpdfstring{$\nu\pi$}{nu pi} phase shift between \texorpdfstring{$\wtwo=1$}{w2=1} and \texorpdfstring{$\wtwo=0$}{w2=0}}
\label{subsec:kpi-shift}

For the toroidal $\B\parallel\hat z$ series, the monopole ring has $\Phiphi=\pi$ by Eq.~\eqref{eq:Phiphi-pi}, hence
\begin{equation}
\gamma^{\rm mono}_\pm=0.
\label{eq:gammamono}
\end{equation}
For the control ring with $\wtwo=0$, Eq.~\eqref{eq:Phiphi-zero} gives
\begin{equation}
\gamma^{\rm triv}_\pm=\frac12.
\label{eq:gammatriv}
\end{equation}
Equations~\eqref{eq:Phiphi-pi} and \eqref{eq:Phiphi-zero} can equivalently be summarized for the toroidal $\B\parallel\hat z$ orbit as $\Phiphi=\wtwo\pi$ modulo $2\pi$, so Eq.~\eqref{eq:Onsager} gives $\gamma_\pm(\wtwo)=1/2-\wtwo/2$ modulo one.
Therefore the $\nu$th Fourier harmonic $e^{2\pi i \nu F/B}$ generated by the Poisson-resummed Landau-level comb, cf. Eq.~\eqref{eq:Poisson-delta-comb}, differs by the phase
\begin{equation}
\Delta\varphi_\nu
=2\pi \nu\bigl(\gamma^{\rm triv}_\pm-\gamma^{\rm mono}_\pm\bigr)
=\nu\pi
\equiv \nu\wtwo\pi\quad (\mathrm{mod}\;2\pi).
\label{eq:kpi-shift}
\end{equation}
This is the precise meaning of the statement in the paper that the $\nu$th harmonic carries a topological phase shift $\nu\wtwo\pi$ in the $\B\parallel\hat z$ geometry.
The 3D curvature phases do \emph{not} distinguish the trivial and monopole cases; they are fixed only by whether the slice is a minimum or a maximum.

\subsection{What is fixed by topology and what is not}

For the spinless real continuum and lattice models analyzed in Secs.~\ref{sec:model-topology}--\ref{sec:lattice}, the Berry phases entering Eq.~\eqref{eq:Onsager} are quantized zero-field quantities.
The exact $\B\parallel\hat z$ Landau spectrum independently confirms the corresponding leading offsets $\gamma_\pm=0$ for the monopole toroidal series.
In this setting the qualitative distinction between the monopole and trivial toroidal series is the change $\gamma=0\leftrightarrow 1/2$, and temperature, disorder broadening, and smooth LK amplitudes do not alter the topological offset in Eq.~\eqref{eq:kpi-shift}.
For generic spinful bands gapped by spin-orbit coupling (SOC), however, the measured LK phase can also receive orbital-magnetic-moment, Zeeman, and non-Abelian band-mixing corrections; Sec.~\ref{sec:soc} discusses this distinction.

%%%%%%%%%%%%%%%%%%%%%%%%%%%%%%%%%%%%%%%%%%%%%%%%%%%%%%%%%%%%%%%%%%%%%%%%%%%%%%
\section{Exact Landau spectrum for \texorpdfstring{$\B\parallel\hat z$}{B parallel z}}
\label{sec:Bz-LL}

In this section we derive the exact Landau spectrum of the continuum model in a uniform magnetic field $\B=B_z\hat z$.
We first keep the sign of the physical charge explicit and then specialize to the algebraic orientation used in the rest of the derivation.

\subsection{Magnetic algebra and ladder operators}

Let $q$ be the carrier charge and define
\[
\mathcal B\equiv \frac{|qB_z|}{\hbar},
\qquad
s_B\equiv \mathrm{sgn}(qB_z).
\]
Minimal coupling replaces $(k_x,k_y)\to(\Pi_x,\Pi_y)$, where
\[
\Pi_i=-i \partial_i-\frac{q}{\hbar}A_i,
\]
with
\begin{equation}
[\Pi_x,\Pi_y]=i  s_B\mathcal B,
\qquad
\Pi_\rho^2\equiv \Pi_x^2+\Pi_y^2.
\label{eq:Pi-comm}
\end{equation}
For a physical electron, $q=-e$, so a positive laboratory field $B_z>0$ gives $s_B=-1$.
The sign reverses the Landau-level chirality and exchanges which circular ladder operator plays the role of the annihilation operator, but it does not change the Landau-level energies, extremal areas, or Onsager offsets used below.
For compactness, after Eq.~\eqref{eq:ladder} we choose the algebraic orientation $s_B=+1$ and write $\mathcal B\to B$.
The opposite orientation is obtained by reversing the ladder chirality.

The sign-adapted Landau ladder operators are
\begin{equation}
a_{s_B}=\frac{\ell_{\mathcal B}}{\sqrt2}(\Pi_x+i  s_B\Pi_y),
\qquad
a_{s_B}^\dagger=\frac{\ell_{\mathcal B}}{\sqrt2}(\Pi_x-i  s_B\Pi_y),
\qquad
[a_{s_B},a_{s_B}^\dagger]=1,
\label{eq:ladder}
\end{equation}
where $\ell_{\mathcal B}=1/\sqrt{\mathcal B}$.
In the algebraic orientation used from now on, $a\equiv a_{+}$ and
\begin{equation}
\Pi_\rho^2=B(2N+1),
\qquad N=a^\dagger a.
\label{eq:PiperpN}
\end{equation}
The field Hamiltonian is
\begin{equation}
H(\bm\Pi,k_z)=(\Pi_x-m_0 t_x)s_x+\Pi_y t_y s_y+k_z s_z.
\label{eq:HBz}
\end{equation}

\subsection{Operator identities and exact reduction to \texorpdfstring{$2\times2$}{2x2} Landau blocks}

Define
\begin{equation}
\Sigma\equiv t_y s_z,
\qquad
M\equiv \Pi_x t_x-\Pi_y t_z s_z.
\label{eq:MSigma}
\end{equation}
A direct Pauli-algebra computation gives the exact identities
\begin{align}
H^2 &= (k_z^2+\Pi_\rho^2+m_0^2)\I_4-2m_0 M-B\Sigma,
\label{eq:H2}\\
M^2 &= \Pi_\rho^2\I_4-B\Sigma,
\label{eq:M2}\\
\{M,\Sigma\}&=0.
\label{eq:anticomm}
\end{align}
Because $M$ anticommutes with $\Sigma$, the spectrum of $H^2$ can be obtained in paired subspaces that mix adjacent Landau indices with opposite $\Sigma$ eigenvalue.

\paragraph*{Explicit $\Sigma$-diagonal representation of $M$.}
To write the block reduction explicitly, rotate the $t$-Pauli matrices by
\begin{equation}
U_1\equiv e^{-i  \frac{\pi}{4}t_x},
\qquad
U_1 t_y U_1^\dagger=t_z,
\qquad
U_1 t_z U_1^\dagger=-t_y.
\label{eq:U1-rotation}
\end{equation}
In this rotated basis,
\begin{equation}
\Sigma'\equiv U_1\Sigma U_1^\dagger=s_z t_z,
\qquad
M'\equiv U_1 M U_1^\dagger=\Pi_x t_x+\Pi_y t_y s_z.
\label{eq:SigmaPrime-MPrime}
\end{equation}
Projecting to the $s_z=\sigma=\pm1$ sector with
\begin{equation}
P_\sigma\equiv \frac{1+\sigma s_z}{2},
\end{equation}
one obtains
\begin{equation}
\Sigma'_\sigma=\sigma t_z,
\qquad
M'_\sigma=\Pi_x t_x+\sigma \Pi_y t_y.
\label{eq:SigmaM-sector}
\end{equation}
Now define sector-adapted Pauli matrices
\begin{equation}
\tau_x^{(\sigma)}\equiv t_x,
\qquad
\tau_y^{(\sigma)}\equiv \sigma t_y,
\qquad
\tau_z^{(\sigma)}\equiv \sigma t_z,
\label{eq:tau-sigma-def}
\end{equation}
which obey the Pauli algebra. In this notation,
\begin{equation}
\Sigma'_\sigma=\tau_z^{(\sigma)},
\qquad
M'_\sigma=\Pi_x\tau_x^{(\sigma)}+\Pi_y\tau_y^{(\sigma)}.
\label{eq:SigmaM-sector-tau}
\end{equation}
In the $\tau_z^{(\sigma)}$ eigenbasis,
\begin{equation}
M'_\sigma=
\begin{pmatrix}
0 & \Pi_x-i \Pi_y\\
\Pi_x+i \Pi_y & 0
\end{pmatrix}
=
\sqrt{2B}
\begin{pmatrix}
0 & a^\dagger\\
a & 0
\end{pmatrix},
\label{eq:Mprime-matrix}
\end{equation}
where
\begin{equation}
a=\frac{\Pi_x+i \Pi_y}{\sqrt{2B}},
\qquad
a^\dagger=\frac{\Pi_x-i \Pi_y}{\sqrt{2B}}.
\label{eq:ladder-Mprime}
\end{equation}
This form is independent of $\sigma$, so the $s_z$ doublet is a spectator.
After a trivial permutation of the internal basis that groups the $\Sigma=\pm1$ sectors together, one may write
\begin{equation}
U M U^\dagger=
\sqrt{2B}
\begin{pmatrix}
0 & a^\dagger\\
a & 0
\end{pmatrix}\otimes \I_2,
\qquad
U\Sigma U^\dagger=
\begin{pmatrix}
\I_2 & 0\\
0 & -\I_2
\end{pmatrix}.
\label{eq:UMUdagger}
\end{equation}
Hence
\begin{equation}
M\ket{n,+,\mu}=\sqrt{2Bn}\ket{n-1,-,\mu},
\qquad
M\ket{n-1,-,\mu}=\sqrt{2Bn}\ket{n,+,\mu},
\label{eq:M-action-explicit}
\end{equation}
with spectator doublet index $\mu=1,2$, while the unpaired state satisfies
\begin{equation}
M\ket{0,+,\mu}=0.
\label{eq:M-zero-state}
\end{equation}
The invariant subspace relevant for $n\ge1$ is therefore
\begin{equation}
\mathcal V_n=\mathrm{span}\bigl\{\ket{n,+,\mu},\ket{n-1,-,\mu}\bigr\}.
\label{eq:Vn-def}
\end{equation}
Restricted to $\mathcal V_n$, one obtains
\begin{equation}
H^2_{(n)}=
\bigl(k_z^2+m_0^2+2Bn\bigr)\I_2-2m_0\sqrt{2Bn}\,\sigma_x.
\label{eq:H2block}
\end{equation}
Diagonalization yields the exact squared energies
\begin{equation}
E_{n,\lambda}^2(k_z)=k_z^2+\bigl(m_0+\lambda\sqrt{2Bn}\bigr)^2,
\qquad n\ge1,\; \lambda=\pm,
\label{eq:LL}
\end{equation}
plus the unpaired $n=0$ branch
\begin{equation}
E_0^2(k_z)=k_z^2+m_0^2.
\label{eq:LL0}
\end{equation}
These expressions are exact within the continuum model.

\subsection{Fermi-level crossings and exact Onsager quantization without a \texorpdfstring{$1/2$}{1/2} shift}
\label{subsec:exact-onsager-Bz}

Assume electron doping with $0<\EF<m_0$ and focus on the only branch that intersects the Fermi level in this regime,
\begin{equation}
E_{n,-}(k_z)=\sqrt{k_z^2+\bigl(m_0-\sqrt{2Bn}\bigr)^2}.
\label{eq:Enminus}
\end{equation}
At fixed $k_z$, the Landau condition $E_{n,-}(k_z)=\EF$ implies
\begin{equation}
\sqrt{2Bn}=m_0\pm\sqrt{\EF^2-k_z^2}.
\label{eq:Landaucondition-kz}
\end{equation}
Using Eq.~\eqref{eq:Azpm}, this is equivalent to
\begin{equation}
\frac{A_\pm(k_z)}{2\pi B}=n,
\label{eq:exact-area-quantization}
\end{equation}
with \emph{no} additional $1/2$ shift.
At $k_z=0$ the two extremal frequencies are
\begin{equation}
F_\pm=\frac{(m_0\pm \EF)^2}{2},
\qquad
n=\frac{F_\pm}{B}.
\label{eq:Fpm}
\end{equation}
Comparing Eq.~\eqref{eq:Fpm} with the standard Landau-fan form $n=F/B-\gamma$ gives
\begin{equation}
\gamma_\pm=0.
\label{eq:gamma0}
\end{equation}
Thus the exact continuum Landau problem reproduces the topological statement $\Phiphi=\pi$ for \emph{both} toroidal series.
The additional 3D curvature phases are fixed by geometry:
\begin{equation}
\varphi_{\rm 3D,+}=-\frac{\pi}{4},
\qquad
\varphi_{\rm 3D,-}=+\frac{\pi}{4}.
\label{eq:phi3dpm}
\end{equation}
No weak-field or thin-torus approximation has entered Eqs.~\eqref{eq:LL}--\eqref{eq:phi3dpm}.
The extra assumption used below is only the many-Landau-level condition $F_\pm/B\gg1$, needed when the exact fan condition is converted into an LK conductivity series.
\subsection{Comparison with the trivial ring}

For the control model Eq.~\eqref{eq:Htriv}, Eq.~\eqref{eq:Phiphi-zero} gives $\Phiphi=0$ on the toroidal orbit.
The extremal areas are still $A_\pm(k_z)$, but the fan condition is now
\begin{equation}
\frac{A_\pm(k_z)}{2\pi B}=n+\frac12,
\label{eq:triv-onsager}
\end{equation}
so $\gamma_\pm^{\rm triv}=1/2$.
The $\pm\pi/4$ curvature phases are the same as in the monopole model.
The change from Eq.~\eqref{eq:exact-area-quantization} to Eq.~\eqref{eq:triv-onsager} is the offset produced by removing the occupied-band thread from the torus hole.

%%%%%%%%%%%%%%%%%%%%%%%%%%%%%%%%%%%%%%%%%%%%%%%%%%%%%%%%%%%%%%%%%%%%%%%%%%%%%%
\section{Oscillatory transverse conductivity for \texorpdfstring{$\B\parallel\hat z$}{B parallel z}}
\label{sec:Bz-sigma}

We next compute the transverse conductivity for the fan above.
The Kubo--Greenwood calculation gives the Landau-index current vertex $W_{NM}$.
In the many-level step, its near-diagonal large-$N$ form is replaced by the orbit-averaged transport weight used in the LK expansion.
The exact spectrum supplies $F_\pm$ and $\gamma_\pm$; the disorder vertex, lifetime, and Dingle broadening enter the amplitude.

\subsection{Microscopic Kubo formula and exact Landau-form-factor kernel}
\label{subsec:Bz-microscopic-kernel}

For $\B\parallel\hat z$ the current operators are
\begin{equation}
J_x=\pdv{H}{\Pi_x}=s_x,
\qquad
J_y=\pdv{H}{\Pi_y}=t_y s_y.
\label{eq:Jxy}
\end{equation}
At $T=0$, a convenient representation of the dc conductivity is the retarded--advanced Kubo--Greenwood formula~\cite{akkermans2007mesoscopic,rammer1986quantum,mahan2000manyparticle}
\begin{equation}
\sigma_{ii}=\frac{e^2}{2\pi V}
\Tr\!\left[\hat v_i \hat{G}^R(\EF)\hat v_i \hat{G}^A(\EF)\right].
\label{eq:KG}
\end{equation}
For short-range scalar disorder, the ladder resummation in the Landau basis can be encoded in a positive Landau-index kernel, so the transverse conductivity takes the form
\begin{equation}
\sigma_{yy}=g_d\,\frac{B_z}{2\pi}\,\frac{1}{2\pi}
\sum_{N,M}\int\!\frac{dk_z}{2\pi}\;
W_{NM}\,\mathcal A_N(\EF,k_z)\,\mathcal A_M(\EF,k_z),
\label{eq:sigyy-kernel}
\end{equation}
where $g_d$ is a model-dependent multiplicity factor. For a single spinless continuum copy of Eq.~\eqref{eq:H0}, $g_d=1$; it may be changed to include additional spin, valley, or symmetry-related pocket multiplicities.
This factor affects only the nonuniversal amplitude, not the frequency or Onsager offset, and
$\mathcal A_N(\varepsilon,k_z)=-2\Im G_N^R(\varepsilon,k_z)$ is the Landau-resolved spectral function.
Here $W_{NM}$ is the branch-projected scalar-disorder current-vertex kernel for the transverse conductivity.
In the regime $0<\EF<m_0$ used here, only the $\lambda=-1$ branch contributes at the Fermi level, so $W_{NM}\equiv W_{M-,N-}$.
The full branch-resolved kernel $W_{M\lambda',N\lambda}$ is defined in Eq.~\eqref{eq:WMNdef} below.

One convenient set of internal eigenvectors of $H^2$ is, for $n\ge1$,
\begin{equation}
\ket{\chi_{n,\lambda}}=\frac{1}{\sqrt2}\Bigl(\ket{n,+}-\lambda\ket{n-1,-}\Bigr),
\label{eq:chinlambda}
\end{equation}
where $\Sigma\ket{n,\pm}=\pm\ket{n,\pm}$.
For scalar disorder the internal structure enters only through the overlaps of these spinors.
Defining the standard Landau form factor
\begin{equation}
J_{mn}(\mathbf q_\perp)\equiv \matrixel{m}{e^{i  \mathbf q_\perp\cdot \mathbf r_\perp}}{n},
\label{eq:Jmn-def}
\end{equation}
one finds immediately
\begin{equation}
\matrixel{\chi_{m,\lambda'}}{e^{i \mathbf q_\perp\cdot\mathbf r_\perp}\I_4}{\chi_{n,\lambda}}
=\frac12\Bigl(J_{mn}+\lambda\lambda' J_{m-1,n-1}\Bigr).
\label{eq:formfactor-superposition}
\end{equation}
Therefore the $q_y^2$-weighted disorder kernel is
\begin{equation}
W_{M\lambda',N\lambda}
\equiv
\int d^2q_\perp\, q_y^2
\Bigl|\matrixel{\chi_{M,\lambda'}}{e^{i \mathbf q_\perp\cdot\mathbf r_\perp}\I_4}{\chi_{N,\lambda}}\Bigr|^2.
\label{eq:WMNdef}
\end{equation}

To evaluate this kernel explicitly, write $u\equiv q_\rho^2/(2B_z)$.
Using polar coordinates $q_x=q_\rho\cos\varphi$, $q_y=q_\rho\sin\varphi$, and $d^2q_\perp=q_\rho\,dq_\rho\,d\varphi$, one obtains
\begin{align}
\int d^2q_\perp\,q_y^2 f(u)
&=\int_0^{\infty} dq_\rho\,q_\rho^3 f\!\left(\frac{q_\rho^2}{2B_z}\right)
\int_0^{2\pi} d\varphi\,\sin^2\varphi \notag\\
&=\pi\int_0^{\infty} dq_\rho\,q_\rho^3 f\!\left(\frac{q_\rho^2}{2B_z}\right).
\end{align}
Since $q_\rho^2=2B_z u$, we have $q_\rho dq_\rho=B_z\,du$ and thus
$q_\rho^3dq_\rho=2B_z^2u\,du$.
Hence
\begin{equation}
\int d^2q_\perp\,q_y^2 f(u)=2\pi B_z^2\int_0^{\infty}du\,u\,f(u).
\label{eq:qangular-reduction}
\end{equation}

For diagonal oscillator matrix elements one uses
\begin{equation}
J_{NN}(u)=e^{-u/2}L_N(u),
\label{eq:JNN-Laguerre}
\end{equation}
with $L_N(u)$ the Laguerre polynomial.
The exact evaluation of the full kernel, however, requires the associated-Laguerre form of the Landau form factors, rather than the diagonal identity in Eq.~\eqref{eq:JNN-Laguerre} alone.
For $M,N\ge0$, define
\begin{equation}
n\equiv \min(M,N),
\qquad
d\equiv |M-N|.
\label{eq:nd-def}
\end{equation}
Then, up to a $u$- and $\varphi$-independent phase that is common to the pair $J_{MN}$ and $J_{M-1,N-1}$,
\begin{equation}
J_{MN}(u,\varphi)=e^{-u/2}\sqrt{\frac{n!}{(n+d)!}}\,(\sqrt u)^d L_n^d(u)\,e^{i (M-N)\varphi}.
\label{eq:JMN-associated}
\end{equation}
Because both $J_{MN}$ and $J_{M-1,N-1}$ carry the same azimuthal dependence $e^{i (M-N)\varphi}$, the angular integration in Eq.~\eqref{eq:WMNdef} remains elementary.
Writing
\begin{equation}
\mathcal F_{MN}(\mathbf q_\perp)
\equiv
\matrixel{\chi_{M,-}}{e^{i \mathbf q_\perp\cdot\mathbf r_\perp}\I_4}{\chi_{N,-}}
=
\frac12\Bigl(J_{MN}+J_{M-1,N-1}\Bigr),
\label{eq:Fmn-def}
\end{equation}
Eq.~\eqref{eq:qangular-reduction} gives
\begin{equation}
W_{MN}=2\pi B_z^2\,\mathcal W_{MN},
\label{eq:WMN-Wcal}
\end{equation}
with
\begin{equation}
\mathcal W_{MN}
=
\frac14\int_0^{\infty}du\,e^{-u}u^{d+1}
\Bigl[a_n L_n^d(u)+a_{n-1}L_{n-1}^d(u)\Bigr]^2,
\label{eq:Wcal-def}
\end{equation}
where
\begin{equation}
a_n\equiv\sqrt{\frac{n!}{(n+d)!}},
\qquad
a_{n-1}\equiv\sqrt{\frac{(n-1)!}{(n+d-1)!}},
\label{eq:an-def}
\end{equation}
and the $a_{n-1}$ term is absent when $n=0$.

The needed integrals follow from the associated-Laguerre orthogonality relation
\begin{equation}
\int_0^{\infty}du\,e^{-u}u^d L_m^d(u)L_n^d(u)
=
\frac{(n+d)!}{n!}\,\delta_{mn},
\label{eq:Laguerre-orthogonality-associated}
\end{equation}
together with the recurrence
\begin{equation}
uL_n^d(u)=(2n+d+1)L_n^d(u)-(n+1)L_{n+1}^d(u)-(n+d)L_{n-1}^d(u).
\label{eq:Laguerre-recurrence-associated}
\end{equation}
They imply
\begin{align}
\int_0^{\infty}du\,e^{-u}u^{d+1}\bigl[L_n^d(u)\bigr]^2
&=(2n+d+1)\frac{(n+d)!}{n!},
\label{eq:Laguerre-square-int}\\
\int_0^{\infty}du\,e^{-u}u^{d+1}L_n^d(u)L_{n-1}^d(u)
&=-\frac{(n+d)!}{(n-1)!}
\qquad (n\ge1).
\label{eq:Laguerre-cross-int}
\end{align}
In the physically relevant regime $M,N\ge1$ (the only one entering the Fermi-level sums for $0<\EF<m_0$), substituting Eqs.~\eqref{eq:Laguerre-square-int} and \eqref{eq:Laguerre-cross-int} into Eq.~\eqref{eq:Wcal-def} yields the exact closed form
\begin{equation}
W_{MN}
=
2\pi B_z^2
\left[
n+\frac{d}{2}-\frac12\sqrt{n(n+d)}
\right],
\qquad
n=\min(M,N),\quad d=|M-N|.
\label{eq:WMN-exact-closed}
\end{equation}
In particular, the diagonal block is
\begin{equation}
W_{NN}=\pi B_z^2 N.
\label{eq:WNN-final}
\end{equation}
Thus the exact kernel is not strictly tridiagonal; rather, it is the dense associated-Laguerre kernel of Eq.~\eqref{eq:WMN-exact-closed}.
For the many-Landau-level semiclassical reduction one only needs its near-diagonal large-$N$ envelope.
If $N,M\gg1$ and $|N-M|\ll N$, then
\begin{equation}
W_{MN}
=
\frac{\pi B_z^2}{2}(N+M)
+O\!\left(B_z^2\frac{(N-M)^2}{N}\right).
\label{eq:WMN-neardiag}
\end{equation}
This is the form actually used in the subsequent Lifshitz--Kosevich asymptotics, because the spectral functions localize both Landau indices near the same Fermi-level Landau number.

\subsection{Equivalent transport form used for the Lifshitz--Kosevich reduction}
\label{subsec:Bz-transport-form}

For the oscillatory asymptotics, we use the equivalent constant-$\tau$ transport form, whose phase content is fixed by the exact Landau spectrum derived in Sec.~\ref{sec:Bz-LL} and whose overall normalization is set by the zero-field Drude limit.

\paragraph*{Assumptions entering the transport form.}
We assume short-range scalar disorder,
\begin{equation}
\overline{V(\mathbf r)}=0,
\qquad
\overline{V(\mathbf r)V(\mathbf r')}=n_{\rm imp}U_0^2\,\delta(\mathbf r-\mathbf r')\,\I_4,
\label{eq:scalar-disorder-assumption}
\end{equation}
and take the disorder self-energy in the self-consistent Born approximation to be momentum independent,
\begin{equation}
\Sigma^{R/A}(\varepsilon)\simeq \mp i  \Gamma(\varepsilon)\,\I_4,
\qquad
\Gamma(\varepsilon)=\frac{\hbar}{2\tau_q(\varepsilon)}.
\label{eq:SCBA-self-energy}
\end{equation}
Starting from the retarded--advanced Kubo formula at finite temperature,
\begin{equation}
\sigma_{yy}(\mu,T;B_z)
=
\frac{e^2}{2\pi V}
\int d\varepsilon\,
\left(-\pdv{f}{\varepsilon}\right)
\Tr\!\left[
\hat{v}_y \hat{G}^R(\varepsilon)\,\hat{\Gamma}_y(\varepsilon)\,\hat{G}^A(\varepsilon)
\right],
\label{eq:Kubo-RA-finiteT}
\end{equation}
we make the standard Lifshitz--Kosevich constant-relaxation-time approximation (CRTA) reduction: (i) retain the dominant intraband contribution in the eigenbasis of $H(B_z)$, (ii) absorb the smooth ladder renormalization of the dressed vertex $\Gamma_y$ into a transport lifetime $\tau_{\rm tr}$, and (iii) use the narrow-width identity
\begin{equation}
\delta_\Gamma(x)\equiv \frac{1}{\pi}\frac{\Gamma}{x^2+\Gamma^2},
\qquad
\int_{-\infty}^{\infty}dx\,[\delta_\Gamma(x)]^2
=
\frac{1}{2\pi\Gamma}
=
\frac{\tau_q}{\pi\hbar}.
\label{eq:deltaGamma-identity}
\end{equation}
If the Fermi window varies slowly on the scale $\Gamma$, then
\begin{equation}
\int d\varepsilon\,
\left(-\pdv{f}{\varepsilon}\right)
[\delta_\Gamma(\varepsilon-E_a)]^2
\simeq
\frac{\tau_q(E_a)}{\pi\hbar}
\left.
\left(-\pdv{f}{\varepsilon}\right)
\right|_{\varepsilon=E_a}.
\label{eq:narrow-width-Fermi-window}
\end{equation}
Thus the conductivity may be written as
\begin{equation}
\sigma_{yy}(\mu,T;B_z)
\simeq
 e^2\int d\varepsilon\,
\left(-\pdv{f}{\varepsilon}\right)
\tau_{\rm tr}(\varepsilon)\,
\mathcal G_{yy}(\varepsilon,B_z).
\label{eq:sigma-transport-function}
\end{equation}
Here the velocity factor is a semiclassical intraband transport weight.
For a nondegenerate zero-field band we define
\begin{equation}
v_i(\bk)
\equiv
\langle u_{\bk}|\hat v_i|u_{\bk}\rangle
=
\frac{\partial E(\bk)}{\partial k_i},
\label{eq:band-velocity-def}
\end{equation}
where $\hbar=1$.
The quantity entering the constant-$\tau$ Lifshitz--Kosevich transport form is the orbit average of the square of this band velocity,
\begin{equation}
\overline{v_i^2}^{\,\rm orb}
\equiv
\frac{\oint dt\, [v_i(\bk(t))]^2}{\oint dt}.
\label{eq:orbital-velocity-square-def}
\end{equation}
The symbol $\oint dt$ denotes integration over one complete semiclassical cyclotron period $T_{\rm cyc}$ of the orbit $\bk(t)$.
Equivalently,
\[
\overline{v_i^2}^{\,\rm orb}
=
\frac{1}{T_{\rm cyc}}\int_0^{T_{\rm cyc}}dt\,[v_i(\bk(t))]^2 .
\]
It should not be confused with the quantum expectation value
$\langle u_{\bk}|\hat v_i^2|u_{\bk}\rangle$, which is generally different from
$[\langle u_{\bk}|\hat v_i|u_{\bk}\rangle]^2$.
This replacement of the exact current-current Kubo vertex by an intraband orbit-averaged transport weight is the standard many-Landau-level CRTA step used for the LK amplitude.

The corresponding velocity-squared-weighted density of states is
\begin{equation}
\mathcal G_{yy}(\varepsilon,B_z)
=
g_d\,\frac{B_z}{2\pi}
\sum_{n\ge1}\int\!\frac{dk_z}{2\pi}\,
\overline{v_y^2}^{\,\rm orb}_{n,k_z}\,
\delta\!\bigl(\varepsilon-E_{n,-}(k_z)\bigr).
\label{eq:Gyy-def}
\end{equation}
Evaluating Eq.~\eqref{eq:sigma-transport-function} at $\varepsilon=\EF$ gives the constant-$\tau$ form used below.
The frequencies and Onsager offsets are controlled entirely by the exact Landau spectrum of Sec.~\ref{sec:Bz-LL}. The disorder model only changes the nonuniversal amplitudes through $\tau_{\rm tr}$ and the usual Dingle factor.

At low temperature, or equivalently for the transport function at $\varepsilon=\EF$, one may therefore write
\begin{equation}
\sigma_{yy}(B_z)=g_d\,\tau_{\rm tr}\,\frac{B_z}{2\pi}
\sum_{n\ge1}\int\!\frac{dk_z}{2\pi}\;
\overline{v_y^2}^{\,\rm orb}_{n,k_z}\,
\delta\!\left(\EF-E_{n,-}(k_z)\right).
\label{eq:sigyy-transport-start}
\end{equation}
The microscopic kernel derived above shows that the exact current-current vertex is smooth and finite near the relevant large-$n$ Landau levels.
Equation~\eqref{eq:sigyy-transport-start} should therefore be understood as the leading many-Landau-level intraband CRTA form used to obtain the LK amplitude.
It is not an identity involving $\langle u_{\bk}|\hat v_y^2|u_{\bk}\rangle$.
The exact ingredients are the Landau spectrum and the resulting fan condition; the transport approximation controls only the nonuniversal amplitude.

At fixed $k_z$, the $n$th Landau tube corresponds semiclassically to a circle in the $(k_x,k_y)$ plane of radius
\begin{equation}
\alpha_n\equiv\sqrt{2B_z n}.
\label{eq:alpha-n-def}
\end{equation}
Introduce
\begin{equation}
\beta(k_z)\equiv\sqrt{\EF^2-k_z^2},
\qquad
Y_r(k_z)\equiv m_0+r\beta(k_z),
\qquad r=\pm.
\label{eq:beta-Yr-kz}
\end{equation}
On the Fermi surface, the two toroidal branches satisfy $\alpha_n=Y_r(k_z)$.
Parametrizing the corresponding orbit by
\begin{equation}
k_x=Y_r(k_z)\cos\phi,
\qquad
k_y=Y_r(k_z)\sin\phi,
\qquad 0\le \phi<2\pi,
\end{equation}
the band velocity is
\begin{equation}
v_y=\pdv{E}{k_y}=\frac{k_\rho-m_0}{E}\frac{k_y}{k_\rho}
= r\,\frac{\beta(k_z)}{\EF}\sin\phi,
\label{eq:vy-on-orbit}
\end{equation}
where we used $E=\EF$ on shell and $k_\rho-m_0=r\beta(k_z)$ on the branch $r$.
A full orbit average therefore gives
\begin{align}
\overline{v_y^2}^{\,\rm orb}_{r,k_z}
&=\frac{1}{2\pi}\int_0^{2\pi}d\phi\, [v_y(\phi)]^2 \notag\\
&=\frac{\beta(k_z)^2}{E_F^2}\frac{1}{2\pi}\int_0^{2\pi}d\phi\,\sin^2\phi
=\frac{\beta(k_z)^2}{2E_F^2}.
\label{eq:vy2-average}
\end{align}
This is what distinguishes the transverse channel: the semiclassical orbit-averaged transport weight is nonzero at the extremal orbit $k_z=0$.

The exact $\lambda=-1$ Landau branch is
\begin{equation}
E_{n,-}(k_z)=\sqrt{k_z^2+\bigl(m_0-\alpha_n\bigr)^2}.
\label{eq:Enminus-sigma}
\end{equation}
At fixed $k_z$, the on-shell condition $E_{n,-}(k_z)=\EF$ has the two solutions
\begin{equation}
\alpha_n=Y_r(k_z)=m_0+r\beta(k_z),\qquad r=\pm.
\label{eq:alpha-on-shell-r}
\end{equation}
To convert the energy delta function into a delta function in Landau index, note that
\begin{equation}
\dv{\alpha_n}{n}=\frac{B_z}{\alpha_n},
\end{equation}
and therefore
\begin{equation}
\left|\dv{E_{n,-}(k_z)}{n}\right|
=\left|\pdv{E_{n,-}}{\alpha_n}\right|\left|\dv{\alpha_n}{n}\right|
=\frac{|m_0-\alpha_n|}{E_{n,-}(k_z)}\frac{B_z}{\alpha_n}.
\end{equation}
Evaluating this on shell gives $|m_0-\alpha_n|=\beta(k_z)$ and $E_{n,-}(k_z)=\EF$, so
\begin{equation}
\left|\dv{E_{n,-}(k_z)}{n}\right|_{\rm on-shell}
=\frac{B_z\,\beta(k_z)}{\alpha_n E_F}.
\label{eq:dEdn-onshell}
\end{equation}
Hence
\begin{equation}
\delta\!\left(\EF-E_{n,-}(k_z)\right)
=\sum_{r=\pm}
\frac{\alpha_n E_F}{B_z\,\beta(k_z)}
\delta\!\left(n-\frac{F_r(k_z)}{B_z}\right),
\label{eq:delta-n-conversion}
\end{equation}
where
\begin{equation}
F_r(k_z)\equiv\frac{Y_r(k_z)^2}{2}
=\frac{1}{2}\Bigl(m_0+r\sqrt{\EF^2-k_z^2}\Bigr)^2
\label{eq:Fr-kz-def}
\end{equation}
is exactly the field-parallel family of Onsager frequencies. The exact Landau-level result of Sec.~\ref{sec:Bz-LL} implies $\gamma_r=0$ for both $r=\pm$.

Substituting Eqs.~\eqref{eq:vy2-average} and \eqref{eq:delta-n-conversion} into the CRTA transport form Eq.~\eqref{eq:sigyy-transport-start}, and then using $\alpha_n=Y_r(k_z)$ on the support of the delta function, we obtain
\begin{equation}
\sigma_{yy}(B_z)=\frac{g_d\,\tau_{\rm tr}}{8\pi^2 E_F}
\sum_{r=\pm}\int_{-E_F}^{E_F}dk_z\;\beta(k_z)Y_r(k_z)
\sum_{n\ge1}\delta\!\left(n-\frac{F_r(k_z)}{B_z}\right).
\label{eq:sigyy-transport-final}
\end{equation}
Because the continuum model is rotationally symmetric in the $x$-$y$ plane, the same formula holds for $\sigma_{xx}=\sigma_{yy}\equiv\sigma_{\perp}$.

The smooth $B_z\to0$ part follows immediately from replacing the Landau comb by its average value $1$:
\begin{align}
\sigma_\perp^{(0)}
&=\frac{g_d\,\tau_{\rm tr}}{8\pi^2 E_F}
\sum_{r=\pm}\int_{-E_F}^{E_F}dk_z\;\beta(k_z)Y_r(k_z) \notag\\
&=\frac{g_d\,\tau_{\rm tr}}{8\pi^2 E_F}
\int_{-E_F}^{E_F}dk_z\;2m_0\sqrt{E_F^2-k_z^2}.
\end{align}
Using
\begin{equation}
\int_{-E_F}^{E_F}dk_z\,\sqrt{E_F^2-k_z^2}=\frac{\pi E_F^2}{2},
\label{eq:semicircle-integral}
\end{equation}
we get
\begin{equation}
\sigma_\perp^{(0)}=g_d\,\tau_{\rm tr}\,\frac{m_0E_F}{8\pi}.
\label{eq:sperp-background}
\end{equation}
Equation~\eqref{eq:sperp-background} is the analytic, nonoscillatory $B_z\to0$ contribution of the continuum CRTA/LK formula.
It should not be identified term by term with the empirical background $\sigma_{xx}^{\rm smooth}(1/B)$ used below in Eq.~\eqref{eq:deltasigma}.
The latter is a finite-window subtraction applied to lattice or experimental data and, in general, also absorbs lattice corrections, finite broadening, slow magnetoresistance, and other nonuniversal smooth contributions.

\subsection{Detailed stationary-phase derivation of the oscillatory Lifshitz--Kosevich series}
\label{subsec:Bz-sigma-detailed-LK}

We now derive the oscillatory part of Eq.~\eqref{eq:sigyy-transport-final} in full detail.
For many occupied Landau levels, the one-sided Poisson formula can be used in the standard Lifshitz--Kosevich form,
\begin{equation}
\sum_{n\ge1}\delta(n-x)=1+2\Re\sum_{\nu=1}^{\infty}e^{2\pi i \nu x}
\quad(x>0),
\label{eq:Poisson-delta-comb}
\end{equation}
where the omitted endpoint terms, which are relevant only near $x=0$, affect only the nonoscillatory piece and are irrelevant for the leading Lifshitz--Kosevich harmonics.
Substituting Eq.~\eqref{eq:Poisson-delta-comb} into Eq.~\eqref{eq:sigyy-transport-final} and keeping only the oscillatory contribution yields
\begin{equation}
\sigma_\perp^{\rm osc}(B_z)=\frac{g_d\,\tau_{\rm tr}}{4\pi^2 E_F}
\sum_{r=\pm}\sum_{\nu=1}^{\infty}
\Re\,\mathcal I_{\nu,r}(B_z),
\label{eq:sperp-osc-Ikr}
\end{equation}
with
\begin{equation}
\mathcal I_{\nu,r}(B_z)
\equiv\int_{-E_F}^{E_F}dk_z\;a_r(k_z)
\exp\!\left[\frac{2\pi i \nu}{B_z}F_r(k_z)\right],
\qquad
 a_r(k_z)\equiv \beta(k_z)Y_r(k_z).
\label{eq:Ikr-def}
\end{equation}

The phase is controlled by the field-parallel dependence of the extremal area; no singular matrix element enters.
Differentiating Eq.~\eqref{eq:Fr-kz-def} gives
\begin{equation}
\dv{F_r}{k_z}=-r\,\frac{Y_r(k_z)}{\beta(k_z)}\,k_z.
\label{eq:Fprime-kz}
\end{equation}
Therefore the only stationary point in the interval $|k_z|<E_F$ is
\begin{equation}
k_z=0.
\label{eq:kz-stationary-point}
\end{equation}
This is precisely the toroidal extremal orbit discussed in the main text.

To evaluate Eq.~\eqref{eq:Ikr-def} by stationary phase, expand every quantity near $k_z=0$.
Since
\begin{equation}
\beta(k_z)=\sqrt{E_F^2-k_z^2}=E_F-\frac{k_z^2}{2E_F}+O(k_z^4),
\label{eq:beta-expansion}
\end{equation}
we obtain
\begin{align}
Y_r(k_z)
&=m_0+r\beta(k_z)
=Y_r-r\frac{k_z^2}{2E_F}+O(k_z^4),
\label{eq:Yr-expansion}\\
F_r(k_z)
&=\frac{Y_r(k_z)^2}{2}
=F_r-r\frac{Y_r}{2E_F}k_z^2+O(k_z^4),
\label{eq:Fr-expansion}\\
a_r(k_z)
&=\beta(k_z)Y_r(k_z)=E_FY_r+O(k_z^2),
\label{eq:ar-expansion}
\end{align}
where, from now on,
\begin{equation}
Y_r\equiv m_0+rE_F,
\qquad
F_r\equiv\frac{Y_r^2}{2}.
\label{eq:Ypm}
\end{equation}
Substituting Eqs.~\eqref{eq:Fr-expansion} and \eqref{eq:ar-expansion} into Eq.~\eqref{eq:Ikr-def} and extending the Gaussian integral to the whole real line gives
\begin{equation}
\mathcal I_{\nu,r}(B_z)
\simeq E_FY_r\,e^{2\pi i \nu F_r/B_z}
\int_{-\infty}^{\infty}dk_z\,
\exp\!\left[-i \pi \nu r\frac{Y_r}{B_zE_F}k_z^2\right].
\label{eq:Ikr-gaussian-pre}
\end{equation}
Now use the standard Fresnel integrals, valid for $\lambda>0$,
\begin{equation}
\int_{-\infty}^{\infty}dx\,e^{-i \lambda x^2}
=\sqrt{\frac{\pi}{\lambda}}\,e^{-i \pi/4},
\qquad
\int_{-\infty}^{\infty}dx\,e^{+i \lambda x^2}
=\sqrt{\frac{\pi}{\lambda}}\,e^{+i \pi/4}.
\label{eq:fresnel-gaussian}
\end{equation}
For the outer toroidal orbit ($r=+$), Eq.~\eqref{eq:Ikr-gaussian-pre} contains the first integral in Eq.~\eqref{eq:fresnel-gaussian} while, for the inner orbit ($r=-$), it contains the second.
Thus
\begin{align}
\mathcal I_{\nu,+}(B_z)
&\simeq E_FY_+\,e^{2\pi i \nu F_+/B_z}
\sqrt{\frac{B_zE_F}{\nu Y_+}}\,e^{-i \pi/4}
\notag\\
&=E_F\sqrt{\frac{B_zE_FY_+}{\nu}}
\exp\!\left[i \left(\frac{2\pi \nu F_+}{B_z}-\frac{\pi}{4}\right)\right],
\label{eq:Iplus-final}\\
\mathcal I_{\nu,-}(B_z)
&\simeq E_FY_-\,e^{2\pi i \nu F_-/B_z}
\sqrt{\frac{B_zE_F}{\nu Y_-}}\,e^{+i \pi/4}
\notag\\
&=E_F\sqrt{\frac{B_zE_FY_-}{\nu}}
\exp\!\left[i \left(\frac{2\pi \nu F_-}{B_z}+\frac{\pi}{4}\right)\right].
\label{eq:Iminus-final}
\end{align}
Taking the real part and substituting into Eq.~\eqref{eq:sperp-osc-Ikr} gives
\begin{equation}
\sigma_\perp^{\rm osc}(B_z)
=g_d\,\tau_{\rm tr}\,\frac{\sqrt{E_F B_z}}{4\pi^2}
\sum_{\nu=1}^{\infty}\frac{1}{\sqrt{\nu}}
\left[
\sqrt{Y_+}\cos\!\left(\frac{2\pi \nu F_+}{B_z}-\frac{\pi}{4}\right)
+\sqrt{Y_-}\cos\!\left(\frac{2\pi \nu F_-}{B_z}+\frac{\pi}{4}\right)
\right].
\label{eq:sperp-leading-T0}
\end{equation}
This is the clean $T=0$ leading Lifshitz--Kosevich result for the transverse channel.
The exact Landau quantization derived in Sec.~\ref{sec:Bz-LL} shows that both toroidal families have $\gamma_\pm=0$, which is consistent with the absence of a nonzero $-2\pi \nu\gamma_\pm$ term in Eq.~\eqref{eq:sperp-leading-T0}. The only remaining phase shift is the universal three-dimensional curvature phase $\varphi_{\rm 3D,\pm}=\mp\pi/4$.

We next derive the standard LK reduction factors in the notation of the two toroidal orbit families. This is the usual Shoenberg/Lifshitz--Kosevich argument~\cite{shoenberg}, written explicitly here to show that the damping factors multiply the clean harmonic without shifting the Onsager offset. For a fixed family $r=\pm$ and harmonic $\nu$, write the clean contribution at energy $E$ as
\begin{equation}
\delta\sigma^{(0)}_{\perp,\nu,r}(E;B_z)
=
\mathcal A^{(0)}_{\nu,r}(E;B_z)
\cos\Theta_{\nu,r}(E;B_z),
\end{equation}
with
\begin{equation}
\Theta_{\nu,r}(E;B_z)
=
2\pi \nu\left[\frac{F_r(E)}{B_z}-\gamma_r\right]
+\varphi_{\rm 3D,r}.
\end{equation}
For the continuum torus, in units $\hbar=e=v=1$,
\begin{equation}
F_r(E)=\frac{(m_0+rE)^2}{2},
\qquad r=\pm .
\end{equation}
The cyclotron mass entering the LK envelope is therefore the positive quantity
\begin{equation}
m_c^{(r)}
\equiv
\left|\left.\pdv{F_r}{E}\right|_{E_F}\right|
=Y_r,
\qquad
Y_r=m_0+rE_F .
\label{eq:mcr-transverse}
\end{equation}
Expanding the phase about the Fermi energy, $E=E_F+\delta E$, gives
\begin{equation}
\Theta_{\nu,r}(E_F+\delta E;B_z)
=
\Theta_{\nu,r}(E_F;B_z)
+a_{\nu,r}\delta E
+O(\delta E^2),
\end{equation}
where
\begin{equation}
a_{\nu,r}
=
\frac{2\pi \nu}{B_z}
\left.\pdv{F_r}{E}\right|_{E_F},
\qquad
|a_{\nu,r}|=\frac{2\pi \nu m_c^{(r)}}{B_z}.
\end{equation}
The sign of $\partial F_r/\partial E$ does not generate an extra phase in the damping factors because the thermal and Lorentzian kernels below are even in their energy variables.
To keep the bookkeeping straight, we use distinct variables below: $\xi$ denotes the thermal energy offset in the Fermi-window convolution, while $\zeta$ denotes the energy offset associated with Lorentzian Landau-level broadening.

First, finite temperature convolves the zero-temperature transport function with the Fermi-window kernel in Eq.~\eqref{eq:sigma-transport-function}. Treating the smooth prefactor $\mathcal A^{(0)}_{\nu,r}$ and $m_c^{(r)}$ as constant over the thermal window,
\begin{align}
R^{(r)}_{T,\nu}(B_z)
&=
\int_{-\infty}^{\infty}d\xi\,
\frac{1}{4k_BT}
\operatorname{sech}^2\!\left(\frac{\xi}{2k_BT}\right)
\cos(a_{\nu,r}\xi)
\notag\\
&=
\frac{X_{\nu,r}}{\sinh X_{\nu,r}},
\qquad
X_{\nu,r}=\frac{2\pi^2 \nu k_BT m_c^{(r)}}{B_z}.
\label{eq:RT-thermal}
\end{align}
In the last step we used
\begin{equation}
\int_{-\infty}^{\infty}du\,
\operatorname{sech}^2 u\,e^{i\omega u}
=
\frac{\pi\omega}{\sinh(\pi\omega/2)} .
\end{equation}
Second, Lorentzian Landau-level broadening with half-width $\Gamma=1/(2\tau_q)$ gives
\begin{align}
R^{(r)}_{D,\nu}(B_z)
&=
\int_{-\infty}^{\infty}d\zeta\,
\frac{1}{\pi}\frac{\Gamma}{\zeta^2+\Gamma^2}
\cos(a_{\nu,r}\zeta)
\notag\\
&=
\exp[-\Gamma |a_{\nu,r}|]
=
\exp\!\left[
-\frac{2\pi^2 \nu k_B T_D m_c^{(r)}}{B_z}
\right],
\label{eq:RD-dingle}
\end{align}
where $T_D=1/(2\pi k_B\tau_q)$ in $\hbar=1$ units. The lifetime $\tau_q$ is the quantum lifetime controlling Landau-level broadening; it need not coincide with the transport lifetime $\tau_{\rm tr}$ entering the overall conductivity prefactor.

The multiplicative form of the thermal and Dingle factors follows because both effects act as independent convolutions in energy on the same Fourier harmonic.
Using complex notation for the clean harmonic, the contribution with both effects included is
\begin{align}
\delta\sigma^{T,D}_{\perp,\nu,r}
&=
\Re\Bigg\{
\mathcal A^{(0)}_{\nu,r}
e^{i \Theta_{\nu,r}(E_F;B_z)}
\left[
\int_{-\infty}^{\infty}d\xi\,
K_T(\xi)e^{i  a_{\nu,r}\xi}
\right]
\left[
\int_{-\infty}^{\infty}d\zeta\,
L_\Gamma(\zeta)e^{i  a_{\nu,r}\zeta}
\right]
\Bigg\} \notag\\
&=
\mathcal A^{(0)}_{\nu,r}
R^{(r)}_{T,\nu}(B_z)
R^{(r)}_{D,\nu}(B_z)
\cos\Theta_{\nu,r}(E_F;B_z),
\label{eq:RT-RD-product}
\end{align}
where
\begin{equation}
K_T(\xi)=
\frac{1}{4k_BT}
\operatorname{sech}^2\!\left(\frac{\xi}{2k_BT}\right),
\qquad
L_\Gamma(\zeta)=
\frac{1}{\pi}\frac{\Gamma}{\zeta^2+\Gamma^2}.
\end{equation}
Equivalently, the combined energy kernel is the convolution $K_T*L_\Gamma$, whose Fourier transform at frequency $a_{\nu,r}$ is the product of the two individual Fourier transforms.
Because both kernels are even and normalized, their Fourier transforms are real damping factors and introduce no additional phase shift.

In SI units, the common replacement in Eqs.~\eqref{eq:RT-thermal} and \eqref{eq:RD-dingle} is
\begin{equation}
\frac{m_c^{(r)}}{B_z}
\rightarrow
\frac{m_{c,r}^{\rm phys}}{e\hbar B_z}.
\end{equation}
Finally, for a spinful system with Zeeman splitting $\Delta_Z=g_s\mu_B B_z$, the two spin branches are shifted in energy by $\pm\Delta_Z/2$. Their average gives
\begin{equation}
\frac{1}{2}\sum_{s=\pm}
\cos\!\left[
\Theta_{\nu,r}
+s\,a^{\rm SI}_{\nu,r}\frac{\Delta_Z}{2}
\right]
=
\cos\!\left(a^{\rm SI}_{\nu,r}\frac{\Delta_Z}{2}\right)
\cos\Theta_{\nu,r},
\end{equation}
where $|a^{\rm SI}_{\nu,r}|=2\pi \nu m^{\rm phys}_{c,r}/(e\hbar B_z)$. This gives the usual spin-splitting factor
\begin{equation}
R^{(r)}_{S,\nu}
=
\cos\!\left(
\frac{\pi \nu g_s m_{c,r}^{\rm phys}}{2m_e}
\right),
\end{equation}
where $g_s$ is the Land\'e $g$ factor and $m_e$ is the bare electron mass. All analytic and numerical results in this work use spinless models, so
\begin{equation}
R^{(r)}_{S,\nu}=1
\end{equation}
throughout the present analysis.

Combining these independent thermal, lifetime, and spin factors, each clean harmonic is multiplied by
\begin{equation}
\mathcal R_{\nu,r}(B_z)
=R_{T,\nu}^{(r)}(B_z)R_{D,\nu}^{(r)}(B_z)R_{S,\nu}^{(r)}.
\label{eq:LK-reduction-combined}
\end{equation}
For the spinless model, $\mathcal R_{\nu,r}(B_z)$ is real and positive, so it changes only the nonuniversal amplitude. It does not shift $\gamma_r$ and therefore does not alter the topological conclusion $\gamma_\pm=0$, equivalently $\Phi_\phi=\pi$, for the monopole toroidal series.
Combining Eqs.~\eqref{eq:sperp-background}, \eqref{eq:sperp-leading-T0}, and \eqref{eq:LK-reduction-combined}, we obtain
\begin{align}
\sigma_{xx}(B_z)=\sigma_{yy}(B_z)
&=\sigma_\perp^{(0)}+\sigma_\perp^{\rm osc}(B_z),
\label{eq:sperp-final-1}\\
\sigma_\perp^{\rm osc}(B_z)
&=g_d\,\tau_{\rm tr}\,\frac{\sqrt{\EF B_z}}{4\pi^2}
\sum_{\nu=1}^{\infty}\frac{1}{\sqrt{\nu}}
\Bigg[
\sqrt{Y_+}\,\mathcal R_{\nu,+}(B_z)
\cos\!\left(\frac{2\pi \nu F_+}{B_z}-\frac{\pi}{4}\right)
\notag\\
&\hspace{6.1cm}
+\sqrt{Y_-}\,\mathcal R_{\nu,-}(B_z)
\cos\!\left(\frac{2\pi \nu F_-}{B_z}+\frac{\pi}{4}\right)
\Bigg].
\label{eq:sperp-final-2}
\end{align}
This is the detailed derivation of the compact formula quoted in the paper.

\subsection{Why the transverse and longitudinal channels have different field scalings}
\label{subsec:Bz-sigma-scaling}

The origin of the transverse scaling is now explicit.
The large parameter in the stationary-phase integral is
\begin{equation}
\lambda_{\nu,r}=\pi \nu\frac{Y_r}{B_zE_F}.
\end{equation}
Because the transverse amplitude $a_r(k_z)=\beta(k_z)Y_r(k_z)$ is smooth and nonzero at the stationary point,
\begin{equation}
a_r(k_z)=E_FY_r+O(k_z^2),
\end{equation}
the Gaussian integral contributes only the textbook stationary-phase factor $\lambda_{\nu,r}^{-1/2}$.
Hence each harmonic carries the universal three-dimensional envelope
\begin{equation}
\sigma_\perp^{\rm osc}\sim B_z^{1/2}\nu^{-1/2}.
\end{equation}

We now derive the corresponding leading longitudinal contribution.
The same stationary-phase method applies, with the only structural change from the transverse channel being the transport weight
\begin{equation}
v_z^2=\frac{k_z^2}{E_F^2},
\end{equation}
which vanishes quadratically on the extremal orbit at $k_z=0$.
Starting from the CRTA expression and using the same delta-function conversion as in Eq.~\eqref{eq:delta-n-conversion}, the longitudinal analog of Eq.~\eqref{eq:sigyy-transport-final} is
\begin{equation}
\sigma_{zz}(B_z)
=
\frac{g_d\tau_{\rm tr}}{4\pi^2 E_F}
\sum_{r=\pm}\int_{-\EF}^{\EF} dk_z\,
\frac{Y_r(k_z)}{\beta(k_z)}\,k_z^2
\sum_{n\ge1}
\delta\!\left(n-\frac{F_r(k_z)}{B_z}\right).
\label{eq:sigzz-transport-final}
\end{equation}
The finite range $-\EF\le k_z\le\EF$ is the physical Fermi-surface range of the torus in the continuum model.
One might worry about the unpaired $n=0$ branch, but it has
\[
E_0(k_z)=\sqrt{k_z^2+m_0^2},
\]
and therefore never crosses the Fermi level in the regime $0<\EF<m_0$ considered here, so the sum starts at $n\ge1$.
The factor $Y_r(k_z)/\beta(k_z)$ is the same Landau-level density factor that appeared in the transverse derivation.
The longitudinal velocity weight is $v_z^2=k_z^2/E_F^2$; together with the factor of $E_F$ from the energy-to-Landau-index delta-function conversion, this produces the net $k_z^2/E_F$ weighting in Eq.~\eqref{eq:sigzz-transport-final}.

Applying Poisson resummation gives the oscillatory part
\begin{equation}
\sigma_{zz}^{\rm osc}
=
\frac{g_d\tau_{\rm tr}}{2\pi^2E_F}
\Re\sum_{r=\pm}\sum_{\nu=1}^{\infty}
e^{2\pi i \nu F_r/B_z}
I^{zz}_{\nu,r}(B_z),
\end{equation}
where the finite-range integral is
\begin{equation}
I^{zz}_{\nu,r}(B_z)
=
\int_{-\EF}^{\EF}dk_z\,
\frac{Y_r(k_z)}{\beta(k_z)}\,k_z^2
\exp\!\left[
\frac{2\pi i \nu}{B_z}\bigl(F_r(k_z)-F_r\bigr)
\right].
\end{equation}
Near the extremal slice,
\[
F_r(k_z)=F_r-r\frac{Y_r}{2E_F}k_z^2+O(k_z^4).
\]
There is a small endpoint subtlety in this longitudinal channel.
Branch by branch, the finite-range integral also has nonstationary endpoint pieces at $|k_z|=E_F$, where the $r=+$ and $r=-$ branches coalesce.
These endpoint pieces do not represent additional extremal orbits.
Using $k_z=E_F\sin t$, one sees that the endpoint amplitudes and phases are the same for the two branches, while the endpoint phase derivatives have opposite signs; the leading endpoint contributions conspire to cancel in the physical sum over $r=\pm$.
After this cancellation, the leading oscillatory contribution is the stationary contribution from $k_z=0$.
The stationary region has width $|k_z|\sim\sqrt{B_z}$, far from the finite endpoints in the weak-field limit, so the extremal-orbit contribution is obtained by replacing $Y_r(k_z)/\beta(k_z)$ by $Y_r/E_F$ and extending the integral to $(-\infty,\infty)$:
\begin{equation}
I^{zz}_{\nu,r}(B_z)
\simeq
\frac{Y_r}{E_F}
\int_{-\infty}^{\infty}dk_z\,k_z^2
\exp\!\left[
-i  r\pi \nu\frac{Y_r}{B_zE_F}k_z^2
\right].
\end{equation}
Thus the relevant stationary-phase integral is of the form
\begin{equation}
\int_{-\infty}^{\infty}dk_z\,k_z^2 e^{\pm i \lambda k_z^2}
\propto \lambda^{-3/2},
\label{eq:kz2-gaussian}
\end{equation}
instead of $\lambda^{-1/2}$. More explicitly, for $\lambda>0$,
\begin{align}
\int_{-\infty}^{\infty}du\,u^2 e^{-i \lambda u^2}
&=
\frac{\sqrt{\pi}}{2}\lambda^{-3/2}e^{-3\pi i/4},
\notag\\
\int_{-\infty}^{\infty}du\,u^2 e^{+i \lambda u^2}
&=
\frac{\sqrt{\pi}}{2}\lambda^{-3/2}e^{+3\pi i/4}.
\label{eq:kz2-fresnel-phase}
\end{align}
Therefore
\begin{equation}
\sigma_{zz}^{\rm osc}\sim B_z^{3/2}\nu^{-3/2}.
\end{equation}
The difference between the transverse and longitudinal channels is therefore not topological.
It comes from whether the measured current component is finite or vanishing on the extremal orbit.
In particular, the transverse curvature phases $\varphi_{\rm 3D,+}=-\pi/4$ and $\varphi_{\rm 3D,-}=+\pi/4$ are supplemented in the longitudinal channel by the extra stationary-phase contribution generated by the quadratic zero of $v_z$ at the extremal slice:
\begin{equation}
\delta\phi^{zz}_{+}=-\frac{\pi}{2},
\qquad
\delta\phi^{zz}_{-}=+\frac{\pi}{2}.
\end{equation}
Equivalently, with $r=\pm1$ labeling the outer and inner toroidal orbits,
\begin{equation}
\phi^{zz}_{r}
=
\varphi_{\rm 3D,r}-r\frac{\pi}{2}
=
-r\frac{3\pi}{4},
\qquad
\phi^{zz}_{+}=-\frac{3\pi}{4},
\quad
\phi^{zz}_{-}=+\frac{3\pi}{4}.
\label{eq:phizz}
\end{equation}
This extra $\mp\pi/2$ is a component-dependent stationary-phase correction, not an additional Berry phase; the Onsager offset remains $\gamma_z=0$.

Combining the preceding factors, the leading longitudinal oscillatory conductivity can be written as
\begin{equation}
\sigma_{zz}^{\rm osc}(B_z)
=\frac{g_d\tau_{\rm tr} B_z^{3/2}}{4\pi^3}
\sum_{\nu=1}^{\infty}\frac{1}{\nu^{3/2}}\sum_{r=\pm}
\frac{R^{(r)}_{T,\nu}(B_z)R^{(r)}_{D,\nu}(B_z)}{\sqrt{\EF Y_r}}
\cos\!\left[
2\pi \nu\left(\frac{F_r}{B_z}-\gamma_z\right)+\phi^{zz}_{r}
\right],
\label{eq:sigzz-final}
\end{equation}
with
\[
\gamma_z=0,
\qquad
\phi^{zz}_{\pm}=\mp\frac{3\pi}{4}.
\]
For the spinless model, $R_S=1$ has been set in Eq.~\eqref{eq:sigzz-final}.
Equation~\eqref{eq:sigzz-final} is not needed for the main diagnostic, but it keeps the distinct $B$ scalings, stationary-phase offsets, and their physical origins clearly separated.

\begin{figure}
\includegraphics[width=\linewidth]{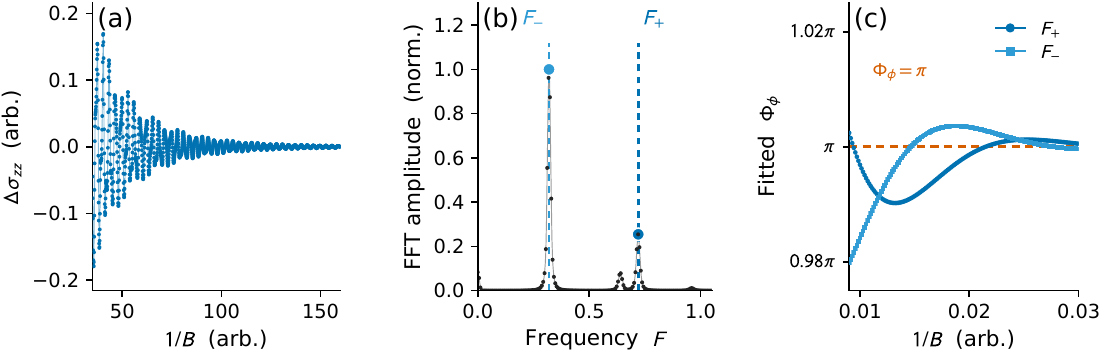}
\caption{Continuum longitudinal-channel oscillations for $\B\parallel\hat z$ in dimensionless continuum units, with arbitrary conductivity normalization.
(a) Oscillatory conductivity $\Delta\sigma_{zz}$ computed from Eq.~\eqref{eq:sigzz-final}, including the quadratic-velocity stationary-phase phases $\phi^{zz}_{\pm}=\mp3\pi/4$.
(b) Hann-windowed FFT showing the same toroidal frequencies $F_{-}$ and $F_{+}$ as the transverse channel.
(c) Berry phases extracted from fixed-frequency fits, zoomed near $\Phi_{\phi}=\pi$, showing convergence to the monopole-ring phase shift.}
\label{figS:BzSigmaZZ}
\end{figure}

Figure~\ref{figS:BzSigmaZZ} visualizes this longitudinal-channel result.
For this longitudinal fit, the fixed stationary phases are
$\phi^{zz}_{+}=-3\pi/4$ and $\phi^{zz}_{-}=+3\pi/4$; after these
component-dependent phases are accounted for, the extracted Onsager offsets still
converge to $\gamma_z=0$, equivalently $\Phi_{\phi}=\pi$.

%%%%%%%%%%%%%%%%%%%%%%%%%%%%%%%%%%%%%%%%%%%%%%%%%%%%%%%%%%%%%%%%%%%%%%%%%%%%%%
\section{Semiclassical torus geometry, component selection, and the \texorpdfstring{$\B\parallel\hat x$}{B parallel x} Lifshitz--Kosevich series}
\label{sec:Bx}

For $\B=B_x\hat x$, cyclotron motion occurs in the $(k_y,k_z)$ plane at fixed $k_x$.
In this geometry, the torus supports both linking and nonlinking extremal orbit families, and the conductivity component determines which family is visible most strongly.

\subsection{Exact \texorpdfstring{$A_{yz}$}{Ayz} as a function of \texorpdfstring{$k_x$}{kx} and the two extremal families}

Write $(x,y,z)\equiv(k_x,k_y,k_z)$ and define the torus radii $R=m_0$ and $r=\EF$.
At fixed $x$, the cross section of Eq.~\eqref{eq:torus} obeys
\begin{equation}
\bigl(\sqrt{x^2+y^2}-R\bigr)^2+z^2=r^2.
\label{eq:torus-cross}
\end{equation}
Defining $s(z)=\sqrt{r^2-z^2}$ and $\Delta_\pm(z)=(R\pm s(z))^2-x^2$, the exact enclosed area in the $yz$ plane is
\begin{equation}
A_{yz}(x)=2\int_{-r}^{r}dz\,
\Bigl[
\Theta\!\bigl(\Delta_+(z)\bigr)\sqrt{\Delta_+(z)}
-\Theta\!\bigl(\Delta_-(z)\bigr)\sqrt{\Delta_-(z)}
\Bigr],
\label{eq:Ayz}
\end{equation}
valid for $|x|<R+r$.
This even function has two symmetry-distinct Lifshitz--Kosevich extremal families.

\paragraph*{Linking family at \texorpdfstring{$x=0$}{x=0}.}
The slice $x=0$ consists of two identical circles, each of area
\begin{equation}
A_{\rm loop}=\pi r^2=\pi \EF^2.
\label{eq:Aloop}
\end{equation}
Each circle links the nodal ring once, so its Berry phase is $\Phitheta=\pi$ and therefore
\begin{equation}
F_0=\frac{A_{\rm loop}}{2\pi}=\frac{\EF^2}{2},
\qquad
\gamma_0=0,
\qquad
\varphi_{\rm 3D,0}=+\frac{\pi}{4}.
\label{eq:F0gamma0}
\end{equation}
The two loops are degenerate in frequency and add coherently in amplitude.

% (moved from main)
This $\B\parallel\hat x$ construction is the component-resolved version of the familiar poloidal Berry-phase diagnostic of nodal rings.
The poloidal $\pi$ phase does not by itself distinguish a monopole ring from a topologically trivial nodal ring, but, in the present toroidal geometry, it complements the $\B\parallel\hat z$ toroidal-orbit diagnostic of $\wtwo=1$.

\paragraph*{Nonlinking family at \texorpdfstring{$x=\pm x_\star$}{x=+-x*}.}
For a thin torus $\EF\ll m_0$, the global maxima of $A_{yz}(x)$ occur at
\begin{equation}
x_\star=m_0-c\,\EF+O(\EF^2/m_0),
\qquad c\simeq0.6522295,
\label{eq:xstar}
\end{equation}
and the maximal area scales as
\begin{equation}
A_{\max}=K\sqrt{m_0}\,\EF^{3/2}+O(\EF^{5/2}/\sqrt{m_0}),
\qquad K\simeq6.19.
\label{eq:Amax}
\end{equation}
These maximal loops do not link the nodal ring, so
\begin{equation}
F_\star=\frac{A_{\max}}{2\pi}=\frac{K}{2\pi}\sqrt{m_0}\,\EF^{3/2}+\cdots,
\qquad
\gamma_\star=\frac12,
\qquad
\varphi_{\rm 3D,\star}=-\frac{\pi}{4}.
\label{eq:Fstar}
\end{equation}
Differentiating $F_\star(\EF)$ gives the thin-torus cyclotron mass
\begin{equation}
m_{c,\star}=\frac{\partial F_\star}{\partial \EF}
=\frac{3K}{4\pi}\sqrt{m_0\EF}
\equiv c_0\sqrt{m_0\EF},
\qquad
c_0\equiv\frac{3K}{4\pi}\simeq1.48.
\label{eq:mcstar}
\end{equation}
The thin-torus constants are determined by the universal equations
\begin{gather}
\int_0^1\frac{y\,dy}{\sqrt{1-y^2}\sqrt{c+y}}
=
\int_0^c\frac{y\,dy}{\sqrt{1-y^2}\sqrt{c-y}},
\label{eq:c-eq}\\
K=4\sqrt2\int_0^1\frac{y\,dy}{\sqrt{1-y^2}}
\Bigl(\sqrt{c+y}-\Theta(c-y)\sqrt{c-y}\Bigr).
\label{eq:K-eq}
\end{gather}

\subsection{Orbit-averaged velocities and the mirror-protected selection rule}
\label{subsec:vel-avgs}

We now evaluate the velocity-squared weights that enter the $\B\parallel\hat x$ conductivity.
Because this is the point at which one can most easily lose track of which quantity is being averaged, we proceed from the semiclassical equations of motion rather than jumping directly to the final orbit averages.
Throughout this subsection we use the thin-torus notation of Sec.~\ref{sec:Bx},
\begin{equation}
R\equiv m_0,
\qquad
r\equiv \EF,
\label{eq:Rr-def}
\end{equation}
so that the Fermi surface is the torus
\begin{equation}
(\rho-R)^2+z^2=r^2,
\qquad
\rho\equiv\sqrt{x^2+y^2}.
\label{eq:torus-Rr}
\end{equation}
For fixed $x=k_x$, a convenient parametrization of a cyclotron orbit is
\begin{equation}
\rho(\theta)=R+r\cos\theta,
\qquad
z(\theta)=r\sin\theta,
\qquad
y(\theta)=\pm\sqrt{\rho(\theta)^2-x^2}.
\label{eq:orbit-param-general}
\end{equation}
The sign distinguishes the two $y$ branches of the same closed orbit.

\paragraph*{Exact time-average measure.}
The quantity entering transport is the semiclassical orbit/time average
\begin{equation}
\overline{f}^{\,\rm orb}_x\equiv \frac{\oint dt\,f}{\oint dt},
\label{eq:time-average-def}
\end{equation}
rather than a uniform average in $\theta$.
The overline in this subsection denotes the same classical orbit/time average of the squared band velocity defined in Eq.~\eqref{eq:orbital-velocity-square-def}, not a quantum expectation value of $\hat v_i^2$.
For $\B=B_x\hat x$, the semiclassical equation of motion is
\begin{equation}
\dot{\bk}=-\bm v\times \B,
\label{eq:semicl-eom-Bx}
\end{equation}
so along the orbit
\begin{equation}
\dot z=B_x v_y.
\label{eq:zdot-vy}
\end{equation}
Using $z(\theta)=r\sin\theta$ and the band velocity derived below, one obtains
\begin{equation}
r\cos\theta\,\dot\theta=B_x v_y.
\label{eq:theta-eom-step}
\end{equation}
Hence
\begin{equation}
dt=\frac{r\,\rho(\theta)}{B_x\,|y(\theta)|}\,d\theta.
\label{eq:dt-exact}
\end{equation}
This formula is exact.
Because the observables needed below are $v_i^2$, which are even under $y\to -y$, the common factor of $2$ from the two $y$ branches cancels between numerator and denominator in Eq.~\eqref{eq:time-average-def}.
Therefore the orbit average may be written on a single branch as
\begin{equation}
\boxed{
\overline{f}^{\,\rm orb}_x=
\frac{\displaystyle\int_{\theta_-}^{\theta_+} d\theta\,\frac{\rho(\theta)}{|y(\theta)|}\,f(\theta)}
{\displaystyle\int_{\theta_-}^{\theta_+} d\theta\,\frac{\rho(\theta)}{|y(\theta)|}}
}
\label{eq:orbit-average-general}
\end{equation}
with $[\theta_-,\theta_+]$ the allowed interval for that orbit.
This is the starting point for all of the velocity averages below.

For the torus branch
\begin{equation}
E(\bk)=\sqrt{k_z^2+(m_0-\rho)^2},
\qquad
\rho\equiv\sqrt{k_x^2+k_y^2},
\label{eq:torus-dispersion}
\end{equation}
the band velocity is $\bm v=\nabla_{\bk}E$, namely
\begin{equation}
v_z=\frac{k_z}{E},
\qquad
v_\rho=\frac{\rho-m_0}{E},
\qquad
v_x=v_\rho\frac{k_x}{\rho},
\qquad
v_y=v_\rho\frac{k_y}{\rho}.
\label{eq:v-components}
\end{equation}
On the Fermi surface $E=r$ this becomes
\begin{equation}
v_z=\sin\theta,
\qquad
v_\rho=\cos\theta,
\qquad
v_x=\frac{x}{\rho(\theta)}\cos\theta,
\qquad
v_y=\frac{y(\theta)}{\rho(\theta)}\cos\theta.
\label{eq:v-parametrized}
\end{equation}
The exact identity
\begin{equation}
v_x^2+v_y^2=v_\rho^2=\cos^2\theta,
\qquad
v_x^2+v_y^2+v_z^2=1.
\label{eq:v-sum-rule}
\end{equation}
provides a quick sanity check on the averages.

\paragraph*{Exact averages for the linking family at $x=0$.}
At $x=0$, the two cyclotron orbits are the two circles
\begin{equation}
y_\pm(\theta)=\pm\bigl(R+r\cos\theta\bigr),
\qquad
z(\theta)=r\sin\theta,
\label{eq:x0-loops}
\end{equation}
which are centered at $y=\pm R$ and each have area $\pi r^2$.
For either loop, $|y|=\rho$, so Eq.~\eqref{eq:dt-exact} reduces to
\begin{equation}
dt=\frac{r}{B_x}\,d\theta.
\label{eq:dt-x0}
\end{equation}
The time average is therefore a uniform average in $\theta$.
From Eq.~\eqref{eq:v-parametrized},
\begin{equation}
v_x(\theta)=0,
\qquad
v_y(\theta)=\pm\cos\theta,
\qquad
v_z(\theta)=\sin\theta.
\label{eq:v-x0}
\end{equation}
Hence
\begin{align}
\overline{v_x^2}^{\,\rm orb}_0
&=\frac{1}{2\pi}\int_0^{2\pi} d\theta\,0=0,
\label{eq:vxavg0-deriv}\\
\overline{v_y^2}^{\,\rm orb}_0
&=\frac{1}{2\pi}\int_0^{2\pi} d\theta\,\cos^2\theta=\frac12,
\label{eq:vyavg0-deriv}\\
\overline{v_z^2}^{\,\rm orb}_0
&=\frac{1}{2\pi}\int_0^{2\pi} d\theta\,\sin^2\theta=\frac12.
\label{eq:vzavg0-deriv}
\end{align}
The result
\begin{equation}
\boxed{
\overline{v_y^2}^{\,\rm orb}_0=\frac12,
\qquad
\overline{v_x^2}^{\,\rm orb}_0=0
}
\label{eq:velavg0}
\end{equation}
is exact at the level of the semiclassical orbit average.
The vanishing of $\overline{v_x^2}^{\,\rm orb}_0$ is a pointwise mirror statement:
on the mirror plane $M_x:k_x\to-k_x$, the band energy is even in $k_x$, so
$v_x=\partial E/\partial k_x=0$ at every point of the $x=0$ orbit.
We are not using the incorrect implication
$\overline{v_x}^{\,\rm orb}_0=0\Rightarrow
\overline{v_x^2}^{\,\rm orb}_0=0$.
Indeed, on the same orbit $v_z(\theta)=\sin\theta$, so
$\overline{v_z}^{\,\rm orb}_0=0$ but
$\overline{v_z^2}^{\,\rm orb}_0=1/2$.
Thus the $F_0$ family is absent from the leading $B_x^{1/2}\nu^{-1/2}$ LK contribution to $\sigma_{xx}$, while it remains visible in $\sigma_{yy}$.

% (moved from main)
If mirror symmetry is weakly broken, the exact pointwise cancellation $v_x=0$ on the $k_x=0$ linking orbit is relaxed.
The $F_0$ series can then leak into $\sigma_{xx}$ with an amplitude controlled by the mirror-breaking perturbation, while the orbit phase remains unchanged as long as the cyclotron orbit remains well-defined and the Berry phase remains protected.

\paragraph*{Thin-torus asymptotics for the maximal family at $x=\pm x_\star$.}
We now turn to the nonlinking maximal orbits.
Write
\begin{equation}
x_\star=R-c r+O(r^2/R),
\qquad
\theta_0\equiv \arccos(-c),
\label{eq:xstar-theta0}
\end{equation}
with the constant $c$ determined in Sec.~\ref{sec:Bx}.
On the positive-$y$ branch of the maximal orbit,
\begin{equation}
y(\theta)^2=\rho(\theta)^2-x_\star^2
=2Rr\bigl(c+\cos\theta\bigr)+O(r^2),
\qquad |\theta|\le \theta_0.
\label{eq:y2-max}
\end{equation}
Therefore
\begin{equation}
y(\theta)=\sqrt{2Rr\bigl(c+\cos\theta\bigr)}\Bigl[1+O(r/R)\Bigr],
\label{eq:y-max-leading}
\end{equation}
and Eq.~\eqref{eq:orbit-average-general} becomes
\begin{equation}
\overline{f}^{\,\rm orb}_\star=
\frac{\displaystyle\int_{-\theta_0}^{\theta_0} d\theta\,\frac{f(\theta)}{\sqrt{c+\cos\theta}}}
{\displaystyle\int_{-\theta_0}^{\theta_0} d\theta\,\frac{1}{\sqrt{c+\cos\theta}}}
+O(r/R).
\label{eq:avg-measure-star}
\end{equation}
Thus the entire problem reduces to weighted one-dimensional integrals with weight
\begin{equation}
w(\theta)=\frac{1}{\sqrt{c+\cos\theta}}.
\label{eq:wtheta}
\end{equation}

The same thin-torus scaling gives a compact area functional for a single maximal loop:
\begin{equation}
A_{\rm one}(c)
=2r\int_{-\theta_0}^{\theta_0} d\theta\,y(\theta)\cos\theta
=\sqrt{Rr^3}\,\mathcal K(c)+O(r^{5/2}/\sqrt R),
\label{eq:Aone-c}
\end{equation}
where
\begin{equation}
\mathcal K(c)
\equiv
2\sqrt2\int_{-\theta_0}^{\theta_0} d\theta\,\cos\theta\sqrt{c+\cos\theta}.
\label{eq:Kofc-def}
\end{equation}
At the maximum, $\mathcal K(c)=K$ and $\mathcal K'(c)=0$.
Differentiating Eq.~\eqref{eq:Kofc-def} with respect to $c$ is straightforward because the boundary term vanishes at $\theta=\pm\theta_0$, where $c+\cos\theta_0=0$.
One finds
\begin{equation}
0=\mathcal K'(c)
=\sqrt2\int_{-\theta_0}^{\theta_0} d\theta\,\frac{\cos\theta}{\sqrt{c+\cos\theta}}.
\label{eq:I1zero}
\end{equation}
This stationarity identity is the main simplification in the velocity averages.

\paragraph*{Leading asymptotics of $v_x^2$, $v_y^2$, and $v_z^2$ on the maximal orbit.}
Using Eqs.~\eqref{eq:v-parametrized} and \eqref{eq:y2-max},
\begin{align}
\left(\frac{x_\star}{\rho(\theta)}\right)^2
&=
\left(\frac{R-cr}{R+r\cos\theta}\right)^2
=1+O(r/R),
\label{eq:xrho-leading}\\
\left(\frac{y(\theta)}{\rho(\theta)}\right)^2
&=
\frac{2r}{R}\bigl(c+\cos\theta\bigr)+O\!\bigl((r/R)^2\bigr).
\label{eq:yrho-leading}
\end{align}
Therefore
\begin{align}
v_x^2(\theta)
&=
\cos^2\theta+O(r/R),
\label{eq:vx2-star-leading}\\
v_y^2(\theta)
&=
\frac{2r}{R}\cos^2\theta\bigl(c+\cos\theta\bigr)+O\!\bigl((r/R)^2\bigr),
\label{eq:vy2-star-leading}\\
v_z^2(\theta)
&=
\sin^2\theta.
\label{eq:vz2-star-leading}
\end{align}
Introduce the weighted integrals
\begin{equation}
I_0\equiv\int_{-\theta_0}^{\theta_0}\frac{d\theta}{\sqrt{c+\cos\theta}},
\qquad
I_1\equiv\int_{-\theta_0}^{\theta_0}\frac{\cos\theta\,d\theta}{\sqrt{c+\cos\theta}},
\qquad
I_2\equiv\int_{-\theta_0}^{\theta_0}\frac{\cos^2\theta\,d\theta}{\sqrt{c+\cos\theta}}.
\label{eq:I012-def}
\end{equation}
Equation~\eqref{eq:I1zero} says simply that $I_1=0$.
To determine $I_2$, consider the total derivative
\begin{equation}
\frac{d}{d\theta}\Bigl[\sin\theta\sqrt{c+\cos\theta}\Bigr]
=
\frac{2c\cos\theta+3\cos^2\theta-1}{2\sqrt{c+\cos\theta}}.
\label{eq:derivative-identity}
\end{equation}
Its integral over $[-\theta_0,\theta_0]$ vanishes because the prefactor $\sqrt{c+\cos\theta}$ vanishes at the endpoints.
Using Eq.~\eqref{eq:I012-def}, one obtains
\begin{equation}
2c I_1+3I_2-I_0=0.
\label{eq:I-identity}
\end{equation}
Since $I_1=0$, this reduces to
\begin{equation}
I_2=\frac{I_0}{3}.
\label{eq:I2eqI0over3}
\end{equation}
Substituting into Eq.~\eqref{eq:avg-measure-star} immediately gives
\begin{equation}
\boxed{\overline{v_x^2}^{\,\rm orb}_\star=\frac13+O(r/R).}
\label{eq:velavgstar-x}
\end{equation}
Likewise,
\begin{equation}
\overline{v_z^2}^{\,\rm orb}_\star
=
1-\overline{v_x^2}^{\,\rm orb}_\star+O(r/R)
=\frac23+O(r/R),
\label{eq:velavgstar-z}
\end{equation}
where we used Eq.~\eqref{eq:v-sum-rule} and the fact that
$\overline{v_y^2}^{\,\rm orb}_\star=O(r/R)$.

For $v_y^2$, Eq.~\eqref{eq:vy2-star-leading} gives
\begin{equation}
\overline{v_y^2}^{\,\rm orb}_\star
=
\frac{2r}{R}
\frac{\displaystyle\int_{-\theta_0}^{\theta_0} d\theta\,\cos^2\theta\sqrt{c+\cos\theta}}
{\displaystyle\int_{-\theta_0}^{\theta_0} d\theta\,(c+\cos\theta)^{-1/2}}
+O\!\bigl((r/R)^2\bigr).
\label{eq:vy2-star-integral}
\end{equation}
Define the universal coefficient
\begin{equation}
\tilde c_y
\equiv
2
\frac{\displaystyle\int_{-\theta_0}^{\theta_0} d\theta\,\cos^2\theta\sqrt{c+\cos\theta}}
{\displaystyle\int_{-\theta_0}^{\theta_0} d\theta\,(c+\cos\theta)^{-1/2}}
\simeq 0.608748,
\label{eq:ctilde-def}
\end{equation}
so that
\begin{equation}
\boxed{
\overline{v_y^2}^{\,\rm orb}_\star
=
\tilde c_y\frac{r}{R}
+O\!\bigl((r/R)^2\bigr)
}.
\label{eq:velavgstar-y}
\end{equation}
Equations~\eqref{eq:velavgstar-x} and \eqref{eq:velavgstar-y} are the leading thin-torus asymptotics of the semiclassical orbit average defined in Eq.~\eqref{eq:time-average-def}.
They are the quantities that control how strongly the maximal orbit appears in $\sigma_{xx}$ and $\sigma_{yy}$.

Putting the results together,
\begin{equation}
\boxed{
\overline{v_y^2}^{\,\rm orb}_0=\frac12,
\qquad
\overline{v_x^2}^{\,\rm orb}_0=0,
\qquad
\overline{v_x^2}^{\,\rm orb}_\star=\frac13+O\!\left(\frac{\EF}{m_0}\right),
\qquad
\overline{v_y^2}^{\,\rm orb}_\star=\tilde c_y\frac{\EF}{m_0}+O\!\left(\frac{\EF^2}{m_0^2}\right)
},
\label{eq:velavg-summary}
\end{equation}
with $\tilde c_y\simeq0.608748$.
The exact mirror statement $v_x\equiv0$ on the $k_x=0$ orbit removes the leading $B_x^{1/2}\nu^{-1/2}$ LK contribution of the linking family from $\sigma_{xx}$.
Subleading stationary-phase terms arise from the expansion away from the mirror plane, where $v_x^2\propto k_x^2$;
these terms scale with an extra power of $B_x$ and are neglected in the leading component-resolved formulas below.
The maximal nonlinking family survives in both components.
Its contribution to $\sigma_{yy}$ is parametrically weaker than that to $\sigma_{xx}$ by a factor of order $\EF/m_0$.

\subsection{Component-resolved Lifshitz--Kosevich forms}

We now combine the geometry of Sec.~\ref{sec:Bx}, the transport weights just derived, and the standard three-dimensional stationary-phase reduction of Lifshitz--Kosevich theory~\cite{shoenberg,mikitik1999,li2018}.
Each extremal closed orbit contributes its own Onsager series, and the conductivity component multiplies that series by the corresponding orbit-averaged $v_i^2$.

\paragraph*{General orbit-by-orbit Lifshitz--Kosevich form.}
For an extremal orbit family $\alpha$ and conductivity component $i=x,y$, the $\nu$th harmonic has the structure
\begin{equation}
\Delta\sigma^{(\alpha)}_{ii}(B_x)
=
2\sum_{\nu=1}^{\infty}
\frac{\mathcal A_{\alpha,ii}\sqrt{B_x}}{\sqrt{\nu}}
R^{(\nu)}_{T,\alpha}R^{(\nu)}_{D,\alpha}R^{(\nu)}_{S,\alpha}
\cos\!\left[2\pi \nu\left(\frac{F_\alpha}{B_x}-\gamma_\alpha\right)+\varphi_{\rm 3D,\alpha}\right].
\label{eq:generic-LK-family}
\end{equation}
The prefactor $2$ is a symmetry factor: for $\alpha=0$ it counts the two identical loops at $x=0$, and for $\alpha=\star$ it counts the two symmetry-related maximal slices at $x=\pm x_\star$.
To make the normalization explicit, start from the velocity-squared-weighted density of states
\begin{equation}
\mathcal G_{ii}(\varepsilon,B_x)
=
\frac{eB_x}{h}
\int\!\frac{dk_x}{2\pi}
\sum_n
\overline{v_i^2}^{\,\rm orb}_{n,k_x}
\delta\!\bigl(\varepsilon-E_n(k_x)\bigr).
\label{eq:Gii-Bx}
\end{equation}
Using Onsager quantization $n+\gamma=F(\varepsilon,k_x)/B_x$ together with
\begin{equation}
m_c(\varepsilon,k_x)=\frac{\hbar^2}{2\pi}\pdv{A}{\varepsilon}=e\hbar\,\pdv{F}{\varepsilon},
\label{eq:mc-F-relation}
\end{equation}
one has
\begin{equation}
\delta\!\bigl(\varepsilon-E_n(k_x)\bigr)
=
\frac{m_c(\varepsilon,k_x)}{e\hbar B_x}
\delta\!\left(n+\gamma-\frac{F(\varepsilon,k_x)}{B_x}\right).
\label{eq:delta-energy-to-n-Bx}
\end{equation}
Poisson resummation of the Landau-level comb and stationary phase over $k_x$ around an extremal slice $x_\alpha$ then yield Eq.~\eqref{eq:generic-LK-family}, with field-independent amplitude
\begin{align}
\mathcal A_{\alpha,ii}
&=\mathcal N\,m_{c,\alpha}\,\mathcal C_\alpha\,
\overline{v_i^2}^{\,\rm orb}_\alpha,
\label{eq:Aiidef}\\
\mathcal N
&\equiv
\frac{e^2\tau_{\rm tr}(\EF)}{2\pi^2\hbar^2}\sqrt{\frac{2\pi e}{\hbar}}.
\label{eq:Ncal-def}
\end{align}
Here the physical cyclotron mass is
\begin{equation}
m_{c,\alpha}=e\hbar\,\frac{\partial F_\alpha}{\partial \varepsilon}\bigg|_{\varepsilon=E_F}.
\label{eq:mc-def-Bx}
\end{equation}
In the natural units used for the explicit orbit data below, this becomes
$m_{c,\alpha}=\partial F_\alpha/\partial E_F$.
The usual three-dimensional curvature factor is
\begin{equation}
\mathcal C_\alpha\equiv
\left.
\left|\frac{\partial^2 A_\alpha}{\partial x^2}\right|^{-1/2}
\right|_{x=x_\alpha}.
\label{eq:curvature-factor-def}
\end{equation}
Since $F_\alpha=(\hbar/2\pi e)A_\alpha$, one also has
\begin{equation}
\left|F_\alpha''\right|^{-1/2}=\sqrt{\frac{2\pi e}{\hbar}}\,\mathcal C_\alpha,
\qquad
F_\alpha''\equiv \left.\frac{\partial^2 F_\alpha}{\partial x^2}\right|_{x=x_\alpha}.
\label{eq:Fpp-Calpha}
\end{equation}
In natural units, the common factor reduces to
\begin{equation}
\mathcal N\big|_{\hbar=e=1}=\frac{\tau_{\rm tr}(\EF)}{2\pi^2}\sqrt{2\pi}.
\label{eq:Ncal-natural}
\end{equation}
Thus the normalization is fully explicit within the same CRTA-Lifshitz--Kosevich approximation; its nonuniversal part resides only in $\tau_{\rm tr}$.
For the linking family, $A_0(x)$ means the \emph{area of one loop}, not the total disconnected area $A_{yz}(x)$.
This distinction is essential: Onsager quantization applies to each closed orbit separately, so the two $x=0$ loops contribute at the same frequency $F_0$ and add in amplitude rather than producing a frequency $2F_0$.

\paragraph*{Linking family at $x=0$.}
From Eq.~\eqref{eq:F0gamma0}, the frequency and phase data are
\begin{equation}
F_0=\frac{r^2}{2},
\qquad
\gamma_0=0,
\qquad
\varphi_{\rm 3D,0}=+\frac{\pi}{4},
\qquad
m_{c,0}=\frac{\partial F_0}{\partial r}=r.
\label{eq:linking-data}
\end{equation}
To obtain the curvature factor, write the area of one loop at small $x$ as
\begin{equation}
A_{\rm loop}(x)
=
r\int_0^{2\pi} d\theta\,\cos\theta\sqrt{\bigl(R+r\cos\theta\bigr)^2-x^2}.
\label{eq:Aloopx}
\end{equation}
Expanding for small $x$,
\begin{equation}
A_{\rm loop}(x)
=
\pi r^2-
\frac{r x^2}{2}
\int_0^{2\pi}d\theta\,\frac{\cos\theta}{R+r\cos\theta}+O(x^4).
\label{eq:Aloop-expand1}
\end{equation}
Using
\begin{equation}
\int_0^{2\pi}\frac{d\theta}{R+r\cos\theta}=\frac{2\pi}{\sqrt{R^2-r^2}},
\qquad
\int_0^{2\pi}\frac{\cos\theta\,d\theta}{R+r\cos\theta}
=
\frac{2\pi}{r}\left(1-\frac{R}{\sqrt{R^2-r^2}}\right),
\label{eq:cos-over-a+bcos}
\end{equation}
one finds
\begin{equation}
A_{\rm loop}''(0)
=
2\pi\left(\frac{R}{\sqrt{R^2-r^2}}-1\right)
=
\pi\frac{r^2}{R^2}+O\!\left(\frac{r^4}{R^4}\right).
\label{eq:Aloop-second}
\end{equation}
Therefore
\begin{equation}
\mathcal C_0
=
\left[A_{\rm loop}''(0)\right]^{-1/2}
=
\frac{R}{\sqrt{\pi}\,r}\Bigl[1+O\!\left(\frac{r^2}{R^2}\right)\Bigr].
\label{eq:C0-curvature}
\end{equation}
Finally, using Eq.~\eqref{eq:velavg0},
\begin{equation}
\mathcal A_{0,yy}=\mathcal N\,m_{c,0}\,\mathcal C_0\,\frac12,
\qquad
\mathcal A_{0,xx}=0.
\label{eq:A0yyA0xx}
\end{equation}
This is the precise leading-order statement: the linking family contributes to $\sigma_{yy}$ but has zero $B_x^{1/2}\nu^{-1/2}$ LK weight in $\sigma_{xx}$. Subleading $F_0$ terms in $\sigma_{xx}$ are suppressed by the extra quadratic factor $v_x^2\propto k_x^2$ for small displacement away from the mirror plane $k_x=0$ and are not kept in the leading formulas below.

\paragraph*{Nonlinking maximal family at $x=\pm x_\star$.}
For the maximal family, the orbit data from Sec.~\ref{sec:Bx} are
\begin{equation}
F_\star=\frac{K}{2\pi}\sqrt{Rr^3}+\cdots,
\qquad
\gamma_\star=\frac12,
\qquad
\varphi_{\rm 3D,\star}=-\frac{\pi}{4},
\qquad
m_{c,\star}=c_0\sqrt{Rr}.
\label{eq:maximal-data}
\end{equation}
To obtain the curvature factor, use Eq.~\eqref{eq:Aone-c} in the form
\begin{equation}
A_{\rm one}(x)=\sqrt{Rr^3}\,\mathcal K(c)+O(r^{5/2}/\sqrt R),
\qquad
c\equiv\frac{R-x}{r}.
\label{eq:Aone-scaling}
\end{equation}
Since $x_\star$ is the maximum, $\mathcal K'(c)=0$ there.
Differentiating twice with respect to $x$ gives
\begin{equation}
A_{\rm one}''(x_\star)
=
\sqrt{\frac{R}{r}}\,\mathcal K''(c)+O\!\left(\sqrt{\frac{r}{R}}\right).
\label{eq:Astar-second-general}
\end{equation}
Defining the positive universal constant
\begin{equation}
\kappa\equiv-\mathcal K''(c)\big|_{c=c_\star}\simeq 8.07888,
\label{eq:kappa-def}
\end{equation}
we obtain
\begin{equation}
\mathcal C_\star
=
\left|A_{\rm one}''(x_\star)\right|^{-1/2}
=
\kappa^{-1/2}\left(\frac{r}{R}\right)^{1/4}
+O\!\left(\left(\frac{r}{R}\right)^{3/4}\right).
\label{eq:Cstar-curvature}
\end{equation}
The transport weights are then fixed by Eq.~\eqref{eq:velavg-summary}:
\begin{align}
\mathcal A_{\star,xx}
&=
\mathcal N\,m_{c,\star}\,\mathcal C_\star
\left[\frac13+O\!\left(\frac{r}{R}\right)\right],
\label{eq:Astarxx-new}\\
\mathcal A_{\star,yy}
&=
\mathcal N\,m_{c,\star}\,\mathcal C_\star
\left[\tilde c_y\frac{r}{R}+O\!\left(\frac{r^2}{R^2}\right)\right].
\label{eq:Astaryy-new}
\end{align}
Hence
\begin{equation}
\frac{\mathcal A_{\star,yy}}{\mathcal A_{\star,xx}}
=
3\tilde c_y\frac{r}{R}+O\!\left(\frac{r^2}{R^2}\right)
\simeq 1.82624\,\frac{\EF}{m_0}+\cdots.
\label{eq:Astarratio}
\end{equation}
Substituting Eqs.~\eqref{eq:C0-curvature}, \eqref{eq:Cstar-curvature}, and \eqref{eq:velavg-summary} into Eq.~\eqref{eq:Aiidef} gives the leading thin-torus absolute amplitudes
\begin{align}
\mathcal A_{0,yy}
&=
\frac{e^2\tau_{\rm tr}(\EF)}{4\pi^2\hbar^2}\sqrt{\frac{2e}{\hbar}}\,
R\Bigl[1+O\!\left(\frac{r^2}{R^2}\right)\Bigr],
\label{eq:A0yy-absolute}\\
\mathcal A_{\star,xx}
&=
\frac{e^2\tau_{\rm tr}(\EF)}{2\pi^2\hbar^2}\sqrt{\frac{2\pi e}{\hbar}}\,
\frac{c_0}{3}\kappa^{-1/2}R^{1/4}r^{3/4}
\Bigl[1+O\!\left(\frac{r}{R}\right)\Bigr],
\label{eq:Astarxx-absolute}\\
\mathcal A_{\star,yy}
&=
\frac{e^2\tau_{\rm tr}(\EF)}{2\pi^2\hbar^2}\sqrt{\frac{2\pi e}{\hbar}}\,
c_0\tilde c_y\kappa^{-1/2}R^{-3/4}r^{7/4}
\Bigl[1+O\!\left(\frac{r}{R}\right)\Bigr].
\label{eq:Astaryy-absolute}
\end{align}
These amplitudes are absolute within the CRTA-Lifshitz--Kosevich approximation, but they are not topologically universal because they depend on the transport lifetime.
The same nonlinking frequency $F_\star$ therefore appears in both tensor components, but it is parametrically weaker in $\sigma_{yy}$ than in $\sigma_{xx}$ in the thin-torus regime.

\paragraph*{Final component-resolved oscillatory conductivities.}
Keeping only the leading $B_x^{1/2}\nu^{-1/2}$ LK terms, the linking family contributes to $\sigma_{yy}$ but not to $\sigma_{xx}$, while the maximal nonlinking family contributes to both components.
Substituting the orbit data into Eq.~\eqref{eq:generic-LK-family} gives
\begin{align}
\sigma_{yy}^{\rm osc}(B_x)
&=
\sum_{\nu=1}^{\infty}
\frac{2\mathcal A_{0,yy}\sqrt{B_x}}{\sqrt{\nu}}
R_{T,0}^{(\nu)}R_{D,0}^{(\nu)}R_{S,0}^{(\nu)}
\cos\!\left(\frac{2\pi \nu F_0}{B_x}+\frac{\pi}{4}\right)
\notag\\
&\quad+
\sum_{\nu=1}^{\infty}
\frac{2\mathcal A_{\star,yy}\sqrt{B_x}}{\sqrt{\nu}}
R_{T,\star}^{(\nu)}R_{D,\star}^{(\nu)}R_{S,\star}^{(\nu)}
\cos\!\left[\frac{2\pi \nu F_\star}{B_x}-2\pi \nu\left(\frac12\right)-\frac{\pi}{4}\right],
\label{eq:sigyy-final}\\
\sigma_{xx}^{\rm osc}(B_x)
&=
\sum_{\nu=1}^{\infty}
\frac{2\mathcal A_{\star,xx}\sqrt{B_x}}{\sqrt{\nu}}
R_{T,\star}^{(\nu)}R_{D,\star}^{(\nu)}R_{S,\star}^{(\nu)}
\cos\!\left[\frac{2\pi \nu F_\star}{B_x}-2\pi \nu\left(\frac12\right)-\frac{\pi}{4}\right].
\label{eq:sigxx-final}
\end{align}
Equation~\eqref{eq:sigxx-final} has no leading $F_0$ term.
On the $k_x=0$ mirror slice, $v_x=0$ at every point of the orbit, and the first nonzero $v_x^2$ contribution appears only when the stationary-phase expansion moves away from that slice.
The $F_0$ series is therefore leading in $\sigma_{yy}$ but not in $\sigma_{xx}$.
Its offset is still $\gamma_0=0$, since the $x=0$ loops link the nodal ring.
The maximal loops at $x=\pm x_\star$ are nonlinking and keep $\gamma_\star=1/2$.
The conductivity component selects which series has leading weight; it does not change the quantization rule.

\subsection{Why the \texorpdfstring{$\B\parallel\hat x$}{B parallel x} problem is treated semiclassically}
\label{subsec:Bx-obstruction}

The $\B\parallel\hat x$ orientation does not give the block reduction used for $\B\parallel\hat z$.
With $[\Pi_y,\Pi_z]=iB$ and $k_x$ conserved, a convenient form of the Hamiltonian is
\begin{equation}
H_{B_x}=k_x s_x t_0+\Pi_y s_y t_y+\Pi_z s_z t_0-m_0 s_x t_x.
\label{eq:HBx}
\end{equation}
Squaring suggests the definitions
\begin{equation}
\Sigma_x\equiv s_x t_y,
\qquad
M_x\equiv k_x t_x-\Pi_y t_z s_z,
\label{eq:MxSigmax}
\end{equation}
and gives
\begin{equation}
H_{B_x}^2=\bigl(k_x^2+\Pi_y^2+\Pi_z^2+m_0^2\bigr)\I_4-2m_0 M_x-B\Sigma_x.
\label{eq:HBx2}
\end{equation}
The obstruction is already visible in the two identities
\begin{equation}
M_x^2=(k_x^2+\Pi_y^2)\I_4,
\label{eq:Mx2}
\end{equation}
and
\begin{equation}
\{M_x,\Sigma_x\}=-2\Pi_y t_x s_y\neq0.
\label{eq:anticomm-x}
\end{equation}
The first equation has no $-B\Sigma_x$ term, and the second shows that $\Sigma_x$ does not label independent two-dimensional blocks.
Thus $H_{B_x}^2$ cannot be reorganized into the same $2\times2$ Landau problems as in Sec.~\ref{sec:Bz-LL}.
The formulas in Sec.~\ref{sec:Bx} are therefore semiclassical LK formulas for this field direction.

%%%%%%%%%%%%%%%%%%%%%%%%%%%%%%%%%%%%%%%%%%%%%%%%%%%%%%%%%%%%%%%%%%%%%%%%%%%%%%
\section{Lattice regularization and numerical analysis protocol}
\label{sec:lattice}

The paper corroborates the $\B\parallel\hat z$ phase pinning with a lattice calculation on a sine-regularized tight-binding model.
This section documents the Hamiltonian, the broadened Kubo observable, and the analysis pipeline used to extract frequencies and phases from the numerical data.

\subsection{Sine-regularized lattice model}

In this subsection, we use the dimensionless lattice convention of the numerical code: the lattice energy scale is set to $\epsilon=1$, momenta appear through $k_i a_i$, and $m_0$ denotes the dimensionless quantity $m_0 a_x$ appearing in the dimensional paper convention.
The lattice regularization used in the paper is
\begin{equation}
H_{\rm lat}(\bk)=
\big[\sin (k_xa_x)-m_0\,t_x\big]s_x+\sin (k_ya_y)\,t_y s_y+\sin (k_za_z)\,s_z.
\label{eq:Hlat}
\end{equation}
Here $a_{i=x,y,z}$ are the lattice constants.
At zero field this model is $PT$ symmetric and realizes the same local ``ring-and-thread'' structure as the continuum theory, up to Brillouin-zone copies required by lattice periodicity.
For small dopings about the nodal ring, its extremal toroidal sections reproduce the continuum frequencies smoothly.

\subsection{Magnetic field and broadened Kubo conductivity}

For $\B\parallel\hat z$, the magnetic field is introduced by the usual Peierls substitution on the lattice hoppings.
We denote the flux per elementary plaquette in units of $h/e$ by $\varphi=p/q$.
The Peierls phase around one plaquette is $2\pi\varphi$, and the dimensionless magnetic-field variable used in the code and in the continuum formulas is
\begin{equation}
B\equiv 2\pi\varphi=\frac{2\pi p}{q}.
\label{eq:lattice-B-from-flux}
\end{equation}
Thus the inverse-field variable used for FFTs and Landau fans is
\begin{equation}
u\equiv\frac{1}{B}=\frac{q}{2\pi p}.
\label{eq:lattice-u-from-q}
\end{equation}
At rational $\varphi=p/q$, the magnetic field enlarges the in-plane magnetic unit cell to $q$ sites in a convenient Landau gauge.
The resulting magnetic Bloch Hamiltonian $H_B(k_\parallel)$ is finite dimensional for each conserved momentum $k_\parallel$.
A noninteracting broadened Kubo conductivity is then evaluated as
\begin{equation}
\sigma_{xx}(\EF,B)=\frac{e^2}{2\pi V}
\Tr\!\left[\hat{v}_x \hat{G}^R(\EF) \hat{v}_x \hat{G}^A(\EF)\right],
\qquad
\hat{G}^{R/A}(\EF)=\frac{1}{\EF-\hat{H}_B\pm i \eta},
\label{eq:lattice-kubo}
\end{equation}
where $\eta$ is a phenomenological broadening that plays the role of a finite lifetime.
The field window in the paper is chosen in the small-flux regime so that the oscillatory signal is periodic in $1/B$ and the lattice Onsager areas can be compared directly with the continuum expectations.

\subsection{Magnetic-supercell convention, velocity operator, and recursive Green-function evaluation}
\label{subsec:magnetic-supercell-current}

We now specify the magnetic-supercell convention used in the lattice Kubo calculation.
For $\B\parallel\hat z$ we use Landau gauge $A_y=Bx$ and rational flux $\varphi=p/q$ per elementary $x$--$y$ plaquette, measured in units of $h/e$.
The Peierls phase around one plaquette is therefore $2\pi p/q$.
Equivalently, in the dimensionless field convention used below, $B=2\pi p/q$ and $u=1/B=q/(2\pi p)$.
The magnetic unit cell contains $q$ pristine lattice spacings in the $x$ direction.
The $y$-direction Peierls phase is implemented as
\begin{equation}
k_y\rightarrow k_y+2\pi(p/q)x,
\qquad x=0,\ldots,q-1 .
\label{eq:landau-gauge-ky}
\end{equation}

Let $T_x$ denote the nearest-neighbor hopping matrix associated with the sine-regularized term $\epsilon\sin(k_xa_x)s_x$; for $a_x=1$, $T_x=-i \epsilon s_x/2$.
In the magnetic-cell basis $x=0,\ldots,q-1$, the code uses the convention
\begin{equation}
H_{x,x+1}=T_x,\qquad x=0,\ldots,q-2,
\qquad
H_{q-1,0}=T_x e^{i  k_x q},
\label{eq:seam-bloch-phase}
\end{equation}
together with the Hermitian-conjugate hoppings.
Thus the magnetic Bloch phase in the $x$ direction is placed only on the hopping that crosses the magnetic-cell seam.

To fix the current convention, introduce an additional dimensionless twist $\theta_x$ on the same seam,
\begin{equation}
H_{q-1,0}^{\theta_x}
=
T_x\exp\!\left[i (k_xq+\theta_x)\right],
\qquad
H_{0,q-1}^{\theta_x}
=
T_x^\dagger\exp\!\left[-i (k_xq+\theta_x)\right].
\label{eq:seam-twist}
\end{equation}
The local current conjugate to this seam twist is defined, up to the conventional charge and $\hbar$ prefactors, by
\begin{equation}
\hat j_{\bar x}
=
\left.
\frac{\partial \hat H^{\theta_x}}{\partial\theta_x}
\right|_{\theta_x=0}.
\label{eq:seam-current}
\end{equation}
This is the seam-current construction used here: the current operator is defined by differentiating with respect to an electromagnetic twist localized on a seam, as in standard lattice treatments of local currents~\cite{watanabe2019proof}.
In the present magnetic Bloch convention the code evaluates the matrix derivative with respect to the magnetic Bloch momentum $k_x$,
\begin{equation}
\hat v_x^{\rm code}
\equiv
\frac{\partial \hat H(k_x)}{\partial k_x}.
\end{equation}
Since $k_x$ enters Eq.~\eqref{eq:seam-twist} only through the combination $k_xq+\theta_x$, the two derivatives are related by the chain rule:
\begin{equation}
\hat v_x^{\rm code}
=
q\,\hat j_{\bar x}.
\label{eq:kx-theta-chain-rule}
\end{equation}
Equivalently,
\begin{equation}
\hat v_x^{\rm code}
=
V_x |q-1\rangle\langle 0|
+
V_x^\dagger |0\rangle\langle q-1|,
\qquad
V_x=
i  q\,T_x e^{i  k_x q}.
\label{eq:boundary-velocity}
\end{equation}
The factor of $q$ in Eq.~\eqref{eq:kx-theta-chain-rule} is part of the implemented $k_x$-derivative convention and affects the conductivity normalization and envelope.
It does not affect the locations of the Landau-level crossings as a function of $1/B$.

For $i=x$, Eq.~\eqref{eq:boundary-velocity} implies that the Kubo trace depends only on Green-function blocks touching the magnetic-cell seam.
Writing $a=q-1$, $b=0$, and denoting the $4\times4$ block of $\hat R=\Im\hat G^R$ between magnetic-cell coordinates $x$ and $y$ by $R_{xy}$, one obtains
\begin{align}
\Tr\!\left[
\hat v_x^{\rm code}\hat R
\hat v_x^{\rm code}\hat R
\right]
=
\Tr_4\Big[
&
V_x R_{ba} V_x R_{ba}
+
V_x R_{bb} V_x^\dagger R_{aa}
\nonumber\\
&
+
V_x^\dagger R_{aa} V_x R_{bb}
+
V_x^\dagger R_{ab} V_x^\dagger R_{ab}
\Big].
\label{eq:boundary-green-trace}
\end{align}
Therefore the $\sigma_{xx}$ calculation in this convention only requires the four selected blocks
\begin{equation}
R_{aa},\quad R_{bb},\quad R_{ab},\quad R_{ba}.
\end{equation}
The numerical implementation exploits this fact by solving the cyclic block-tridiagonal linear system
\begin{equation}
\left[z-\hat H_B(\bk;p/q)\right]G(z,\bk)=1
\end{equation}
only for right-hand sides localized at the two seam blocks.
This selected-block recursive Green-function evaluation is a computational consequence of the boundary support of Eq.~\eqref{eq:boundary-velocity}; it is not an additional physical assumption.
For fixed internal block dimension, the selected-block calculation scales much more favorably with $q$ than dense inversion of the full $4q\times4q$ matrix.
Recursive Green-function and block-decomposition methods do the numerical heavy lifting in such selected Green-function-block calculations~\cite{lewenkopf2013recursive,drouvelis2006parallel}.

The other velocity components in the same $\B\parallel\hat z$ implementation are not seam-supported.
They are block diagonal throughout the magnetic unit cell:
\begin{equation}
v_y(x)
=
\epsilon\cos\!\left[k_y+2\pi(p/q)x\right]\,t_y s_y,
\qquad
v_z(x)
=
\epsilon\cos k_z\,s_z .
\label{eq:vy-vz-blockdiag}
\end{equation}
Their Kubo traces involve
\begin{equation}
\Tr\!\left[
\hat v_i\hat R\hat v_i\hat R
\right]
=
\sum_{x,y=0}^{q-1}
\Tr_4\!\left[
v_i(x)R_{xy}v_i(y)R_{yx}
\right],
\qquad i=y,z .
\label{eq:blockdiag-green-trace}
\end{equation}
Thus, in the present implementation, $\sigma_{yy}$ and $\sigma_{zz}$ require all $R_{xy}$ blocks and are evaluated by the denser Green-function path.
This is why they are substantially more expensive than the selected-block $\sigma_{xx}$ calculation.

Finally, the velocity matrix used here belongs to the seam-gauge magnetic cell.
It is the derivative of the seam-gauge Hamiltonian defined by the Bloch phase at the boundary of the magnetic unit cell.
If the same magnetic Hamiltonian is written with the Bloch phase spread across the sites of the magnetic cell, the two Hamiltonian matrices are related by a $k_x$-dependent change of basis.
In formulas,
\begin{equation}
\tilde H(k_x)=U^{-1}(k_x)H(k_x)U(k_x),
\qquad
U(k_x)=\sum_{x=0}^{q-1}e^{i k_x x}|x\rangle\langle x|.
\label{eq:k-dependent-unitary}
\end{equation}
The spectra of $H$ and $\tilde H$ are the same at fixed $k_x$.
The matrix derivative is not the same object.
Differentiating gives
\begin{equation}
\frac{\partial \tilde H}{\partial k_x}
=
U^{-1}\frac{\partial H}{\partial k_x}U
+
\left(\frac{\partial U^{-1}}{\partial k_x}\right)HU
+
U^{-1}H\left(\frac{\partial U}{\partial k_x}\right).
\label{eq:velocity-convention-dependence}
\end{equation}
Thus the current matrix used in a lattice Kubo trace has to be specified together with the electromagnetic-coupling convention.
In practice this means specifying the orbital-position convention, or using an equivalent covariant velocity prescription~\cite{esteveparedes2023velocity}.

In the numerical analysis below we keep the seam convention fixed.
The conductivities are used to read off SdH periods and Onsager offsets from the magnetic spectrum.
We do not use the absolute normalization, the smooth envelope, or the ratio of current components as calibrated material transport data.
Those quantities can change with current form factors, with the $q$ factor in Eq.~\eqref{eq:kx-theta-chain-rule}, or with a different distribution of the Peierls coupling inside the magnetic unit cell.

\subsection{Background subtraction, FFT, and phase extraction}

For each field orientation the stored output is a list of pairs $(B_j,\sigma_{ii}(B_j))$, with $i=x$ or $y$ chosen according to the current component being analyzed.
The fluxes are rational,
\begin{equation}
\varphi_j=\frac{p_j}{q_j},
\qquad
B_j=2\pi\varphi_j=\frac{2\pi p_j}{q_j},
\end{equation}
and the inverse-field coordinate used in the FFT and fan plots is
\begin{equation}
u_j=\frac{1}{B_j}=\frac{q_j}{2\pi p_j}.
\end{equation}
For the $p=1$ data used here, uniform sampling in $q$ is therefore uniform sampling in $u$, but the two variables differ by the factor $2\pi$.
We subtract an empirical smooth background $\sigma_{ii}^{\rm smooth}(u)$ over the analysis $u$ window.
This background is defined operationally by the chosen smooth fit to the nonoscillatory part of the raw data, for example, a low-order polynomial in $u$ or another comparably smooth curve whose variation scale is much longer than the relevant oscillation periods $1/F_\alpha$.
It contains the analytic continuum contribution in Eq.~\eqref{eq:sperp-background} in the appropriate weak-field continuum limit, but for lattice or experimental data it also absorbs lattice corrections, finite broadening, slow magnetoresistance, and other smooth nonuniversal contributions.
The oscillatory signal is then
\begin{equation}
\Delta\sigma_{ii}(u)=\sigma_{ii}(u)-\sigma_{ii}^{\rm smooth}(u),
\qquad u=\frac{1}{B}.
\label{eq:deltasigma}
\end{equation}
A Hann-windowed fast Fourier transform (FFT) of $\Delta\sigma_{ii}(u)$ identifies the dominant frequencies.
Explicitly, after uniform sampling in $u=1/B$, the finite data segment is multiplied by
\begin{equation}
W_j=\frac{1}{2}\left[1-\cos\!\left(\frac{2\pi j}{N-1}\right)\right],
\qquad j=0,\ldots,N-1,
\end{equation}
before taking the FFT. This taper suppresses spectral leakage from the finite $1/B$ interval. The FFT is applied only after representing the signal on a uniform $u$ grid; for nonuniform experimental field values, one should first interpolate the background-subtracted signal onto such a grid. For $\B\parallel\hat z$, the dominant peaks are associated with the lattice analogs of the inner and outer toroidal contours $F_\pm$. For $\B\parallel\hat x$, the same FFT step is applied component by component: $\sigma_{xx}$ is dominated by the nonlinking $F^\ast$ series, while $\sigma_{yy}$ contains both the linking $F_0$ series and the nonlinking $F^\ast$ series.

To extract phases, we do not fit an arbitrary global envelope function. Instead, we use a local quadrature form in sliding inverse-field windows. In a window $w$ centered at $u_w$, after fixing the frequencies $F_r$ from the FFT, we fit
\begin{equation}
\Delta\sigma^{(w)}_{xx}(u)
=
\sum_{r=\pm}\sum_{\nu=1}^{\nu_{\max}}
\left[
C^{(w)}_{\nu,r}\cos\theta_{\nu,r}(u)
+S^{(w)}_{\nu,r}\sin\theta_{\nu,r}(u)
\right].
\label{eq:fitform}
\end{equation}
Here
\begin{equation}
\theta_{\nu,r}(u)=2\pi \nu F_r u+\varphi_{\rm 3D,r},
\qquad
\varphi_{\rm 3D,+}=-\pi/4,\quad
\varphi_{\rm 3D,-}=+\pi/4 .
\end{equation}
The coefficients $C^{(w)}_{\nu,r}$ and $S^{(w)}_{\nu,r}$ are constants within the window $w$.
For the Fig.~3(c) phase extraction, the fit basis includes harmonics $\nu=1,\ldots,6$.
The Berry phase plotted in the paper is extracted from the fundamental $\nu=1$ quadrature coefficients; the higher harmonics stabilize the local fit and provide internal consistency checks.
The fit form in Eq.~\eqref{eq:fitform} is written explicitly for the $\B\parallel\hat z$ two-frequency fit, where $r=\pm$ and the fitted offset tests the toroidal Berry phase $\Phiphi$.
The $\B\parallel\hat x$ phase extraction in Fig.~4(c) uses the identical sliding-window quadrature construction after replacing the orbit label $r$ by $\alpha\in\{0,\star\}$ and the conductivity component by the measured component $i$.
Thus $\Delta\sigma_{xx}$ is fit with the nonlinking family $\alpha=\star$, while $\Delta\sigma_{yy}$ is fit with both $\alpha=0$ and $\alpha=\star$.
The fixed curvature phases and expected offsets are
\begin{equation}
\varphi_{\rm 3D,0}=+\frac{\pi}{4},
\qquad
\gamma_0=0
\quad(\Phitheta=\pi),
\qquad
\varphi_{\rm 3D,\star}=-\frac{\pi}{4},
\qquad
\gamma_\star=\frac{1}{2}
\quad(\Phi=0),
\label{eq:Bx-phase-fit-targets}
\end{equation}
as derived in Sec.~\ref{sec:Bx}.
In this way, the same local quadrature protocol gives the fitted Berry phases in Fig.~3(c) for $\B\parallel\hat z$ and in Fig.~4(c) for $\B\parallel\hat x$.
Equivalently,
\begin{equation}
\Delta\sigma^{(w)}_{xx}(u)
=
\sum_{r=\pm}\sum_{\nu=1}^{\nu_{\max}}
A^{(w)}_{\nu,r}
\cos\!\left[
2\pi \nu\left(F_ru-\gamma^{(w)}_{\nu,r}\right)
+\varphi_{\rm 3D,r}
\right],
\end{equation}
with the local positive envelope
\begin{equation}
A^{(w)}_{\nu,r}
=
\sqrt{
\left(C^{(w)}_{\nu,r}\right)^2+
\left(S^{(w)}_{\nu,r}\right)^2
}
\end{equation}
and local offset
\begin{equation}
\gamma^{(w)}_{\nu,r}
=
\frac{1}{2\pi \nu}
\operatorname{atan2}\!\left(
S^{(w)}_{\nu,r},
C^{(w)}_{\nu,r}
\right)
\quad \mathrm{mod}\ \frac{1}{\nu}.
\end{equation}
Here $\operatorname{atan2}(y,x)$ denotes the argument of $x+iy$.
With this convention,
\[
C^{(w)}_{\nu,r}=A^{(w)}_{\nu,r}\cos(2\pi \nu\gamma^{(w)}_{\nu,r}),
\qquad
S^{(w)}_{\nu,r}=A^{(w)}_{\nu,r}\sin(2\pi \nu\gamma^{(w)}_{\nu,r}).
\]
Thus the envelope is operationally defined as the quadrature magnitude $A^{(w)}_{\nu,r}$, not as an arbitrary unknown function of $B$. The LK phase information is extracted from the quadrature angle, equivalently from $\gamma^{(w)}_{\nu,r}$; for the $\B\parallel\hat x$ fits one uses the same definitions with $r$ replaced by $\alpha=0,\star$.

The same procedure applies to experimental data. One subtracts a smooth nonoscillatory background, identifies the Onsager frequencies from the FFT or independently known extremal areas, fixes the curvature phases from the extremal geometry, and extracts the local quadrature coefficients in sliding $1/B$ windows. For the fundamental harmonic we write $\gamma_\alpha^{(w)}\equiv\gamma_{1,\alpha}^{(w)}$, with $\alpha=r=\pm$ for $\B\parallel\hat z$ and $\alpha=0,\star$ for $\B\parallel\hat x$; higher harmonics provide consistency checks after the $1/\nu$ ambiguity is unwrapped. We define the effective phase inferred from the fitted LK offset as
\begin{equation}
\Phi^{(w)}_{{\rm eff},\alpha}
=2\pi\left[\frac{1}{2}-\gamma^{(w)}_\alpha\right]
\quad (\mathrm{mod}\ 2\pi).
\end{equation}
In the spinless quantized-Berry-phase model used for Figs.~3 and 4 of the paper, $\Phi_{\rm eff}$ equals the orbit Berry phase.
In an SOC-gapped spinful material, $\Phi_{\rm eff}$ should instead be interpreted as the full LK phase combination, including geometric Berry, orbital-moment, Zeeman, and possible non-Abelian corrections.
We assess fit stability by varying the background subtraction, the window width and position, the number of harmonics, and the local amplitude model. In the numerical data shown in the paper, the $\B\parallel\hat z$ toroidal series converge to $\gamma_\pm=0$, i.e., $\Phiphi=\pi$, while the $\B\parallel\hat x$ component-resolved fits converge to $\gamma_0=0$ for the linking $F_0$ series and to $\gamma_\star=1/2$ for the nonlinking $F_\star$ series, denoted $F^\ast$ in Fig.~4.

\subsection{Numerical parameters for Figs.~3 and 4}
\label{subsec:fig34-numerical-params}

The lattice-Kubo data in Figs.~3 and 4 of the paper use the dimensionless sine-regularized model with $m_0=1$ and $\EF=0.20$.
The inverse-field variable is $u=1/B$; for rational-flux data $\varphi=p/q$ this equals $u=q/(2\pi p)$, as described above.
The graphdiyne reference conversion to SI units is applied only after the dimensionless data are generated, using $B_{\rm SI}^{-1}=u/B_0$ and $F_{\rm SI}=B_0F$.
Unless otherwise stated, the production finite-field Kubo runs use $p=1$.
Thus the magnetic unit cell contains $q\simeq2\pi u$ sites in the enlarged direction selected by the Landau gauge.
The plotted windows therefore correspond to approximately $q=220$--$1005$ for Fig.~3 and $q=251$--$1634$ for Fig.~4.
For $\B\parallel\hat z$, the production magnetic-Brillouin-zone sampling is $(N_{\theta_x},N_{k_y},N_{k_z})=(2,8,1001)$, where $\theta_x$ is the seam twist of the $q$-site magnetic cell.
For $\B\parallel\hat x$, we use the same code convention after the cyclic relabeling $(x_{\rm code},y_{\rm code},z_{\rm code})=(y,z,x)_{\rm phys}$; the same code-frame grid therefore corresponds in physical coordinates to dense sampling along the field-parallel momentum, $(N_{k_x},N_{\theta_y},N_{k_z})=(1001,2,8)$.
We checked the stability of the dominant FFT peaks and the extracted offsets against denser sampling up to $(4,16,2002)$ in the same code-frame convention, i.e., up to $2002$ points along the field-parallel momentum.

For Fig.~3, the continuum reference frequencies are
\begin{equation}
F_-=\frac{(1-\EF)^2}{2}=0.320,
\qquad
F_+=\frac{(1+\EF)^2}{2}=0.720 .
\end{equation}
The plotted $\B\parallel\hat z$ conductivity trace is analyzed on the inverse-field interval $u\in[35,160]$.
The FFT is performed after a Hann window and uses zero padding to locate the two dominant peaks.
The local quadrature phase scan uses window widths
\begin{equation}
\Delta u=15,20,25,30,35,40,50,
\end{equation}
with $60$ window centers for each width.
The fit basis includes harmonics $\nu=1,\ldots,6$, while the Berry phase reported in Fig.~3(c) is extracted from the fundamental $\nu=1$ quadrature coefficients.
The displayed Fig.~3(c) curve uses the $\Delta u=40$ scan, with smoothing applied only as a visual guide.
The damping parameter entering the Fig.~3 oscillatory envelope is $B_d=0.028$; its graphdiyne reference conversion is given below in the graphdiyne reference-scale subsection.

For Fig.~4, the same dimensionless values $m_0=1$ and $\EF=0.20$ are used.
The linking in-plane frequency is
\begin{equation}
F_0=\frac{\EF^2}{2}=0.020,
\end{equation}
and the nonlinking maximal orbit gives $F^\ast\simeq0.0893$.
The plotted $\B\parallel\hat x$ data use the inverse-field interval $u\in[40,260]$.
The $\sigma_{xx}$ analysis is dominated by the nonlinking $F^\ast$ series, whereas the $\sigma_{yy}$ phase fit contains both $F_0$ and $F^\ast$.
The curvature phases and expected offsets are those in Eq.~\eqref{eq:Bx-phase-fit-targets}.
The Fig.~4(c) quadrature windows have centers $u_w\in[95,205]$ and width $\Delta u=110$, with smoothing again used only for the plotted guide curve.

\subsection{Smaller effective lattice constant check for \texorpdfstring{$\B\parallel\hat z$}{B parallel z}}
\label{subsec:smaller-lattice-check}

The paper figures use $m_0^{\rm lat}=1$ and $\EF^{\rm lat}=0.20$ in the dimensionless sine-regularized model.
This choice keeps the magnetic unit cells moderate and makes the numerical convergence easy to inspect, but if one takes the graphdiyne ring radius $k_0=0.0202~\text{\AA}^{-1}$ literally, it corresponds to an effective sine-lattice length $1/k_0\simeq4.95$ nm.
As a separate unit-scale check, we therefore also computed the $\B\parallel\hat z$ $\sigma_{xx}$ signal with
\begin{equation}
m_0^{\rm lat}=0.1,\qquad \EF^{\rm lat}=0.01 .
\end{equation}
Keeping the same physical graphdiyne ring radius gives the effective sine-lattice constant
\begin{equation}
a_{\rm eff}
=
\frac{m_0^{\rm lat}}{k_0}
\simeq 4.95~\text{\AA},
\label{eq:small-lattice-aeff}
\end{equation}
and hence the magnetic-field unit
\begin{equation}
B_0^{\rm eff}
=
\frac{\hbar}{e a_{\rm eff}^2}
=
\frac{\hbar k_0^2}{e(m_0^{\rm lat})^2}
\simeq 2.69\times10^3~{\rm T}.
\label{eq:small-lattice-B0}
\end{equation}
With $v=7.0\times10^5~{\rm m/s}$, the same conversion gives
\begin{equation}
E_0^{\rm eff}=\frac{\hbar v}{a_{\rm eff}}\simeq0.932~{\rm eV},
\qquad
\EF=\EF^{\rm lat}E_0^{\rm eff}\simeq9.3~{\rm meV}.
\end{equation}
The continuum toroidal frequencies in this run are
\begin{equation}
F_-^{\rm lat}=\frac{(0.1-0.01)^2}{2}=0.00405,
\qquad
F_+^{\rm lat}=\frac{(0.1+0.01)^2}{2}=0.00605,
\end{equation}
or, after Eq.~\eqref{eq:small-lattice-B0},
\begin{equation}
F_-\simeq10.9~{\rm T},
\qquad
F_+\simeq16.2~{\rm T}.
\end{equation}
This run is still a cubic sine-regularized $\bk\cdot\mathbf{p}$ calculation, not a full atomistic graphdiyne tight-binding one.
Its purpose is to check that the field and frequency scales remain realistic when the sine-regulator length is reduced to an angstrom-scale value.

\begin{figure*}[t]
\centering
\includegraphics[width=\textwidth]{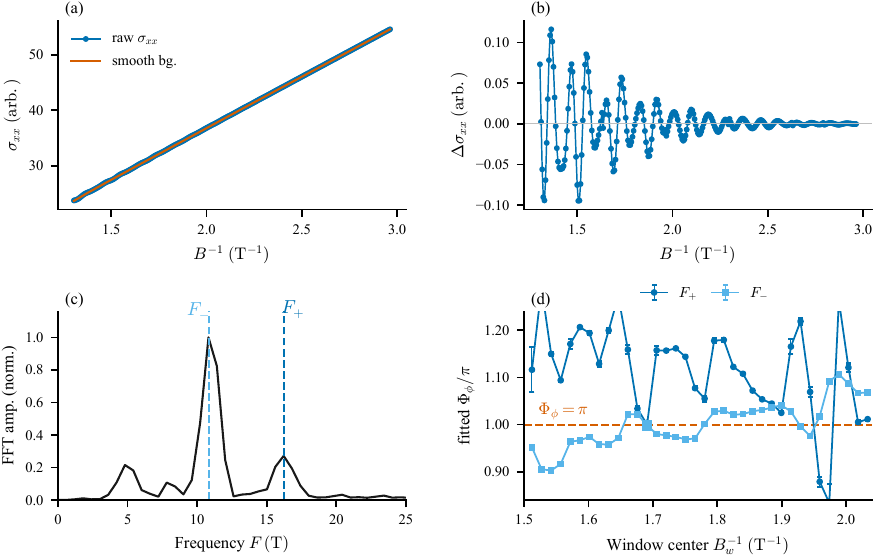}
\caption{
Smaller-effective-lattice-constant $\B\parallel\hat z$ unit-scale check for $\sigma_{xx}$.
The run uses $m_0^{\rm lat}=0.1$, $\EF^{\rm lat}=0.01$, $p=1$, $q=22000$--$50000$ in steps of $100$, and the graphdiyne-scale conversion in Eqs.~\eqref{eq:small-lattice-aeff} and \eqref{eq:small-lattice-B0}.
(a) Raw conductivity and smooth background.
(b) Background-subtracted oscillatory signal.
(c) Hann-windowed FFT, showing peaks at the expected toroidal frequencies $F_-\simeq10.9$ T and $F_+\simeq16.2$ T.
(d) Representative local-quadrature phase check near the best simultaneous window, with width $\Delta B^{-1}\simeq0.298~{\rm T}^{-1}$ and center $B_w^{-1}\simeq1.69~{\rm T}^{-1}$.
At this window, the fitted central values are $\Phi_-/\pi\simeq0.998$ and $\Phi_+/\pi\simeq1.000$.
The broader window dependence is more sensitive to background and endpoint choices than in the main paper data, so this run is used as a supplementary unit-scale check rather than as the primary phase extraction.
}
\label{fig:smaller-lattice-bz}
\end{figure*}

The rational-flux interval $q=22000$--$50000$ corresponds to
$B^{-1}\simeq1.30$--$2.96~{\rm T}^{-1}$, or $B\simeq0.34$--$0.77$ T, after the graphdiyne-scale conversion above.
The much larger magnetic-cell length compared with the paper calculation is the main numerical cost of reducing the effective lattice constant.

\subsection{Window dependence and phase-error estimate}
\label{subsec:phase-error}

We quantify the stability of the Fig.~3(c) phase extraction by repeating the local quadrature fit over a set of sliding-window widths and centers.
For each branch we define the percent deviation from the quantized toroidal Berry phase by
\begin{equation}
\epsilon_r^{(w)}
=100\,\frac{\left|\Phi_{B,r}^{(w)}-\pi\right|}{\pi},
\qquad r=\pm,
\label{eq:phase-error-percent}
\end{equation}
after unwrapping the fitted phase about $\Phiphi=\pi$.
Equivalently, for the fundamental harmonic in the notation of Eq.~\eqref{eq:fitform}, $\epsilon_r^{(w)}=200|\gamma_r^{(w)}|$ when $\gamma_r^{(w)}$ is chosen near zero.
We also use the simultaneous error
\begin{equation}
\epsilon_{\max}^{(w)}=\max\!\left(\epsilon_+^{(w)},\epsilon_-^{(w)}\right)
\label{eq:phase-error-max}
\end{equation}
because the claim $\Phiphi=\pi$ should hold for both toroidal branches in the same fitting window.

\begin{figure*}[t]
\centering
\includegraphics[width=\textwidth]{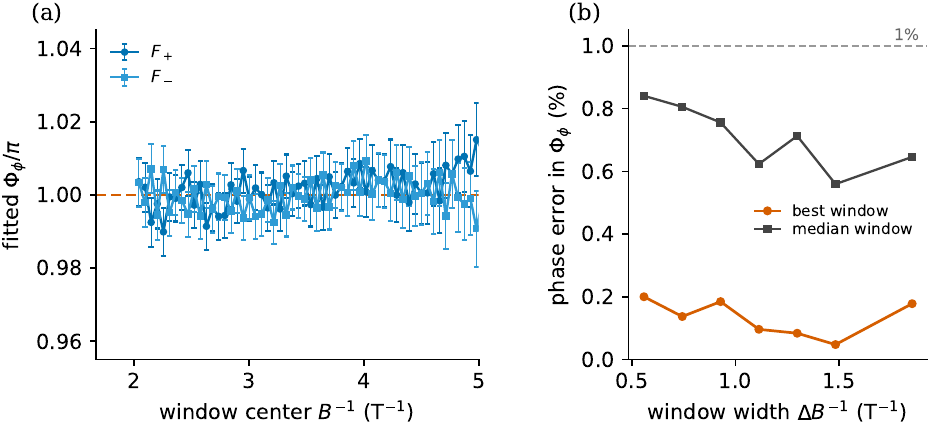}
\caption{
Window dependence of the local quadrature phase extraction used for Fig.~3(c) of the paper.
(a) Fitted Berry phases for the two toroidal frequencies as a function of the graphdiyne-scaled sliding-window center $B_w^{-1}=u_w/B_0$ at fixed width $\Delta B^{-1}=40/B_0\simeq1.48~{\rm T}^{-1}$.
The dashed horizontal line marks $\Phiphi=\pi$.
Error bars are the least-squares standard errors from the local quadrature fit and do not include systematic variation with the window choice.
(b) Dependence of the central-value phase error on the window width.
The orange curve gives the best simultaneous error $\epsilon_{\max}$ available at each width, while the gray curve gives the median $\epsilon_{\max}$ over window centers.
The dashed horizontal line marks a $1\%$ error.
}
\label{fig:phase-convergence}
\end{figure*}

The best simultaneous window occurs at $\Delta u=40$ and $u_w\simeq121.27$ in dimensionless units, equivalently
$\Delta B^{-1}\simeq1.48~{\rm T}^{-1}$ and
$B_w^{-1}\simeq4.50~{\rm T}^{-1}$ in the graphdiyne reference scaling, where
\begin{equation}
\epsilon_+=0.0474\%,\qquad
\epsilon_-=0.0415\%,\qquad
\epsilon_{\max}=0.0474\%.
\end{equation}
This best-window central value should not be taken as the conservative uncertainty of the method.
As shown in Fig.~\ref{fig:phase-convergence}(b), the median simultaneous error over window centers is $0.56\%$ for $\Delta u=40$ and remains of order $0.6$--$0.8\%$ for nearby window widths.
The fit standard errors in Fig.~\ref{fig:phase-convergence}(a) are also typically of the same order.
We therefore quote the lattice phase extraction as consistent with $\Phiphi=\pi$ at about the $1\%$ level, while noting that the best-window central value is much closer to quantization, at the $5\times10^{-2}\%$ level.
The small drift visible near the edge of the sliding-window range in Fig.~\ref{fig:phase-convergence}(a) is a windowing/systematic effect rather than a distinct phase: such windows are more sensitive to endpoint truncation, residual background curvature, finite Dingle damping, and the slowly varying local envelope.
Accordingly, the quoted phase is taken from the stable window range and checked against variations of the window width and center.

\subsection{Landau-fan consistency checks}
\label{subsec:landau-fan-checks}

The sliding-window quadrature fit is the phase-extraction method used in the paper for both field geometries because it fits the oscillatory signal directly after fixing the FFT frequencies and curvature phases.
As an independent consistency check, we also construct Landau fans from conductivity maxima of the band-pass-filtered oscillatory signal.
This is the conductivity counterpart of the common Shubnikov--de Haas (SdH) convention in which resistance minima are indexed by integers; for small scalar oscillations, a resistance minimum corresponds to a conductivity maximum.
The compact fans shown in Figs.~3(d) and 4(d) of the paper are the \emph{raw} fans, before subtracting the three-dimensional curvature phase.
They are included there only as visual consistency checks.
The quantitative fan analysis is kept here: we show both the raw and curvature-corrected fan conventions, list the fitted slopes and intercepts, and use the corrected intercepts only as secondary checks on the FFT/local-quadrature phase extraction.
For a single dominant frequency $F_\alpha$ with curvature phase $\varphi_{\rm 3D,\alpha}$, conductivity maxima of the fundamental harmonic are indexed by an integer $n$ and approximately satisfy
\begin{equation}
n-\frac{\varphi_{\rm 3D,\alpha}}{2\pi}
=
F_\alpha u-\gamma_\alpha,
\qquad u=\frac{1}{B}.
\label{eq:fan-fit-convention}
\end{equation}
Thus the intercept of the curvature-corrected fan is $-\gamma_\alpha$.
In this convention $\gamma_\alpha=0$ gives a corrected intercept at zero, whereas $\gamma_\alpha=1/2$ gives a corrected intercept at $-1/2$ modulo an integer.
Using conductivity minima instead would replace $n$ by $n+1/2$; this shifts the auxiliary raw index by one half but gives the same $\gamma_\alpha$ if the indexing convention is treated consistently.
Because a fan diagram uses only selected extrema and then extrapolates the fitted line to $1/B\to0$, it is more sensitive than the quadrature fit to endpoint effects, damping, residual background curvature, and extremum-picking choices.

\begin{figure*}[t]
\centering
\includegraphics[width=\textwidth]{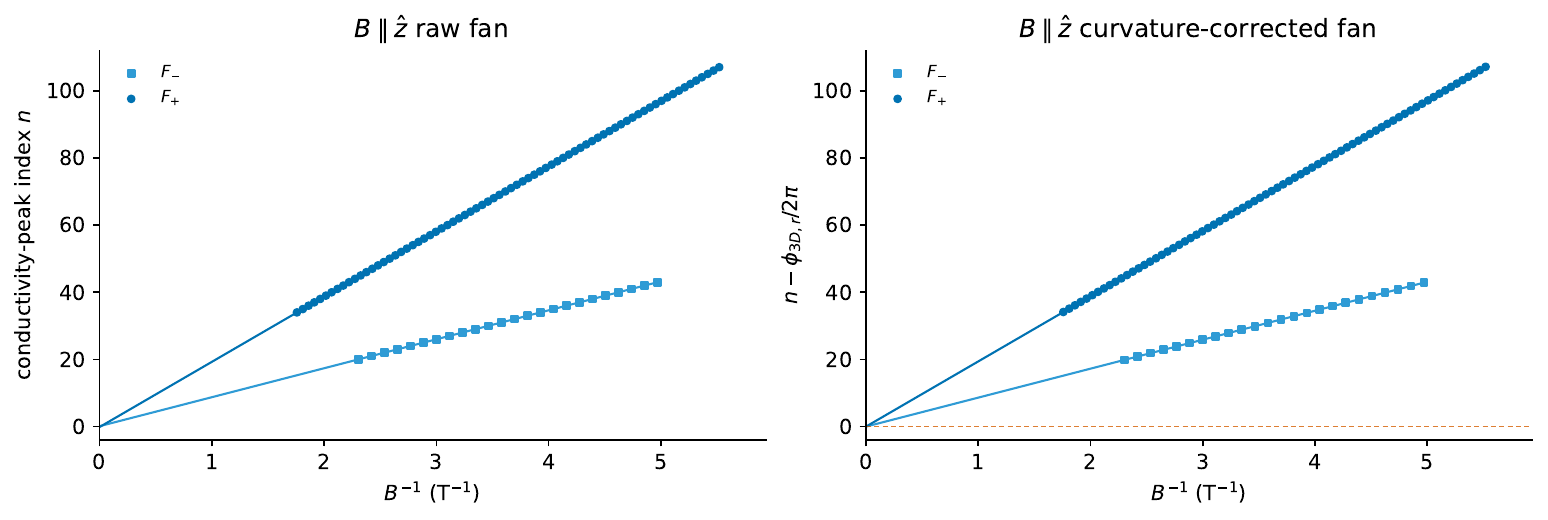}
\caption{
Conductivity-peak Landau-fan consistency check for the $\B\parallel\hat z$ lattice signal in Fig.~3 of the paper.
The left panel indexes maxima of the two band-pass-filtered oscillatory components using the integer $n$ and uses the same raw convention as the compact fan in Fig.~3(d) of the paper.
The right panel subtracts the three-dimensional curvature phase, so that Eq.~\eqref{eq:fan-fit-convention} has intercept $-\gamma_r$.
The horizontal axes are shown in the graphdiyne reference inverse-field units of Sec.~\ref{subsec:graphdiyne-figure-units}.
The fit parameters are listed in the text above; the curvature-corrected intercepts are close to zero for both branches.
These values are consistent with $\gamma_\pm=0$, equivalently $\Phiphi=\pi$, and the fitted slopes reproduce the FFT frequencies.
}
\label{fig:landau-fan-bz}
\end{figure*}

Figure~\ref{fig:landau-fan-bz} shows the corresponding fan construction for $\B\parallel\hat z$.
The left panel uses the same raw convention as the compact paper fan in Fig.~3(d), while the right panel shows the curvature-corrected fan.
The raw intercepts contain the expected curvature contribution $\varphi_{\rm 3D,\pm}/2\pi=\mp1/8$.
After subtracting this contribution, the corrected intercepts are within about $5\times10^{-3}$ of zero for both branches.
This agrees with the exact continuum fan relation and with the local quadrature result in Fig.~\ref{fig:phase-convergence}, but it should still be viewed as a check rather than the primary quantitative phase estimate.

\begin{samepage}
\medskip
\noindent\textit{$\B\parallel\hat z$ fan fit parameters.}
The raw fan fits the conductivity-maximum index $n$ to $m u+b$.
The curvature-corrected fan fits $n-\varphi_{\rm 3D,r}/2\pi$ to $m u+b$, so the corrected intercept is $b=-\gamma_r$.
\begin{center}
\begin{tabular}{lcccc}
\hline\hline
Series & Fan & Slope $m$ (T) & Intercept $b$ & $b\ {\rm mod}\ 1$ \\
\hline
$F_-$ & raw & $8.6239\pm0.0013$ & $+0.1299\pm0.0049$ & $0.1299\pm0.0049$ \\
$F_+$ & raw & $19.4064\pm0.0010$ & $-0.1300\pm0.0039$ & $0.8700\pm0.0039$ \\
$F_-$ & corrected & $8.6239\pm0.0013$ & $+0.0049\pm0.0049$ & $0.0049\pm0.0049$ \\
$F_+$ & corrected & $19.4064\pm0.0010$ & $-0.0050\pm0.0039$ & $0.9950\pm0.0039$ \\
\hline\hline
\end{tabular}
\end{center}
\medskip
\end{samepage}

\begin{figure*}[t]
\centering
\includegraphics[width=\textwidth]{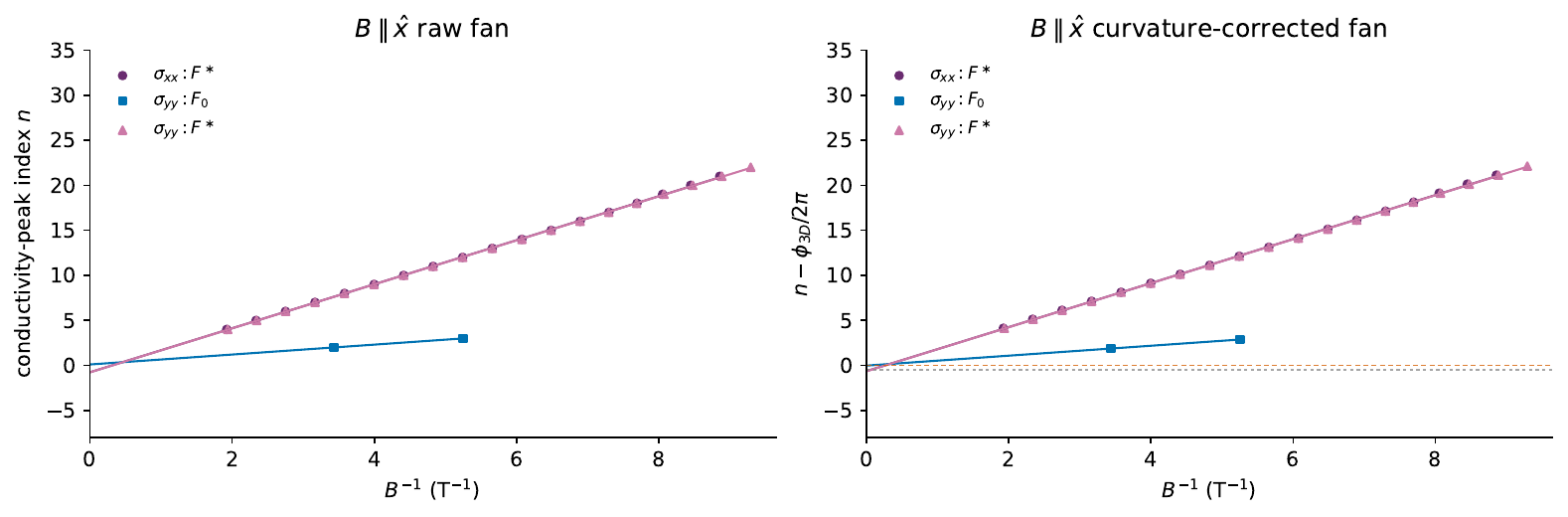}
\caption{
Conductivity-peak Landau-fan consistency check for the $\B\parallel\hat x$ component-resolved signals in Fig.~4 of the paper.
The left panel uses the same raw convention as the compact fan in Fig.~4(d) of the paper.
The right panel uses the curvature-corrected convention of Eq.~\eqref{eq:fan-fit-convention}; the orange dashed line marks the $\gamma=0$ intercept and the gray dashed line marks the $\gamma=1/2$ intercept.
The horizontal axes are shown in the graphdiyne reference inverse-field units of Sec.~\ref{subsec:graphdiyne-figure-units}.
The linking $F_0$ series in $\sigma_{yy}$ is close to the $\gamma=0$ line, while the nonlinking $F^\ast$ series in $\sigma_{xx}$ and $\sigma_{yy}$ lies closer to the $\gamma=1/2$ line.
The fit parameters are listed in the text above.
The larger deviations compared with Fig.~\ref{fig:landau-fan-bz} come from the much smaller $\B\parallel\hat x$ frequencies: the same finite field window contains fewer resolved oscillations, making the extrapolated intercept more sensitive to peak-picking, endpoint, damping, and slow-envelope effects.
}
\label{fig:landau-fan-bx}
\end{figure*}

The $\B\parallel\hat x$ fan in Fig.~\ref{fig:landau-fan-bx} is less quantitative.
Here the relevant frequencies are $F^\ast\simeq2.4$ T and $F_0\simeq0.55$ T in graphdiyne reference units, corresponding to dimensionless values $F^\ast\simeq0.089$ and $F_0\simeq0.020$.
They are much smaller than the $\B\parallel\hat z$ frequencies $F_+\simeq19.4$ T and $F_-\simeq8.6$ T.
For a fixed inverse-field interval $\Delta u$, the number of resolved oscillations scales as $F_\alpha\Delta u$.
Thus the $F^\ast$ fan contains roughly $3.5$--$8$ times fewer periods than the two $\B\parallel\hat z$ toroidal fans, while the $F_0$ fan contains roughly $16$--$36$ times fewer periods.
The fan intercept is then obtained by extrapolating a line fitted at finite $u=1/B$ back to $u=0$.
With fewer extrema, small systematic shifts of the picked peak positions---from band-pass filtering, endpoint truncation, Dingle damping, or a slowly varying envelope---produce a larger change in the fitted intercept.
This makes the $\B\parallel\hat x$ fan intercept less precise than the FFT/local-quadrature analysis.
For this reason we use the $\B\parallel\hat x$ Landau fan only to check consistency with the component-resolved FFT and local phase fits in Fig.~4(c), not as a precision estimate of the Berry phase.

\begin{samepage}
\medskip
\noindent\textit{$\B\parallel\hat x$ fan fit parameters.}
The same raw and curvature-corrected conventions are used.
The $F_0$ row has a small formal statistical error because only a few nearly collinear extrema are selected; this does not imply a small systematic uncertainty.
\begin{center}
\begin{tabular}{lcccc}
\hline\hline
Series & Fan & Slope $m$ (T) & Intercept $b$ & $b\ {\rm mod}\ 1$ \\
\hline
$\sigma_{xx}:F^\ast$ & raw & $2.4430\pm0.0079$ & $-0.7589\pm0.0459$ & $0.2411\pm0.0459$ \\
$\sigma_{yy}:F_0$ & raw & $0.5509\pm{<}10^{-4}$ & $+0.1082\pm{<}10^{-4}$ & $0.1082\pm{<}10^{-4}$ \\
$\sigma_{yy}:F^\ast$ & raw & $2.4424\pm0.0053$ & $-0.7649\pm0.0320$ & $0.2351\pm0.0320$ \\
$\sigma_{xx}:F^\ast$ & corrected & $2.4430\pm0.0079$ & $-0.6339\pm0.0459$ & $0.3661\pm0.0459$ \\
$\sigma_{yy}:F_0$ & corrected & $0.5509\pm{<}10^{-4}$ & $-0.0168\pm{<}10^{-4}$ & $0.9832\pm{<}10^{-4}$ \\
$\sigma_{yy}:F^\ast$ & corrected & $2.4424\pm0.0053$ & $-0.6399\pm0.0320$ & $0.3601\pm0.0320$ \\
\hline\hline
\end{tabular}
\end{center}
\medskip
\end{samepage}

\clearpage

\subsection{Numerical reproducibility checks}

For a stable phase extraction, the following checks are essential.
\begin{itemize}
  \item The FFT peak positions and fitted $\gamma_\pm$ should be insensitive to the choice of the smooth background.
  \item The fitted phases should remain stable as the broadening $\eta$ is varied within the regime where the oscillations remain well resolved.
  \item The extracted frequencies should converge as the set of rational fluxes is refined and the magnetic unit cell is enlarged.
  \item The fixed-frequency local quadrature fit should be consistent with the FFT peak positions and with the expected curvature phases: $\varphi_{\rm 3D,\pm}=\mp\pi/4$ for the $\B\parallel\hat z$ toroidal series, $\varphi_{\rm 3D,0}=+\pi/4$ for the linking $\B\parallel\hat x$ series, and $\varphi_{\rm 3D,\star}=-\pi/4$ for the nonlinking $\B\parallel\hat x$ series.
  \item The extracted offsets should be stable against replacing the constant local quadrature amplitudes $C^{(w)}_{\nu,r},S^{(w)}_{\nu,r}$ by a weakly varying local envelope, for example a first-order polynomial in $u-u_w$. If the phase changes appreciably under this replacement, the chosen window is too wide or the background subtraction is inadequate.
\end{itemize}
These checks are the natural lattice counterpart of the continuum derivation: the frequencies are set by extremal sections, while the fitted offsets distinguish the $\B\parallel\hat z$ monopole-ring phase $\Phiphi=\pi$ from $\Phiphi=0$ and the $\B\parallel\hat x$ linking phase $\Phitheta=\pi$ from the nonlinking $\Phi=0$ series.

%%%%%%%%%%%%%%%%%%%%%%%%%%%%%%%%%%%%%%%%%%%%%%%%%%%%%%%%%%%%%%%%%%%%%%%%%%%%%%
\section{Unit conversion and graphdiyne frequency estimates}
\label{sec:units}

The continuum formulas in the main text are often written in units $\hbar=e=v=1$.
This section restores SI units and makes the graphdiyne estimates more explicit.

\subsection{Onsager conversion to tesla}

For an extremal orbit of area $A$ in momentum space (units m$^{-2}$), the Onsager frequency is
\begin{equation}
F[\mathrm T]=\frac{\hbar}{2\pi e}A.
\label{eq:Fsi}
\end{equation}
For the toroidal $\B\parallel\hat z$ orbits,
\begin{equation}
A_\pm=\pi(k_0\pm k_F)^2,
\qquad
F_\pm=\frac{\hbar}{2e}(k_0\pm k_F)^2.
\label{eq:Fpm-SI}
\end{equation}
Here $k_0$ is the nodal-ring radius.
The quantity $k_F$ is the radial Fermi-surface minor radius, i.e., the Fermi wave number measured from the ring.
For a linearized dispersion $E=\hbar v\sqrt{k_z^2+(k_\rho-k_0)^2}$,
\begin{equation}
k_F=\frac{\EF}{\hbar v}.
\label{eq:deltak}
\end{equation}

\subsection{ABC-stacked graphdiyne}

Using the parameters quoted in the paper and in Ref.~\cite{nomura2018},
\begin{equation}
k_0\simeq 2.02\times10^8\;\mathrm{m}^{-1}\quad (0.0202\,\text{\AA}^{-1}),
\qquad
v\simeq 7.0\times10^5\;\mathrm{m/s},
\label{eq:graphdiyne-params}
\end{equation}
Eq.~\eqref{eq:deltak} gives, at $\EF=30$ meV,
\begin{equation}
k_F\simeq
\frac{30\times10^{-3}\times1.602\times10^{-19}\;\mathrm J}
{(1.055\times10^{-34}\;\mathrm{J\,s})(7.0\times10^5\;\mathrm{m/s})}
\simeq 6.5\times10^7\;\mathrm{m}^{-1}.
\label{eq:deltak-30meV}
\end{equation}
Substituting into Eq.~\eqref{eq:Fpm-SI} yields frequencies of order tens of tesla.
For convenience, Table~\ref{tab:graphdiyne} shows the resulting $F_\pm$ values for several representative dopings.

\begin{table}[t]
\caption{Representative $\B\parallel\hat z$ toroidal frequencies for ABC-stacked graphdiyne using Eqs.~\eqref{eq:Fpm-SI} and \eqref{eq:deltak} with $k_0=0.0202$ \AA$^{-1}$ and $v=7.0\times10^5$ m/s.}
\label{tab:graphdiyne}
\begin{ruledtabular}
\begin{tabular}{c c c c}
$\EF$ (meV) & $k_F$ ($10^7$ m$^{-1}$) & $F_+$ (T) & $F_-$ (T) \\
\colrule
20 & 4.34 & 19.82 & 8.28 \\
30 & 6.51 & 23.48 & 6.17 \\
40 & 8.68 & 27.45 & 4.37 \\
\end{tabular}
\end{ruledtabular}
\end{table}

At $\EF=30$ meV, the same estimate gives
\begin{equation}
F_+=23.48\;\mathrm T,
\qquad
F_-=6.17\;\mathrm T,
\end{equation}
which rounds to $F_+\simeq23.5$ T and $F_-\simeq6.2$ T as quoted in the paper.
The table also makes clear that the inner and outer frequencies move in opposite directions with doping, which is useful for experimental cross-checks.

\subsection{Graphdiyne reference scale used for the numerical figures}
\label{subsec:graphdiyne-figure-units}

The lattice-Kubo figures in the paper are dimensionless calculations with
$m_0=1$ and $\EF=0.2$.
To display the field and frequency axes in SI units, we use the graphdiyne
ring radius as a reference scale,
\begin{equation}
k_0=0.0202~\text{\AA}^{-1},
\qquad
B_0=\frac{\hbar k_0^2}{e}\simeq 26.9~{\rm T}.
\label{eq:graphdiyne-B0}
\end{equation}
Thus a dimensionless frequency $F$ in the numerical model is plotted as
\begin{equation}
F_{\rm SI}=B_0F,
\label{eq:graphdiyne-F-convert}
\end{equation}
and a dimensionless inverse field $u=1/B$ is plotted as
\begin{equation}
B_{\rm SI}^{-1}=\frac{u}{B_0}.
\label{eq:graphdiyne-invB-convert}
\end{equation}
For rational-flux data this is $B_{\rm SI}^{-1}=q/(2\pi pB_0)$; therefore the plotted inverse-field coordinate is not the magnetic-cell denominator $q$ itself.
With the representative graphdiyne velocity
$v=7.0\times10^5~{\rm m/s}$, the same scale gives
\begin{equation}
E_0=\hbar v k_0\simeq 93.2~{\rm meV},
\qquad
\EF=0.2E_0\simeq18.6~{\rm meV}.
\label{eq:graphdiyne-E0}
\end{equation}
This conversion is used only to attach graphdiyne reference units to the
dimensionless lattice data; it is not a separate full-band graphdiyne
transport calculation.

\begin{table}[t]
\caption{Graphdiyne reference conversion for the dimensionless numerical
figures in the paper, using $B_0=\hbar k_0^2/e\simeq26.9$ T.}
\label{tab:graphdiyne-figure-units}
\begin{ruledtabular}
\begin{tabular}{c c c}
Quantity & Dimensionless value & Graphdiyne reference value \\
\colrule
$\EF$ & $0.2$ & $18.6$ meV \\
$F_-$ in paper Fig.~3 & $0.320$ & $8.6$ T \\
$F_+$ in paper Fig.~3 & $0.720$ & $19.4$ T \\
$F_0$ in paper Fig.~4 & $0.020$ & $0.54$ T \\
$F^\ast$ in paper Fig.~4 & $0.0893$ & $2.4$ T \\
$B_d$ & $0.028$ & $0.755$ T \\
\end{tabular}
\end{ruledtabular}
\end{table}

The damping parameter used for paper Fig.~3 is a model damping field,
entering the oscillatory amplitude as $\exp[-\nu B_d/B]$.
After the graphdiyne reference conversion, it corresponds to
$B_D=B_0B_d\simeq0.755$ T.
The equivalent Dingle temperature is branch dependent because
$B_D=2\pi^2k_BT_Dm_c^{\rm phys}/(e\hbar)$.
For the two $\B\parallel\hat z$ toroidal branches of the numerical model,
the conversion gives values of order $1$--$2$ K, specifically about
$1.3$ K for the outer branch and $1.9$ K for the inner branch.
We therefore quote $B_d$ or $B_D$ as the primary damping parameter for
the numerical figures, and quote $T_D$ only after specifying the
graphdiyne reference mass.

%%%%%%%%%%%%%%%%%%%%%%%%%%%%%%%%%%%%%%%%%%%%%%%%%%%%%%%%%%%%%%%%%%%%%%%%%%%%%%
\section{Finite spin-orbit coupling, geometric versus LK phase, and observability criteria}
\label{sec:soc}

The main paper notes that weak spin-orbit coupling (SOC) gaps both the semimetallic ring and the occupied-band thread, and that the geometric Berry phase of a local gapped orbit then deviates from $\pi$ by corrections of order $\Delta_{\rm SOC}/\EF$.
For measured quantum oscillations, however, this geometric Berry phase is not by itself the full LK phase.
Orbital-magnetic-moment, Zeeman, and possible non-Abelian spinful corrections can enter at the same order.
Because this distinction is central to any candidate-material assessment, here we derive the Berry-only massive-Dirac contribution and then separate it from the measured LK offset.
Throughout this section $\Delta_{\rm SOC}>0$ denotes the local Dirac mass on the orbit under consideration, i.e., one half of the local SOC-induced band gap.

\subsection{Local massive-Dirac reduction around a weakly gapped orbit}
\label{subsec:soc-dirac}

Near any small patch of a nodal line, weak SOC converts the massless real crossing into a massive two-band problem of the form
\begin{equation}
H_{\rm loc}(\bm q)=\hbar v_1 q_1\sigma_x+\hbar v_2 q_2\sigma_y+\Delta_{\rm SOC}\sigma_z,
\label{eq:Hloc-soc}
\end{equation}
where $q_1$ and $q_2$ are momenta measured in the two directions transverse to the line and $v_{1,2}$ are the corresponding velocities.
The eigenvalues are
\begin{equation}
E_{\pm}(\bm q)=\pm\sqrt{\Delta_{\rm SOC}^2+\hbar^2 v_1^2 q_1^2+\hbar^2 v_2^2 q_2^2}.
\label{eq:Eloc-soc}
\end{equation}
At fixed positive energy $E>\Delta_{\rm SOC}$, the constant-energy contour is an ellipse,
\begin{equation}
\frac{q_1^2}{\bigl(E^2-\Delta_{\rm SOC}^2\bigr)/(\hbar^2 v_1^2)}+
\frac{q_2^2}{\bigl(E^2-\Delta_{\rm SOC}^2\bigr)/(\hbar^2 v_2^2)}=1.
\label{eq:ellipse-soc}
\end{equation}
Parametrizing this ellipse by
\begin{equation}
q_1=\frac{\sqrt{E^2-\Delta_{\rm SOC}^2}}{\hbar v_1}\cos\varphi,
\qquad
q_2=\frac{\sqrt{E^2-\Delta_{\rm SOC}^2}}{\hbar v_2}\sin\varphi,
\label{eq:ellipse-param}
\end{equation}
the corresponding $\bm d$-vector of Eq.~\eqref{eq:Hloc-soc} sweeps a latitude circle on the Bloch sphere with fixed polar angle determined by
\begin{equation}
\cos\vartheta=\frac{\Delta_{\rm SOC}}{E}.
\label{eq:costheta-soc}
\end{equation}
For the upper band, the Berry connection along the orbit is the usual spin-\(1/2\) connection
\begin{equation}
\mathcal A_\varphi=\frac{1-\cos\vartheta}{2}.
\label{eq:Aphi-soc}
\end{equation}
Integrating around the closed orbit gives the Berry phase magnitude
\begin{equation}
\Phib(E)=\oint d\varphi\,\mathcal A_\varphi
=\pi\left(1-\frac{\Delta_{\rm SOC}}{E}\right).
\label{eq:Berry-massive-Dirac}
\end{equation}
This is the explicit local form of the Berry-only statement quoted in Ref.~\cite{li2018}: finite SOC moves a formerly quantized $\pi$ geometric Berry phase away from $\pi$ by a correction of order $\Delta_{\rm SOC}/E$.
If one retained only this geometric Berry phase in the Onsager phase, the corresponding Berry-only contribution to the offset would be
\begin{equation}
\gamma_B(E)\equiv\frac12-\frac{\Phib(E)}{2\pi}=\frac{\Delta_{\rm SOC}}{2E}.
\label{eq:gamma-berry-only}
\end{equation}
It is tempting to read Eq.~\eqref{eq:gamma-berry-only} as the measured LK or Landau-fan offset.
It is not: in a generic spinful band, the semiclassical quantization also contains the orbital magnetic moment, Zeeman coupling, and possible non-Abelian band-mixing terms~\cite{xiao2010berry,fuchs2010topological}.
The pure massive-Dirac model in Eq.~\eqref{eq:Hloc-soc} is the simplest way to see the distinction.
For a magnetic field perpendicular to the local orbit plane, its exact Landau levels are
\begin{equation}
E_n^2=\Delta_{\rm SOC}^2+2n\,e\hbar |B|\,v_1v_2,
\qquad n=1,2,\ldots ,
\label{eq:massive-dirac-LL}
\end{equation}
where $e>0$.
Combining Eq.~\eqref{eq:massive-dirac-LL} with the orbit area derived in the next subsection gives
\begin{equation}
\frac{\hbar A(E,\Delta_{\rm SOC})}{2\pi e |B|}=n.
\label{eq:pure-dirac-gamma-zero}
\end{equation}
Thus the full LK offset of the isolated massive-Dirac problem remains $\gamma_{\rm LK}=0$ for the nonzero Landau levels.
Semiclassically, the linear Berry-only shift in Eq.~\eqref{eq:gamma-berry-only} is canceled by the orbital-magnetic-moment contribution, with the same orientation convention.
Measured linear phase shifts in an SOC-gapped material therefore require physics beyond the isolated particle-hole-symmetric two-band Dirac reduction, such as remote-band corrections, quadratic terms, Zeeman splitting, or non-Abelian spinful mixing.
For the present material screen, $\Delta_{\rm SOC}/E\ll1$ remains the necessary condition that the local orbit is a weakly gapped descendant of the spinless nodal-line orbit, but it is not by itself a formula for the measured phase error.

\subsection{Orbit area, cyclotron mass, and Landau-level spacing in the weak-SOC regime}
\label{subsec:soc-area-mass}

Equation~\eqref{eq:ellipse-soc} gives the $k$-space area enclosed by the gapped orbit directly:
\begin{equation}
A(E,\Delta_{\rm SOC})=\pi\frac{E^2-\Delta_{\rm SOC}^2}{\hbar^2 v_1 v_2},
\qquad E>\Delta_{\rm SOC}.
\label{eq:A-massive-Dirac}
\end{equation}
Hence the Onsager frequency is
\begin{equation}
F(E,\Delta_{\rm SOC})=\frac{\hbar}{2\pi e}A(E,\Delta_{\rm SOC})
=\frac{E^2-\Delta_{\rm SOC}^2}{2e\hbar v_1 v_2}.
\label{eq:F-massive-Dirac}
\end{equation}
Compared with the massless result, finite SOC reduces the frequency only at second order in $\Delta_{\rm SOC}/E$.
By contrast, the geometric Berry phase in Eq.~\eqref{eq:Berry-massive-Dirac} changes linearly in $\Delta_{\rm SOC}/E$.
As discussed above, this Berry-only linear change is not automatically the measured LK offset because orbital-moment and spinful corrections enter the full semiclassical phase.

The corresponding cyclotron mass is
\begin{equation}
m_c(E)=\frac{\hbar^2}{2\pi}\pdv{A}{E}=\frac{E}{v_1 v_2},
\label{eq:mc-massive-Dirac}
\end{equation}
which reduces to $m_c=E/v^2$ in the isotropic case $v_1=v_2=v$.
The semiclassical Landau-level spacing is then
\begin{equation}
\hbar\omega_c(E,B)=\frac{\hbar e B}{m_c(E)}=\frac{\hbar eB v_1 v_2}{E},
\label{eq:hwc-general}
\end{equation}
and in the isotropic case
\begin{equation}
\hbar\omega_c(E,B)=\frac{\hbar eBv^2}{E}.
\label{eq:hwc-isotropic}
\end{equation}
Equations~\eqref{eq:criterion-orbit-exists}--\eqref{eq:criterion-lk-phase} are used for different parts of the material screen.
Equation~\eqref{eq:criterion-orbit-exists} checks only that the local gapped orbit exists:
\begin{align}
\frac{\Delta_{\rm SOC}}{E} &< 1. \label{eq:criterion-orbit-exists}
\end{align}
Equation~\eqref{eq:criterion-frequency} is the stronger requirement for using the massless orbit area as a frequency proxy, since Eq.~\eqref{eq:F-massive-Dirac} changes the area only at order $(\Delta_{\rm SOC}/E)^2$:
\begin{equation}
\left(\frac{\Delta_{\rm SOC}}{E}\right)^2\ll 1.
\label{eq:criterion-frequency}
\end{equation}
Equation~\eqref{eq:criterion-lk-phase} concerns the measured LK offset itself:
\begin{equation}
\left|\delta\gamma_{\rm remote}+\delta\gamma_Z+\delta\gamma_{\rm NA}+\cdots\right|\ll 1,
\label{eq:criterion-lk-phase}
\end{equation}
where the terms denote remote-band, Zeeman, and non-Abelian spinful corrections.
The ratio $\Delta_{\rm SOC}/\hbar\omega_c$ is kept only as a finite-field scale comparison.
A value that is not very small does not by itself mean that Landau quantization is lost, and it is not a formula for the LK phase error.

\subsection{Oscillation-count criterion and low-field damping in a finite field window}
\label{subsec:oscillation-count}

For a finite field window, the relevant count is the number of periods accumulated in inverse field,
\begin{equation}
N_{{\rm osc},\alpha}=F_\alpha\left(\frac{1}{B_{\min}}-\frac{1}{B_{\max}}\right).
\label{eq:Nosc}
\end{equation}
In the estimates below we use $N_{{\rm osc},\alpha}\gtrsim 5$ as a working target for a Landau-fan fit.
The actual requirement depends on the background subtraction, nearby frequencies, and the noise level.
No condition $B>F$ follows from this count.

The low-field cutoff is controlled by damping.
With
\begin{equation}
R_T=\frac{X}{\sinh X},
\qquad
X=\frac{2\pi^2 k_B T m_c}{\hbar eB},
\qquad
R_D=\exp\left[-\frac{2\pi^2 k_B T_D m_c}{\hbar eB}\right],
\label{eq:RT-RD}
\end{equation}
the field dependence enters through $m_c/B$.
For this reason the material tables keep $F_\alpha$, $m_c$, $\Delta_{\rm SOC}/E_F$, and $\Delta_{\rm SOC}/\hbar\omega_c$ as separate entries.

%%%%%%%%%%%%%%%%%%%%%%%%%%%%%%%%%%%%%%%%%%%%%%%%%%%%%%%%%%%%%%%%%%%%%%%%%%%%%%
\section{Material estimates and screening tables}
\label{sec:master}

The tables record the inputs used for the estimates quoted in the paper.
For graphdiyne, $k_0$ and $v$ come from the fitted three-dimensional model.
For the Si/Ge structures, the ring radii are read from the candidate figures, while $v$ and $\Delta_{\rm SOC}$ are monolayer proxies.
For the TMD rows, the SOC scale is only a family-level proxy.
When $\Delta_{\rm SOC}$ is entered in this proxy sense, it is not a computed gap on the relevant three-dimensional ring and is not used as a direct LK phase shift.

\subsection{General formulas used in the material screening tables}
\label{subsec:master-formulas}

The screening estimates use the same ring-axis torus geometry as the continuum calculation, with SI units restored.
For a nodal ring of radius $k_0$ and transverse velocity $v$,
\begin{equation}
k_F=\frac{E_F}{\hbar v}
\label{eq:kF-master}
\end{equation}
is the minor radius of the doped torus.
The two ring-axis extremal radii are
\begin{equation}
k_\pm=k_0\pm k_F,
\end{equation}
and the corresponding Onsager frequencies and cyclotron masses are
\begin{equation}
F_\pm=\frac{\hbar}{2e}(k_0\pm k_F)^2,
\label{eq:Fpm-master}
\end{equation}
\begin{equation}
m_{c,\pm}=\frac{\hbar(k_0\pm k_F)}{v}.
\label{eq:mcpm-master}
\end{equation}
These are the ring-axis quantities tabulated below.
Full-band geometry or SOC corrections are included only where candidate-specific data are available.

For the in-plane geometry used in the estimates of the LK damping scales, the linking orbit at $k_x=0$ has
\begin{equation}
F_0=\frac{\hbar k_F^2}{2e},
\label{eq:F0-master}
\end{equation}
and the nonlinking maximal family uses
\begin{equation}
F_\star\simeq \frac{\hbar K}{2\pi e}\sqrt{k_0 k_F^3},
\qquad K\simeq 6.19.
\label{eq:Fstar-master}
\end{equation}

\subsection{ABC-stacked graphdiyne: derivation of \texorpdfstring{$k_0$}{k0}, full-model Lifshitz scale, and corrected low-field estimates}
\label{subsec:graphdiyne-master}

ABC-stacked graphdiyne is at present the best-controlled electronic platform because all relevant parameters entering the weak-SOC diagnostic can be traced to the DFT-fitted model of Ref.~\cite{nomura2018}, while the SOC scale is suppressed by the light carbon atomic number and by the approximate mirror decoupling between $\pi$ and $\sigma$ sectors~\cite{huertas2006,min2006,nomura2018}.
The model Hamiltonian can be cast in the form of Eq.~\eqref{eq:Hlat} by a basis rotation:
\begin{gather}
H_{\rm ABC}({\bk})=
u_0\cos(k_zd)
+(\hbar vk_x-u_2\sin(k_zd)\,t_x)s_x
+\hbar vk_y\,t_y s_y
+m(k_z)\,s_z,\notag\\
m(k_z)=\Delta+u_1\cos(k_z d).
\end{gather}
The effective parameters quoted by Ref.~\cite{nomura2018} are
\begin{gather}
v=7.0\times 10^5\;\mathrm{m/s},\qquad
\Delta=0.213\;\mathrm{eV},\notag\\
u_0=0.125\;\mathrm{eV},\qquad
u_1=0.233\;\mathrm{eV},\qquad
u_2=0.230\;\mathrm{eV},
\label{eq:gdy-parameters}
\end{gather}
with interlayer spacing $d=3.29$ \AA.
The nodal ring occurs where the mass term vanishes,
\begin{equation}
m(k_z)=0
\quad\Rightarrow\quad
\cos(k_z^{\star} d)=-\frac{\Delta}{u_1}=-0.9142.
\label{eq:gdy-kzstar}
\end{equation}
The ring radius then follows from the transverse term,
\begin{equation}
\hbar v k_0=u_2\sin(k_z^{\star}d)=u_2\sqrt{1-\frac{\Delta^2}{u_1^2}},
\label{eq:gdy-k0-derive-1}
\end{equation}
which gives
\begin{equation}
k_0=\frac{u_2}{\hbar v}\sqrt{1-\frac{\Delta^2}{u_1^2}}
\simeq 0.0202\;\text{\AA}^{-1}.
\label{eq:gdy-k0-final}
\end{equation}
This is the value used in the graphdiyne tables below.

For the ABC-stacked graphdiyne estimates, the physical cyclotron masses entering the LK thermal, Dingle, and Zeeman factors are obtained by restoring units in Eqs.~\eqref{eq:mcpm-master}, \eqref{eq:F0-master}, and \eqref{eq:Fstar-master}. With $k_F=E_F/(\hbar v)$ they are
\begin{align}
\frac{m^{\rm GDY}_{c,\pm}}{m_e}
&=\frac{\hbar(k_0\pm k_F)}{v m_e},
\label{eq:gdy-mc-pm}\\
\frac{m^{\rm GDY}_{c,0}}{m_e}
&=\frac{E_F}{v^2 m_e},
\label{eq:gdy-mc-zero}\\
\frac{m^{\rm GDY}_{c,\star}}{m_e}
&=c_0\frac{\hbar\sqrt{k_0k_F}}{v m_e},
\qquad c_0=\frac{3K}{4\pi}\simeq1.48 .
\label{eq:gdy-mc-star}
\end{align}
The corresponding Zeeman spin-reduction factors for the $\nu$th harmonic are therefore explicitly
\begin{align}
R^{\pm}_{S,\nu}
&=\cos\!\left[\frac{\pi\nu g_\pm}{2}
\frac{\hbar(k_0\pm k_F)}{v m_e}\right],
\label{eq:gdy-RS-pm}\\
R^{0}_{S,\nu}
&=\cos\!\left[\frac{\pi\nu g_0}{2}
\frac{E_F}{v^2m_e}\right],
\label{eq:gdy-RS-zero}\\
R^{\star}_{S,\nu}
&=\cos\!\left[\frac{\pi\nu g_\star}{2}
 c_0\frac{\hbar\sqrt{k_0k_F}}{v m_e}\right].
\label{eq:gdy-RS-star}
\end{align}
Here the generic spin factor $g_s$ used above has been replaced by orbit-resolved effective values $g_\alpha$, which are material parameters. The spinless analytic and numerical models set the spin-reduction factors $R_{S,\nu}^{(\alpha)}$ to unity. If one uses the simple estimate $g_\alpha\simeq2$, the fundamental graphdiyne Zeeman factors remain close to one for the masses in Table~\ref{tab:graphdiyne-master}; for example at $E_F=8$ meV, $R^+_{S,1}\simeq0.9935$ and $R^-_{S,1}\simeq0.9954$.

The same DFT-fitted model also makes clear when the simple torus picture breaks down.
Because the full band energy contains the tilt term $u_0\cos(k_z d)$, the Fermi surface undergoes a Lifshitz-type reconstruction once the conduction-band minimum at $k_z=\pi/d$ reaches the nodal-ring energy.
The ring energy is
\begin{equation}
E_{\rm ring}=u_0\cos(k_z^{\star}d)=-u_0\frac{\Delta}{u_1}\simeq -114.3\;\mathrm{meV},
\label{eq:gdy-ering}
\end{equation}
whereas the $k_z=\pi/d$ conduction-band edge is
\begin{equation}
E_{\rm cond}(\pi/d)=-u_0+|\Delta-u_1|\simeq -105.0\;\mathrm{meV}.
\label{eq:gdy-econd}
\end{equation}
The corresponding offset above the ring is therefore
\begin{equation}
\delta E_{\rm hg}=E_{\rm cond}(\pi/d)-E_{\rm ring}\simeq 9.3\;\mathrm{meV}.
\label{eq:gdy-hourglass-threshold}
\end{equation}
Below this scale the full model has a simple torus and both $F_+$ and $F_-$ are literal Lifshitz--Kosevich extrema.
Above this scale the full Fermi surface is distorted by the $u_0$ term and the smaller branch should be re-checked numerically in the full band structure.
For this reason, the continuum torus formulas remain the correct local benchmark for the ring-axis diagnostic, but the simplest full-model two-frequency regime is expected at small-to-moderate doping.

\begin{table*}[t]
\caption{Doping dependence of the parameters for ABC-stacked graphdiyne using $k_0=0.0202$ \AA$^{-1}$ and $v=7.0\times10^5$ m/s. The weak-SOC screening uses $\Delta_{\rm SOC}=0.01$ meV, representative of the carbon-based estimate~\protect\cite{huertas2006,min2006,nomura2018}. The oscillation counts use the window $[B_{\min},B_{\max}]=[1,45]$ T.}
\label{tab:graphdiyne-master}
\begin{center}
\scriptsize
\setlength{\tabcolsep}{4pt}
\begin{tabular}{c c c c c c c c c c c}
\toprule
$E_F$ (meV) & $k_F$ (\AA$^{-1}$) & $F_+$ (T) & $F_-$ (T) & $F_0$ (T) & $m_{c,+}/m_e$ & $m_{c,-}/m_e$ & $m_{c,0}/m_e$ & $m_{c,\star}/m_e$ & $N_+$ & $N_-$ \\
\midrule
3  & 0.00065 & 14.31 & 12.58 & 0.014 & 0.0345 & 0.0323 & 0.0011 & 0.0089 & 13.99 & 12.30 \\
5  & 0.00109 & 14.91 & 12.02 & 0.039 & 0.0352 & 0.0316 & 0.0018 & 0.0114 & 14.58 & 11.76 \\
7  & 0.00152 & 15.52 & 11.48 & 0.076 & 0.0359 & 0.0309 & 0.0025 & 0.0135 & 15.18 & 11.23 \\
8  & 0.00174 & 15.84 & 11.22 & 0.099 & 0.0363 & 0.0305 & 0.0029 & 0.0145 & 15.48 & 10.97 \\
9  & 0.00195 & 16.15 & 10.96 & 0.126 & 0.0366 & 0.0302 & 0.0032 & 0.0154 & 15.79 & 10.71 \\
10 & 0.00217 & 16.47 & 10.70 & 0.155 & 0.0370 & 0.0298 & 0.0036 & 0.0162 & 16.10 & 10.46 \\
15 & 0.00326 & 18.11 &  9.45 & 0.349 & 0.0388 & 0.0280 & 0.0054 & 0.0198 & 17.70 &  9.24 \\
20 & 0.00434 & 19.82 &  8.28 & 0.620 & 0.0406 & 0.0262 & 0.0072 & 0.0229 & 19.38 &  8.09 \\
25 & 0.00543 & 21.61 &  7.18 & 0.969 & 0.0424 & 0.0244 & 0.0090 & 0.0256 & 21.13 &  7.02 \\
28 & 0.00608 & 22.72 &  6.56 & 1.215 & 0.0435 & 0.0234 & 0.0101 & 0.0271 & 22.22 &  6.42 \\
30 & 0.00651 & 23.48 &  6.17 & 1.395 & 0.0442 & 0.0226 & 0.0108 & 0.0280 & 22.96 &  6.03 \\
\bottomrule
\end{tabular}
\end{center}
\end{table*}

The graphdiyne numbers illustrate two experimentally useful points.
First, the ring-axis frequencies remain in the standard high-field range across the weakly doped torus regime, while the smaller frequency still provides many oscillation periods in a broad inverse-field window.
Second, the SOC scales are very well separated.
For example, at $E_F=30$ meV one finds $m_{c,+}\simeq 0.044\,m_e$, $F_+\simeq 23.5$ T, and therefore
\begin{equation}
\hbar\omega_c(B=F_+)\simeq 61.5\;\mathrm{meV},
\qquad
\frac{\Delta_{\rm SOC}}{\hbar\omega_c}\simeq 1.6\times10^{-4}.
\label{eq:gdy-hwc-ratio}
\end{equation}
The same estimate gives $\Delta_{\rm SOC}/E_F\simeq 3.3\times10^{-4}$ and $(\Delta_{\rm SOC}/E_F)^2\simeq 1.1\times10^{-7}$.
Thus orbit existence, frequency stability, and spinless-descendant phase robustness are all satisfied by wide margins in graphdiyne.
As in Sec.~\ref{sec:soc}, the ratio $\Delta_{\rm SOC}/\hbar\omega_c$ serves only as a finite-field scale comparison, not as a direct formula for the LK phase error.

A more subtle question is the low-field cutoff set by thermal and Dingle damping.
Using the representative graphdiyne mass $m_c\simeq 0.036\,m_e$ together with Eq.~\eqref{eq:RT-RD}, one finds at $T=2$ K and $T_D=5$ K that
\begin{equation}
R_T(1\,\mathrm T)\simeq 0.84,
\qquad
R_T(2\,\mathrm T)\simeq 0.95,
\label{eq:gdy-RT-numbers}
\end{equation}
and
\begin{equation}
R_D(1\,\mathrm T)\simeq 0.07,
\qquad
R_D(2\,\mathrm T)\simeq 0.27,
\qquad
R_D(5\,\mathrm T)\simeq 0.59.
\label{eq:gdy-RD-numbers}
\end{equation}
Even with this low-field cutoff, graphdiyne remains favorable. For realistic Dingle temperatures, a few-tesla lower cutoff is the more conservative choice; a one-tesla window would require an exceptionally clean sample.

\subsection{Expanded wurtzite Si and Ge: plausible screening targets, but not yet material-specific predictions}
\label{subsec:sige-master}

Reference~\cite{li2025general} considerably broadened the list of candidates by proposing fourteen Si/Ge structures with $\mathbb Z_2$-monopole nodal lines.
For the present transport proposal, however, the numerical certainty is much lower than in graphdiyne.
The reason is that the currently available literature establishes the existence of the nodal rings in the expanded structures, but not yet a dedicated set of ring-specific $k_0$, $v$, and SOC values extracted from the fully relaxed three-dimensional weak-SOC band structure of each candidate.
In the present screening tables we therefore adopt the following hierarchy.
The ring radii $k_0$ are read off from Figs.~27--29 of Ref.~\cite{li2025general}, while the velocities and SOC scales are proxied from the better-established monolayer silicene/germanene values of Ref.~\cite{liu2011}.
This is sufficient to decide whether the family is intrinsically plausible, but not sufficient to claim a definitive material prediction.

Two further practical points should be kept separate.
First, the known hexagonal (wurtzite-like) phase of Si and the hexagonal 4H phase of Ge have been reported experimentally~\cite{jennings1976,kiefer2010}, so these systems are not purely abstract.
Second, the nodal-line phase discussed in Ref.~\cite{li2025general} requires weakened interlayer coupling relative to the standard crystal; one possible route is an intercalated or van der Waals heterostructure engineering strategy analogous in spirit to the silicene/h-BN platform of Ref.~\cite{wiggers2019}.
Thus the Si/Ge family is a realistic secondary target rather than an established platform.

\begin{table*}[t]
\caption{List of model parameters for the fourteen Si/Ge structures of Ref.~\protect\cite{li2025general}. The tabulated frequencies use $E_F=20$ meV and the oscillation counts use $[B_{\min},B_{\max}]=[5,45]$ T. Here SG denotes the space group, $(\mu,\nu)$ labels the topological indices used in Ref.~\protect\cite{li2025general}, $k_0$ is figure-estimated from the candidate paper, and $v$ and $\Delta_{\rm SOC}$ are proxied from monolayer silicene/germanene~\protect\cite{liu2011}. This table is therefore a screening table rather than a material-specific final prediction.}
\label{tab:sige-master}
\begin{center}
\scriptsize
\setlength{\tabcolsep}{4pt}
\begin{tabular}{c c c c c c c c c c c}
\toprule
Crystal & SG & $(\mu,\nu)$ & $k_0$ (\AA$^{-1}$) & $\hbar v k_0$ (meV) & $F_+$ (T) & $F_-$ (T) & $\Delta_{\rm SOC}/\hbar\omega_c$ & $N_+$ & $N_-$ & Element \\
\midrule
1 & 194 & 0,0 & 0.057 & 206 & 129 &  87 & 0.014 & 23 & 16 & Si \\
2 & 15  & 0,0 & 0.066 & 239 & 168 & 120 & 0.012 & 30 & 21 & Si \\
3 & 2   & 0,0 & 0.066 & 239 & 168 & 120 & 0.012 & 30 & 21 & Si \\
4 & 15  & 0,1 & 0.076 & 275 & 219 & 163 & 0.011 & 39 & 29 & Si \\
5 & 2   & 0,1 & 0.076 & 275 & 219 & 163 & 0.011 & 39 & 29 & Si \\
6 & 2   & 1,1 & 0.076 & 275 & 219 & 163 & 0.011 & 39 & 29 & Si \\
7 & 2   & 1,1 & 0.066 & 239 & 168 & 120 & 0.012 & 30 & 21 & Si \\
1 & 194 & 0,0 & 0.054 & 135 & 127 &  70 & 0.308 & 22 & 12 & Ge \\
2 & 15  & 0,0 & 0.054 & 135 & 127 &  70 & 0.308 & 22 & 12 & Ge \\
3 & 2   & 0,0 & 0.054 & 135 & 127 &  70 & 0.308 & 22 & 12 & Ge \\
4 & 15  & 0,1 & 0.063 & 158 & 166 & 100 & 0.269 & 29 & 18 & Ge \\
5 & 2   & 0,1 & 0.063 & 158 & 166 & 100 & 0.269 & 29 & 18 & Ge \\
6 & 2   & 1,1 & 0.063 & 158 & 166 & 100 & 0.269 & 29 & 18 & Ge \\
7 & 2   & 1,1 & 0.054 & 135 & 127 &  70 & 0.308 & 22 & 12 & Ge \\
\bottomrule
\end{tabular}
\end{center}
\end{table*}

With the proxy values used in Table~\ref{tab:sige-master}, the Si rows pass this screening level: the frequencies remain in a measurable field range, and the silicene SOC proxy is small compared with both $E_F$ and $\hbar\omega_c$.
This should not be read as a material-specific prediction.
The ring radius, velocity, and SOC gap still have to be extracted from the relaxed three-dimensional structures.

The Ge rows do not pass the same proxy test at $E_F=20$ meV.
Using $\Delta_{\rm SOC}=23.9$ meV from germanene gives $\Delta_{\rm SOC}>E_F$, so the weakly gapped local-orbit condition is already lost.
A smaller actual ring SOC, or a larger carrier density, would be needed before this family could be treated as controlled.

\subsection{Transition-metal dichalcogenides: family-level SOC screen and the special status of Cr-based compounds}
\label{subsec:tmd-master}

Table~\ref{tab:tmd-soc-filter} uses screening-level SOC scales.
Mo/W entries use monolayer spin splittings.
Cr entries use order-of-magnitude estimates.
The table has one purpose: decide whether a ring-resolved SOC calculation is worth doing for a TMD family.
\begin{table*}[t]
\caption{Screening-level SOC scales used for the TMD candidates. Mo/W entries use representative monolayer spin splittings; Cr entries use order-of-magnitude estimates.}
\label{tab:tmd-soc-filter}
\begin{center}
\scriptsize
\setlength{\tabcolsep}{5pt}
\begin{tabular}{c c c c c}
\toprule
Material & SOC proxy (meV) & Ring scale (meV) & SOC/ring & Screening verdict \\
\midrule
CrS$_2$  &  30 & 45 & 0.7 & uncertain / marginal \\
CrSe$_2$ &  40 & 45 & 0.9 & likely eliminated \\
CrTe$_2$ &  50 & 45 & 1.1 & eliminated \\
MoS$_2$  & 148 & 50 & 3.0 & eliminated \\
MoSe$_2$ & 186 & 50 & 3.7 & eliminated \\
MoTe$_2$ & 219 & 50 & 4.4 & eliminated \\
WS$_2$   & 426 & 50 & 8.5 & eliminated \\
WSe$_2$  & 466 & 50 & 9.3 & eliminated \\
WTe$_2$  & 484 & 50 & 9.7 & eliminated \\
\bottomrule
\end{tabular}
\end{center}
\end{table*}

CrS$_2$ inputs:
\[
k_0\simeq 0.041\;\mathrm{\AA}^{-1},\qquad
v\simeq 3.5\times10^5\;\mathrm{m/s},\qquad
E_F=3\text{--}10\;\mathrm{meV}.
\]

Eq.~\eqref{eq:Fpm-master}:
\[
F_+\simeq 59\text{--}68\;\mathrm T,
\qquad
F_-\simeq 52\text{--}44\;\mathrm T.
\]

The sign of the $F_-$ slope follows from $k_0-k_F$.
Frequency is acceptable.
The SOC scale is the weak point.
With $\Delta_{\rm SOC}=15$ meV,
\[
\frac{\Delta_{\rm SOC}}{\hbar\omega_c}\simeq 0.31\text{--}0.29.
\]

With $\Delta_{\rm SOC}=30$ meV,
\[
\frac{\Delta_{\rm SOC}}{\hbar\omega_c}\simeq 0.62\text{--}0.57.
\]

On this basis, CrS$_2$ stays only as a follow-up case for a ring-resolved SOC calculation.
The Mo/W rows are set aside for the weak-SOC SdH screen used here.

\subsection{Other candidates and scope boundaries}
\label{subsec:other-candidates-master}

% (moved from main)
The phase assignment used above assumes closed semiclassical cyclotron orbits.
Open sheets, saddle-point orbits, and magnetic-breakdown networks need their own orbit analysis before an Onsager offset is assigned.
If doping later produces closed pockets with extremal sections that link the relevant nodal structure, the Berry-phase bookkeeping used here can be applied to those pockets.

Layered polysilane and alternating-twist multilayer graphene are not entries in the electronic torus table.
In the polysilane case the proposal is phononic, so it does not supply an electronic Fermi surface for this magnetotransport test~\cite{li2025general}.
In alternating-twist multilayer graphene the $\mathbb Z_2$ charge belongs to Dirac points rather than to a nodal ring.
Neither case gives the paired ring-axis toroidal frequencies $F_+$ and $F_-$ used for the present diagnostic~\cite{qian2023stable}.

\subsection{Practical hierarchy of candidates for the present proposal}
\label{subsec:summary-master}

Table~\ref{tab:master-summary} is the end point of the screening arithmetic above.
A useful entry must have a closed toroidal pocket, enough periods in the chosen inverse-field window, and a small SOC scale on the same local orbit.

\begin{table*}[t]
\caption{Practical hierarchy of candidate platforms for the weak-SOC Shubnikov--de Haas diagnostic of a $\mathbb Z_2$ monopole nodal ring. Representative frequencies are the ring-axis toroidal frequencies used for the direct $\wtwo$ test.}
\label{tab:master-summary}
\centering
\scriptsize
\renewcommand{\arraystretch}{1.00}
\setlength{\tabcolsep}{2pt}
\begin{tabular*}{\textwidth}{@{\extracolsep{\fill}}>{\raggedright\arraybackslash}p{0.15\textwidth}>{\raggedright\arraybackslash}p{0.24\textwidth}>{\raggedright\arraybackslash}p{0.18\textwidth}>{\raggedright\arraybackslash}p{0.18\textwidth}>{\raggedright\arraybackslash}p{0.20\textwidth}@{}}
\toprule
Platform & Parameter status & Representative frequencies & Weak-SOC status & Assessment for the present proposal \\
\midrule
ABC-stacked graphdiyne &
DFT-fitted $k_0$ and $v$ from Ref.~\cite{nomura2018}; carbon SOC estimate from Refs.~\cite{huertas2006,min2006,nomura2018} &
$F_+\sim 14$--$23$ T and $F_-\sim 13$--$6$ T for $E_F=3$--$30$ meV &
$\Delta_{\rm SOC}/E_F\ll1$ and $\Delta_{\rm SOC}/\hbar\omega_c\sim10^{-4}$ &
Baseline platform in this screen \\
Expanded wurtzite Si family &
$k_0$ read from Ref.~\cite{li2025general}; $v$ and $\Delta_{\rm SOC}$ taken from silicene as proxies~\cite{liu2011} &
$F_+\sim 130$--$220$ T and $F_-\sim 87$--$163$ T at $E_F=20$ meV &
Proxy gives $\Delta_{\rm SOC}/E_F\sim10^{-2}$ &
Secondary case; needs structural realization and fitted three-dimensional $k_0$, $v$, and SOC \\
Expanded wurtzite Ge family &
Same geometric estimate as Si; SOC taken from germanene as a proxy~\cite{liu2011} &
$F_+\sim 127$--$166$ T and $F_-\sim 70$--$100$ T at $E_F=20$ meV &
The literal 23.9 meV proxy is larger than $E_F=20$ meV &
Use only if the actual ring SOC is much smaller than the monolayer proxy \\
Cr-based TMD subset (especially CrS$_2$) &
Rough $k_0$ and $v$; Cr SOC values are order-of-magnitude proxies &
CrS$_2$: $F_+\sim 59$--$68$ T and $F_-\sim 52$--$44$ T for $E_F=3$--$10$ meV &
Proxy SOC is comparable with the low-$E_F$ window &
Follow up only after a SOC-included ring calculation \\
Mo/W TMD families &
Heavy-element SOC dominates the monolayer and higher-order-topology literature~\cite{wang2019,zhu2011,kormanyos2015} &
No weak-SOC ring-frequency table used here &
Outside the controlled weak-SOC limit &
Excluded from the nearly spinless SdH test \\
SiH phononic / alternating-twist multilayer graphene &
Outside the electronic toroidal nodal-ring setting~\cite{li2025general,qian2023stable} &
Different oscillation problem &
Not applicable &
Nearby topology, but not this toroidal-ring proposal \\
\bottomrule
\end{tabular*}
\renewcommand{\arraystretch}{1.0}
\end{table*}

With the inputs available now, graphdiyne is the only case where the geometry and the weak-SOC scale both come from a material-specific model.
The Si rows remain a concrete calculation to do next.
The Ge and TMD rows do not yet give the same controlled LK-phase limit.

\clearpage

% Enable titles in references via apsrev4-2 CONTROL entry.
\bstctlcite{apsrev4-2Control}

\bibliographystyle{apsrev4-2}
\bibliography{sm_refs}